# Opportunities in Electrically Tunable 2D Materials Beyond Graphene: Recent Progress and Future Outlook

Tom Vincent[1§], Jiayun Liang[2§], Simrjit Singh[3§], Eli G. Castanon[1], Xiaotian Zhang[2], Amber McCreary[4], Deep Jariwala[3*], Olga Kazakova[1†], and Zakaria Y. Al Balushi[2‡]

[1] Department of Quantum Technology, National Physical Laboratory, Hampton Road, Teddington TW11 0LW, U.K.

[2] Department of Materials Science and Engineering, University of California, Berkeley, CA 94720, United States of America

[3] Department of Electrical and Systems Engineering, University of Pennsylvania, Philadelphia, PA 19104, United States of America

[4] Nanoscale Device Characterization Division, Physical Measurement Laboratory, National Institute of Standards and Technology, Gaithersburg, MD 20899, United States of America

**Abstract:**

The interest in two-dimensional and layered materials continues to expand, driven by the compelling properties of individual atomic layers that can be stacked and/or twisted into synthetic heterostructures. The plethora of electronic properties as well as the emergence of many different quasiparticles, including plasmons, polaritons, trions and excitons with large, tunable binding energies that all can be controlled and modulated through electrical means has given rise to many device applications. In addition, these materials exhibit both room-temperature spin and valley polarization, magnetism, superconductivity, piezoelectricity that are intricately dependent on the composition, crystal structure, stacking, twist angle, layer number and phases of these materials. Initial results on graphene exfoliated from single bulk crystals motivated the development of wide-area, high purity synthesis and heterojunctions with atomically clean interfaces. Now by opening this design space to new synthetic two-dimensional materials "beyond graphene", it is possible to explore uncharted opportunities in designing novel heterostructures for electrical tunable devices. To fully reveal the emerging functionalities and opportunities of these atomically thin materials in practical applications, this review highlights several representative and noteworthy research directions in the use of electrical means to tune these aforementioned physical and structural properties, with an emphasis on discussing major applications of beyond graphene 2D materials in tunable devices in the past few years and an outlook of what is to come in the next decade.

§ equal first author contributions

* dmj@seas.upenn.edu

† olga.kazakova@npl.co.uk

‡ albalushi@berkeley.edu

Review Contents

## I. INTRODUCTION

Ever since the successful isolation of a single sheet of graphene from bulk graphite in 2004[1], research into two-dimensional (2D) materials, where the interlayer interactions are governed by van der Waals (vdW) forces, has surged. Graphene, with its linear electronic dispersion, has been the most well-studied thus far, with significant progress towards commercialization[2,3]. Graphene, however, is just the tip of the iceberg in terms of vdW materials, where a plethora of materials have vdW bonding between the layers and thus can be isolated down to the atomically thin limit[4]. Depending on the composing elements, the crystal structure, and even relative angles between two layers, these materials can display a wide variety of electrical properties, including semi-metallic, metallic, semiconducting and insulating behavior. They even support exotic phenomena such as charge density waves and superconductivity[5–10]. Furthermore, magnetism at the monolayer limit has now been discovered[11,12], opening applications in nm-scale magneto-optoelectronic and spintronic devices[13,14]. The transition metal dichalcogenide family (TMD) ($MoS_2$, $WS_2$, $WSe_2$, etc.) has been the most well studied after graphene, but large strides are being made regarding other 2D materials as well.

    2D vdW materials display some truly fascinating properties that are inherently not possible with their thicker, three-dimensional (3D) counterparts. For one, it has been well-documented that the electrical, optical, and magnetic properties can dramatically change at the few-layer thickness regime, or even between even versus odd number of layers[15–18]. Additionally, 2D materials display extremely large mechanical flexibility, where large amounts of strain can be applied to tune their properties before rupture[19,20]. The vdW bonding between the layers allows for these materials to be grown or transferred onto a wide variety of substrates without the difficulty of epitaxial requirements, enabling easy integration with silicon technology that is technologically mature. Multi-layered heterostructures can be created with a variety of 2D materials, even with sub-degree control of the rotational alignment between the layers[9,21].

    Another immense benefit for using 2D materials in next-generation electronics and opto-electronics, and the focus of this review, is that they are able to be strongly tuned using electric fields in ways that are not possible for 3D materials due to their reduced dielectric screening[22]. Furthermore, the existence of

2D insulating materials such as few-layered h-BN allows researchers to bring materials within atomic-scale proximity to the backgate, allowing for extremely efficient tuning *via* electric fields. To fully reveal the emerging functionalities and opportunities of these atomically thin materials in practical applications, this review thoroughly details several representative and noteworthy research directions regarding the use of electrical means to tune both the physical and structural properties of layered vdW materials, with an emphasis on discussing the major advancements in 2D materials beyond graphene in the past 5 years and an outlook of what is to come in the next decade. A schematic overview of current and perspective applications of 2D materials beyond graphene is presented in Figure 1.

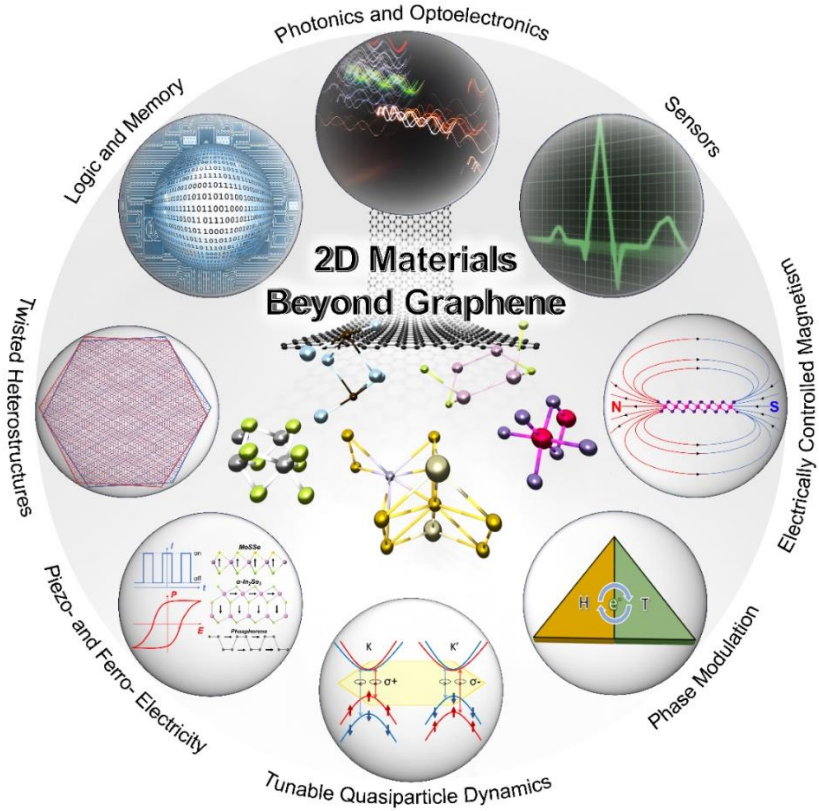

**FIG. 1.** Current and prospective applications of 2D materials beyond graphene

First, a broad overview of the materials beyond graphene is presented, detailing crystal structure, optical and electrical properties, and other relevant features that are important for the review. Next, we outline the current status of electrical tunability in materials beyond graphene, including in applications for logic and memory devices, photonics and optoelectronic devices, and sensors (gas, bio, and strain sensors). We then turn our attention to more emerging areas and application opportunities that are showing exciting potential. These novel research areas include tunable quasiparticle dynamics, electrically-controlled magnetism, piezo- and ferro-electricity, phase modulation, and electrically-tunable twisted heterostructures.

## II. MATERIALS BEYOND GRAPHENE

Since the first discovery of graphene, 2D vdW materials have attracted immense research interest thanks to their unique layered structures. As shown in Figure 2, common vdW materials can be divided into groups based on their crystal systems, structure, and tunable properties. Atoms are bonded with each other in each layer but the interactions between neighboring layers are mostly dominated by vdW forces. Thanks to the weak vdW interactions, pristine surfaces of atomic, ultrathin layers can be easily preserved through mechanical exfoliation and other synthesis methods. Weak vdW interactions also allow the stacking of different 2D vdW materials without the limitation of lattice matching, which provides researchers more degrees of freedom in the design of complex vdW heterostructures with desired functionalities. In addition, the surface properties become extremely significant due to the ultrathin geometry of vdW materials and can be tuned with surface modification.

| Crystal System | Material | Structure | | Space group | Tunable Electronic Property | Crystal System | Material | Structure | | Space group | Tunable Electronic Property |
|---|---|---|---|---|---|---|---|---|---|---|---|
| | | Side view | Top view | | | | | Side view | Top view | | |
| Hexagonal | 2H-TMDC (2H-MoTe$_2$, 2H-MoSe$_2$, 2H-MoS$_2$, 2H-WTe$_2$, etc.) + Janus Structure (MoSSe, etc.) | | | P6$_3$/mmc (#194) | Charge density, Exciton binding energy, Correlated states | Trigonal | 1T-TMDC (1T-MoS$_2$, etc.) | | | P$\bar{3}$m1 (#164) | Charge density |
| | h-boron nitride (h-BN) | | | P6$_3$/mmc (#194) | Charge density, Correlated states | | Cr$_2$Ge$_2$Te$_6$ | | | R$\bar{3}$ (#148) | Magnetization |
| | MXenes (Ti$_3$C$_2$, etc.) | | | P6$_3$/mmc (#194) | Polarization | Orthorhombic | g-C$_3$N$_4$ | | | Cmcm (#63) | Charge density |
| | Fe$_3$GeTe$_2$ | | | P6$_3$/mmc (#194) | Magnetization | | Black Phosphorus | | | Cmca (#64) | Charge density |
| | III-VI compound family (α-2H-In$_2$Se$_3$, etc.) | | | P6$_3$mc (#186) | Charge density | | Group-IV monochalcogenides (SnS, GeS, SnSe, GeSe, etc.) | | | Pnma (#62) | Polarization |
| | β-InSe | | | P6$_3$/mmc (#194) | Charge density | Monoclinic | 1T'-TMDC (1T'-MoS$_2$, 1T'-WTe$_2$, etc.) | | | P2$_1$/m (#11) | Charge density, Polarization |
| | Silicene/ Germanene | | | P6/mmm (#191) | Charge density | | CrI$_3$, CrBr$_3$ | | | C2/m (#12) | Magnetization Curie Temperature |
| Trigonal | III-VI compound family (α-3R-In$_2$Se$_3$, etc.) | | | R3m (#160) | Polarization | | CuInP$_2$S$_6$ | | | Cc (#9) | Polarization |

**FIG. 2.** Summary of 2D vdW materials mentioned in this work [23–34].

***Transition metal dichalcogenides (TMDs) and Janus structures*** are of the type $MX_2$ (M = Mo, W, *etc,*. and X = S, Se or Te) and for each monolayer, a layer of M-atoms is sandwiched between two layers of X atoms. While conventional TMDs possess the same chalcogen layer on the two sides of the transition metal layer, the chalcogen layers have two different chalcogen compositions in the Janus structure. Based on the varied stacking orders of M-atom and X-atom layers, TMDs can be categorized into 2H (ABA) and 1T (ABC) phase. The 1T phase can be further divided into the 1T' and the $T_d$ phase as a result of structural distortion.

***Hexagonal boron nitride (h-BN)*** is an insulator isostructural to graphene, where lattice points of the honeycomb structure are alternatively occupied by B-atom and N-atom.

***MXenes*** are layered structures resulting from selectively etching of the A atom layer of the parent phase MAX, where M is an early transition metal, A is a group 13 or 14 element and X is C or N.

***$Fe_3GeTe_2$*** consists of a $Fe_3Ge$ layer sandwiched between two Te layers. The $Fe_3Ge$ layer is composed of a hexagonal network formed by Fe atoms while the remaining Fe and Ge atoms are bonded to it covalently.

***III-VI compounds*** are of the type $A_2X_3$ or AX, where A and X are group 13 and 16 elements, respectively. Among them, $In_2Se_3$ is one of the most well-studied materials. It consists of a Se–In–Se–In–Se quintuple layer and the difference between its 2H- and 3R- phases originates from the stacking of the quintuple layers. Similarly, InSe consists of Se-In-In-Se quadruple layers. For its β-phase, the second quintuple layer is rotated by 60° around the [001] direction[34].

***Elemental 2D materials*** include silicene, germanene and black phosphorus. Silicene and germanene are analogous to graphene. The major difference is that silicene and germanene are buckled structures while graphene is a flat structure. Black phosphorus has a pleated honeycomb-like structure.

***$Cr_2Ge_2Te_6$*** consists of $CrTe_6$ octahedra arranged in a honeycomb-like structure, where each center of the honeycomb-like network is occupied by the Ge-Ge dimer.

***Graphitic carbon nitride*** has a monolayer form which consists of energetically preferred tri-s-triazine structure.

***Group-IV monochalcogenides*** are of the type AX, where A and X are group 14 and 16 elements, respectively. They have a highly-distorted rock-salt structure, where each A atom is placed in the distorted octahedron and bonded with six X atoms.

***Chromium trihalides*** are of the type $CrX_3$ (X=Cl/Br/I). For each monolayer, $CrX_6$ octahedra pairs are arranged in a honeycomb-like structure, where a pair of $CrX_6$ octahedra is formed by two $CrX_6$ octahedra sharing an edge with each other.

***Transition metal thiophosphates*** are denoted by the chemical formula $A^{II}B^{II}P_2X_6$ or $A^{I(III)}B^{III(I)}P_2X_6$, where both A and B are transition metals and X refers to group 16 elements. Among them, $CuInP_2S_6$ is one of the most important compounds. Its monolayer consists of a sulfur framework where the octahedral voids are filled by the Cu, In, and P-P triangular patterns.

## III. CURRENT STATUS IN ELECTRICAL TUNABILITY OF 2D MATERIALS BEYOND GRAPHENE

Strong electrostatic tunability is the highlight feature of all 2D semiconductor-based field-effect devices. This is also the principle reason for their suitability as a channel material in extremely scaled transistors[35,36]. The suitability of vdW 2D channels for post Si electronics or more than Moore electronics is well known and elaborately discussed in reviews exclusively focusing on transistor devices[35–37]. Here, we will focus on the following question: What does this electrostatic or gate tunability enable in 2D semiconductor devices that is unattainable or not demonstrated yet in other known semiconductors?

### A. Logic and memory devices
#### 1. *Gate tunable transport in diodes and switching devices*

In the context of electronic devices, the impact of electrostatic tunability of 2D materials can be looked at by first examining the simplest semiconductor device which is the p-n junction. When a p-n junction is comprised of atomically-thin semiconductors, it can be electrostatically tuned across both p and n layers that are vertically stacked resulting in unusual electrical characteristics. This was first reported on n-type 2D $MoS_2$ and p-type 1D carbon nanotube junctions[38]. Both a tunable rectification ratio of the diode and the first evidence of anti-ambipolar (Λ-shaped) transfer characteristics were demonstrated, with two OFF states at either extremes of gate bias and one ON state in the center [Figure 3(a)][38]. This observation has since been generalized to several other 2D-2D[39], 2D-organic[40], all-organic[41] and other unconventional, mixed-dimensional[42,43] heterojunction systems, emanating from strong electrostatic modulation in 2D materials-based device examples.

Numerous reports on various applications of such anti-ambipolar and tunable diode characteristics in 2D heterojunction devices have now been reported, as summarized in other reviews[44]. One of the most prominent applications is in tunable, multivalued logic. A ternary inverter with three logic states has been demonstrated using an anti-ambipolar heterojunction between small gap black phosphorus (BP) and larger gap $ReS_2$, which form a type III junction, connected in series with a p-type BP FET device[45] [Figure 3(b) and 3(c)]. Similar results have also been obtained in gated type II p-n heterostructures of $BP/MoS_2$[46] $MoS_2/WSe_2$[47] and $SnS_2/WSe_2$[48]. Anti-ambipolarity provides both positive and negative transconductance with a steep transition between them that can be tuned either actively with dual gates[49] [Figure 3(d)], as seen in $MoS_2$/BP self-aligned dual gated junctions, or passively with device geometry[40]. This allows frequency modulation[42], multiplication applications and opens room for constructing spiking neurons[49,50].

The above listed semiconductor combinations and device applications are promising for enhanced complementary metal–oxide–semiconductor (CMOS) or even neuromorphic and analog computing. However, all of them are comprised of materials that are still far from the maturity level of Si or III-Vs. Among 2D materials beyond graphene, only $MoS_2$, $WS_2$, $MoSe_2$, $WSe_2$, and h-BN have been synthesized at wafer-scale uniformity on substrates that are CMOS- compatible[51]. Therefore, a more practical way to take advantage of 2D materials in heterojunctions is to combine them with commercial-scale 3D materials such as Si and III-Vs. Along this direction, several recent breakthroughs have been made[43,52]. The earliest reports of gate-tunable 2D/3D heterojunctions include integration of graphene on Si to make gate-tunable Schottky diodes or barristors[53]. A key disadvantage of 2D/3D semiconductor junctions is that the 3D material does not allow significant electric field tunability or even a dual-gated structure. On the other hand, complementary doping type and density in 3D semiconductors is already well established and controlled, leading to high tunability of rectification and ON/OFF current ratios. Recent reports on junctions of $MoS_2$ with Si and GaN have established superior performance, as seen in Figure 3(e) and 3(f)[54]. The large rectification and resistance tunability suggest that these tunable heterojunctions can be dynamically configured to operate as both diode and transistors, renewing the opportunity to explore the long-abandoned concept of diode-transistor logic in a new way.

Finally, 2D/3D junctions are also an important research avenue for tunneling field effect transistors (T-FETs). The controlled degenerate doping in the 3D case combined with strong gate-tunability in the 2D layer makes a compelling case for T-FETs. However, all T-FET demonstrations in this regard thus far have been limited by low ON/OFF ratios for sub-thresholds < 60 mV/dec or low ON currents. To overcome this, several attributes are desired from the 2D/3D combination: (1) Near-intrinsic doping and clean semiconductors with large density of states at band edges, (2) clean and defect-free interfaces and (3) superior coupling to the gate with a gate dielectric that is both thin and has high breakdown strength. Meeting all these criteria is challenging and therefore presents plenty of space for exploration of material combinations and interface design/engineering. Likewise, there is also opportunity for more complex designs, such as p-n-p or n-p-n bipolar junction transistors. While some attempts have been made in this regard, the performance and properties remain far from optimal[55,56]. Controlled interfaces and control over doping once again will hold the key towards higher performance and more mature devices.

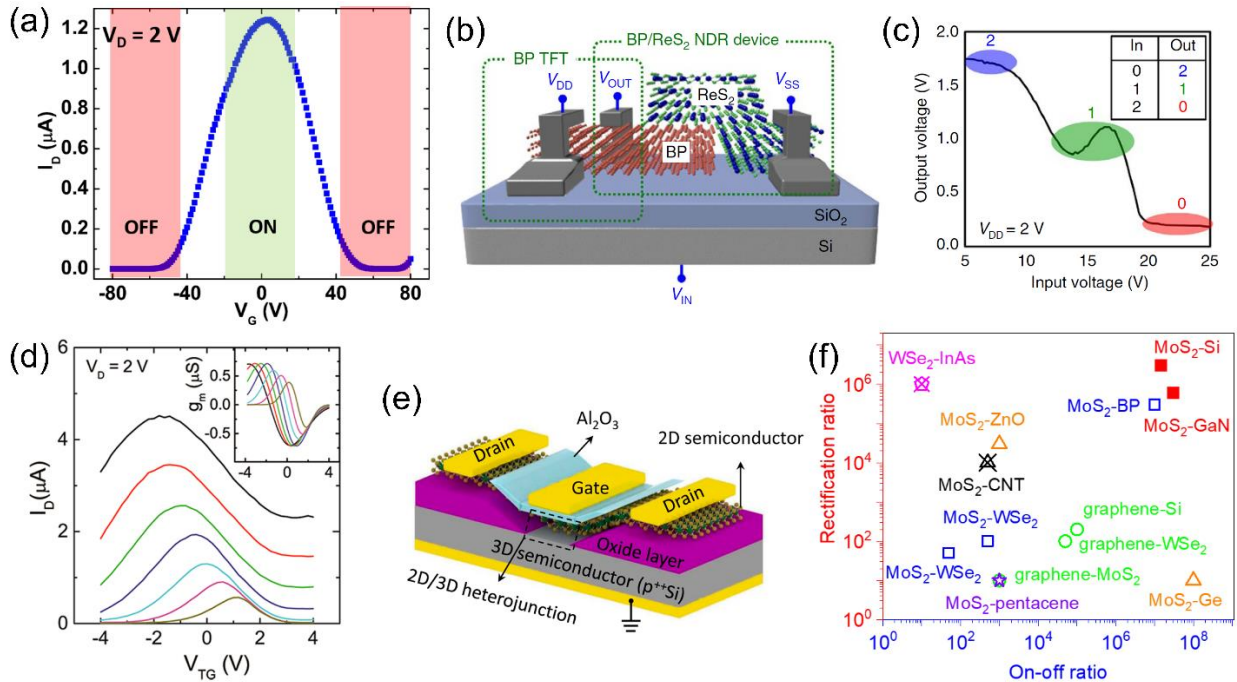

**FIG. 3.** (a) Anti-ambipolar transfer characteristic of a carbon nanotube-$MoS_2$ p-n heterojunction diode showing characteristic Λ-shaped curve. Adapted with permission from [38] (b) Schematic of an $ReS_2$-BP gated heterojunction diode in series with a p-type BP FET that forms a tri-state inverter. (c) Inverter voltage transfer characteristic showing three distinct logic states. Adapted with permission from [45]. Copyright © 2016. (d) Tunable anti-ambipolar transfer characteristics and transconductance (inset) of a dual gated $MoS_2$-BP heterojunction Adapted with permission from [49]. Copyright © 2018 American Chemical Society. (e) Schematic of a gated 2D($MoS_2$)/3D (Si) semiconductor heterojunction. (f) Comparison of various gate-tunable p-n heterojunction diodes comprising at least one 2D semiconductor on the metrics of ON-OFF ratio and rectification ratio both of which signify the degree of tunability Adapted with permission from [54]. Copyright © 2020 American Chemical Society

## 2. Gate-tunable memristive phenomena in atomically thin materials

As silicon-based, non-volatile memory (NVM) devices are anticipated to approach their fundamental scaling limits, new technologies and devices are continuously being investigated to meet ever-increasing demand of high-density data storage. Resistive random-access memory (RRAM) devices have emerged as potential candidates in the pursuit of cheaper and smaller NVM devices with high write/erase speed, greater endurance and retention.[57–59] Owing to their superiority over flash memory devices in terms of scalability without compromising on data storage, lower power consumption and CMOS compatibility, RRAM technology is foreseen to play a leading role in the future of memory industry. So far, the most widely used functional materials for RRAM are based on insulating transition metal oxides however, these materials are facing the bottleneck of scalability.[60–62] Therefore, extensive research efforts have been devoted in exploring new functional materials for RRAM devices. Two-dimensional vdW materials-based RRAMs have received tremendous research attention over the past few years,[63–65] with the advantage of highly-scalable memory cells with low power consumption and fast switching speed. Several reports have explored the use of graphene either as a highly conductive electrode material or as an active layer in RRAM based NVM devices.[66–68] Likewise, other 2D materials such as TMDs have also been used in order to improve the performance of NVMs. Here, we provide a comprehensive overview of the most recent advancements in NVM devices based on 2D materials beyond graphene with a focus on electrical tunability of memory device property.

Among 2D TMDs, $MoS_2$ is the most widely explored semiconductor as the channel material in NVM devices due to the combined effect of excellent mechanical flexibility and high charge-carrier mobility (> 20 $cm^2$ $V^{-1}$ $s^{-1}$ for $N \sim 10^{11}$ $cm^{-2}$).[69] Memristive phenomena in monolayer $MoS_2$-based lateral devices through the inclusion of grain boundaries (GBs) has been investigated.[70] Resistance switching ratios of $\sim 10^3$ at small bias and small set fields ($\sim 10^4$ $Vcm^{-1}$) [Figure 4(a)] were observed in devices with GBs connected to just one electrode. Owing to the atomically thin nature, these $MoS_2$ GB memristors exhibit gate-tunability of the set voltage from 3.5 to 8.0 V [Figure 4(b)]. The same authors have shown multi-terminal memtransistors using polycrystalline monolayered $MoS_2$ for complex neuromorphic computing operations.[71] These memtransistors also exhibit high cycling endurance (~ 475 cycles) and long retention times (24 hours). However, these devices rely on GBs with site-specific orientation, which is difficult to control. Recently, a focused helium ion beam was used for precisely introducing sulphur vacancy-related, site-specific defects into TMDs ($MoS_2$). The device exhibited high endurance (> 1180 cycles) and retention (for $>10^3$ s), as well as gate-tunable smaller set voltages.[72] Another recent study has shown a high gate tunability of the switching ratio from $\sim 10^0$ to $10^5$ using a few-layer $MoS_2$-based transistor.[73]

A quasi-non-volatile memory utilizing a semi-floating gate architecture was also used, achieving ultrahigh writing speeds of ~15 ns and refresh times of ~10 s, which is comparable to or higher than commercial dynamic random access memory [Figure 4(c)].[74]. The high writing performance was obtained due to the presence of two different charge transport mechanisms, e.g., switch path and flash memory path, in the device. Furthermore, the gate tunability of the device resulted in a high resistive switching ratio exceeding $10^3$ [Figure 4(d)]. In another report, a vertical stack of a $MoS_2$/h-BN/Graphene heterostructure exhibited a significantly low off-state current $\sim 10^{-14}$ A, resulting in a ultrahigh switching ratio over $10^9$[75].

Another prototype approach for obtaining tunable memory operation has been demonstrated by adding redox-active molecules on a 2D channel layer. By controlling molecular configurations through the gate voltage, carrier concentration in the 2D material could be modulated, resulting in the multistate memory operation.[76] The device exhibited a decent switching ratio of $\sim 10^3$ but showed relatively poor endurance

(~50 cycles) performance. Thus, further efforts to improve the retention and endurance of such kind of devices are required to be useful for memory applications.

### B. Ferroelectric Memristors

Recently, research in the area of NVM devices has focused on designing 2D ferroelectric field effect transistors (FeFETs), where a ferroelectric layer with high polarization field and dielectric constant is used as the gate dielectric material for producing efficient gating effect in the semiconductor channel layer.[77,78] Due to switchable electric dipoles and large retention properties, FeFETs are considered as the potential candidates for building NVM with low-power consumption and ultrafast logic operation. Earlier studies on 2D FeFETs were focused on using graphene as the conducting channel with bulk ferroelectric materials such as lead zirconate titanate (PZT) or organic poly(vinylidenefluoride-trifluoroethylene) (PVDF-TrFE) thin films whereby, the resistance of the conducting channel was effectively tuned by the reversible electrical polarization of the ferroelectric thin film.[68,79,80] In line with the conventional approach, integration of graphene channel with atomically thin 2D ferroelectric as the gate dielectric has also been demonstrated. The use of ultrathin ferroelectric materials increases the effective gate field, thereby reducing the required writing or erasing voltage to flip the electric polarization and leading to a low-power consumption memory device.[81,82]

The fabrication of a 2D FeFET consisting of graphene as the conducting channel and ferroelectric 2D α-$In_2Se_3$ as the top gate dielectric has been recently demonstrated, where the resistance states of the graphene channel were efficiently modulated by sweeping the ferroelectric gate voltage due to switching of the polarization direction in α-$In_2Se_3$ layer. The device exhibited excellent endurance for ~$10^5$ switching cycles and retention performance for ≈1000s.[81] Another recent study achieved a giant electroresistance switching ratio ≈$10^6$ through the modulation of Schottky barrier height and width *via* ferroelectric switching in large area α-$In_2Se_3$ grown on graphene using molecular beam epitaxy [Figure 4(e) and 4(f)].[82] Devices with semiconducting 2D materials have also been investigated. Ferroelectric memory transistors using monolayer to few layer $MoS_2$ as channel and P(VDF-TrFE) as the ferroelectric top gate insulator exhibited an ON/OFF ratio of ≈$10^3$ concurrently with retention properties for more than 1000 s.[83] A similar approach has been adopted for BP[77] and $MoSe_2$[84] FE-FETs, which showed similar or higher ON/OFF ratios, retention and endurance. However, slow dipole dynamics of ferroelectric polymers and low thermal durability in comparison to its inorganic counterparts can hamper the practical applicability of organic ferroelectric-based FETs. Thus, FeFETs integrating 2D materials with inorganic ferroelectric materials were also investigated. Non-volatile memory devices using mono- to few-layer TMDs ($WSe_2$ for p-type and $MoS_2$ for n-type) as the channels and epitaxial PZT thin film as the FE layer have been fabricated [Figure 4(g)].[85] The ferroelectric gate tunability of the device (3L-$WSe_2$) was demonstrated and an ON/OFF ratio was tuned from ≈$10^4$ to ≈$10^5$ with positive gate bias [Figure 4(h)]. Furthermore, the device exhibited endurance over 400 switching cycles and retention for up to $10^4$ s. Another recent work reported high-performance FE-FET using $MoS_2$ channel on top of AlScN dielectric and achieved high ON/OFF ratio of ≈$10^6$ and retention up to $10^4$ s.[86] A 2D/2D vdW heterostructure FeFETs for non-volatile ReRAM applications has also been designed, wherein 2D $CuInP_2S_6$ was integrated on top of the $MoS_2$ channel as a ferroelectric insulator.[87] The $MoS_2$/$CuInP_2S_6$ FeFET showed ON/OFF ratio of >$10^4$ *via* ferroelectric polarization switching-induced changes in the band alignment between the metal and $CuInP_2S_6$.

So far, the most intensively explored device concept for FeFETs consists of a ferroelectric gate dielectric integrated with a semiconductor channel layer. However, this conventional approach suffers from charge trapping and gate leakage issues, resulting in short retention times. Recently, a few 2D vdW materials have been shown to exhibit ferroelectricity as well as semiconducting properties, e.g., α-In$_2$Se$_3$[88] and CuInP$_2$S$_6$[89]. A few reports have realized FeFETs using ferroelectric 2D materials as the channel layer and the resistance states of the devices were controlled by ferroelectric polarization switching.[90,91] In a recent study, planar as well as vertical memristors using α-In$_2$Se$_3$ were investigated, demonstrating memristive phenomena based on both in-plane and out-of-plane polarization.[92] Both the devices exhibited a resistive switching ratio >10$^3$ due to modulation of the Schottky barrier height with ferroelectric gate bias and showed excellent endurance and retention properties for over 100 cycles and up to 1000 s, respectively. Another recent study has also observed similar performance (ON/OFF ratio >10$^3$) using a α-In$_2$Se$_3$ FeFET.[93] To provide an overview of the performance of different 2D materials based NVM devices, a comparison between the ON/OFF ratio and the retention time is plotted in Figure 5.

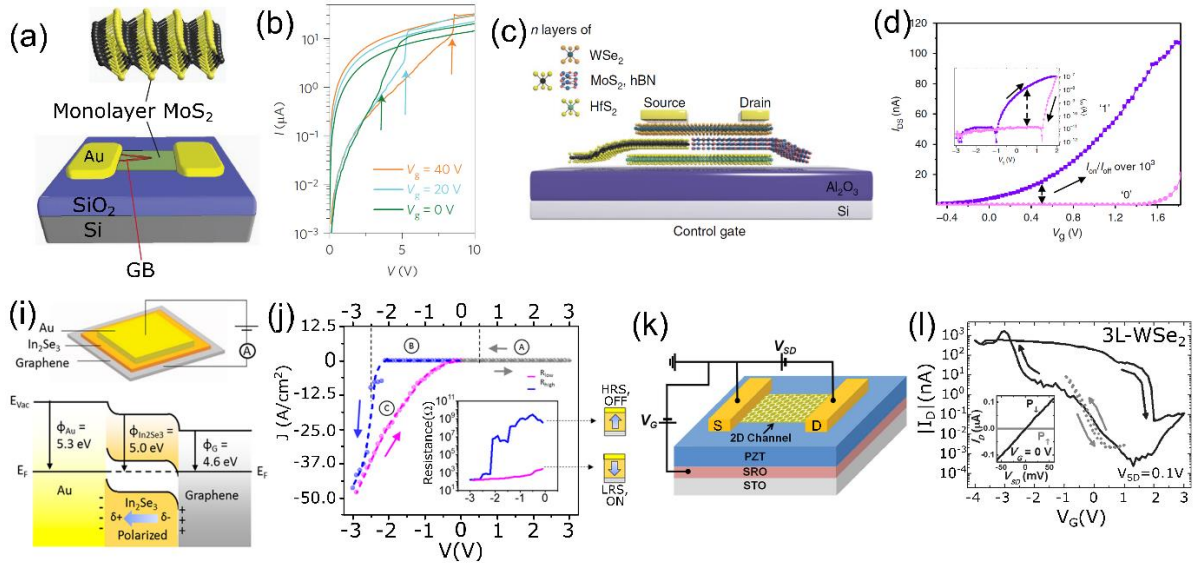

**FIG. 4.** (a) Schematic of an intersecting-GB MoS$_2$ memristor with two GBs connected to one of the electrodes. (b) Log-linear plot of the *I-V* characteristics of an intersecting-GB memristor at different gate voltage ($V_g$) which shows a shift in the set voltage with $V_g$. Adapted with permission from [70]. Copyright © 2015, Nature Publishing Group. (c) Schematic structure of the 2D semi floating gate memory. The WSe$_2$, h-BN and HfS$_2$ serve as the channel, blocking layer and floating gate, respectively. In addition, the inserted MoS$_2$ layer and the WSe$_2$ channel form the p–n-junction switch. (d) Transfer characteristics showing a shift in threshold voltage resulting in a large $I_{On}/I_{Off}$ ratio exceeding 10$^3$. Adapted with permission from [74] Copyright © 2018 (e) Schematic of the ferroelectric memory junction fabricated on the as-grown 6 nm α-In$_2$Se$_3$ and its corresponding band alignments at equilibrium. (f) *J-V* characteristics measured from the ferroelectric memory device. Showing the modulation of resistance state with ferroelectric polarization direction. Adapted with permission from [82] Copyright © 2018 American Chemical Society (g) Schematic of the 2D TMD/PZT heterostructure device with a circuit diagram for the back-gating. (h) Hysteresis loop of $I_D$-$V_G$, measured from representative p-type 3L-WSe$_2$ channel based FeFET. Adapted with permission from [85] CC-BY Wiley Company

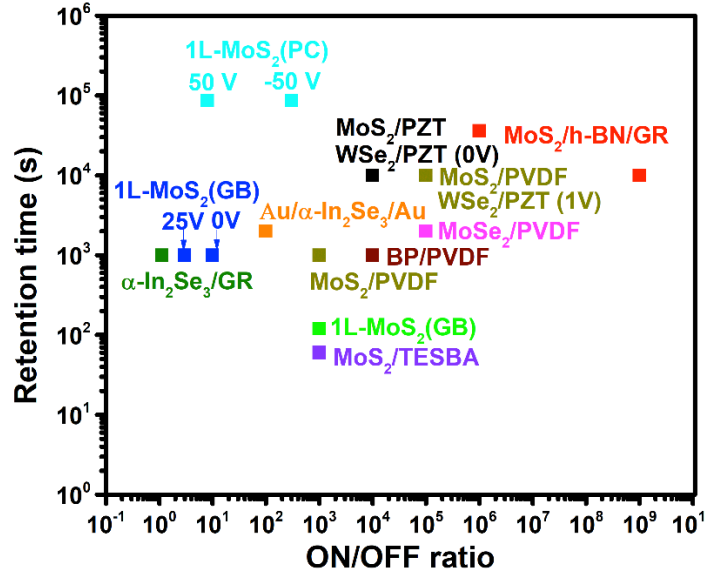

**FIG. 5.** Comparison between ON/OFF ratio and retention time of different 2D materials-based memory devices.[70,71,92,94,95,72,75,76,78,81,83–85] *Note: GB (Grain boundary), PC (Polycrystalline), GR (Graphene), PZT(Pb[Zr$_{0.2}$Ti$_{0.8}$]O$_3$), BP (Black phosphorus), TESBA (4-(Triethoxysilyl)butanal), PVDF (Polyvinylidene fluoride).

## C. Photonics and electro-optic devices
### 1. Light harvester

Harnessing solar energy using photovoltaic (PV) cells is by far the most promising strategy to fulfil the ever-increasing demand of renewable energy. PV cells using Si, III-V compounds and, very recently, organic/inorganic hybrid perovskites have been investigated and significant conversion efficiency has also been achieved.[96–98] However, the fabrication of these devices is associated with certain complexities, such as the requirement of epitaxial growth techniques, limited abundance of some elements and chemical stabilities, which raise concerns about their long-term economic viability. Recently, 2D TMDs, which have bandgaps of energies ranging from the ultraviolet to the near infrared (NIR) regions, have been considered as a new class of functional materials to potentially serve as thin absorbers for high performance PV cells, due to their high absorption coefficient and appropriate bandgaps in the bulk (1.3-1.4 eV) for attaining maximum power conversion efficiency per the Shockley-Quiesser (SQ) limit.[99] Two-dimensional TMDs are inherently flexible and the lack of dangling bonds at the surface encourages the creation of high-quality heterointerfaces, resulting in the low cost, flexible and light weight PV devices operating in a large region of the solar spectrum. PV cells based on 2D TMD heterostructures are also expected to exhibit high power conversion efficiency (PCE) as the maximum photovoltage is not limited by the built-in electrical potential energy, as is the case for conventional p-n junctions. The charge generation process in 2D TMDs is associated with strongly bound excitons instead of weakly bound electron-hole pairs. The exciton generation and dissociation occur simultaneously in a narrow region near the heterojunction, resulting in a large carrier concentration gradient that acts as a powerful PV driving force. The atomically thin nature offers a unique platform for achieving electrically tunable PV properties for high efficiency. The tuning of carrier density in active layers of a p-n junction also has strong implications for photodetectors. For instance, increasing (decreasing) of carrier density in p and/or n layers results in an increase (decrease) of built-in potential (depletion width) which, in turn, results in a decrease (increase) in the reverse saturation current, directly impacting the sensitivity and responsivity of the photodiode detectors. In this section, we highlight p-n junctions based on 2D semiconductors where electric-field induced tuning of the semiconductor induces modulation of electrical response of the junction to light absorption.

*Homojunctions*

PV energy conversion was first reported in lateral p-n homojunctions using several electrostatically-doped monolayered 2D materials such as $WSe_2$, $MoS_2$ and BP etc.[100–103] Biasing a pair of gate electrodes with opposite polarities, the carrier type and density was controlled in the channel, resulting in the formation of a p-n junction. The maximum PCE achieved using such junctions is ≈ 0.5%. However, it can be further enhanced to ~14 % under standard air mass (AM)-1.5 solar spectrum using multilayered (≈10 atomic layers) $MoSe_2$ crystals stacked onto dielectric h-BN.[104] Recently, using high-resolution angle-resolved photoemission spectroscopy, the surface PV effect has been observed in a $β$-InSe semiconductor that allows for further tuning by *in situ* surface potassium doping. This study can be further extended to engineer the photovoltaic effect in InSe-based p-n devices.[105] Furthermore, vertical p-n homojunctions have also been designed. A vertical homojunction using thin and thick flakes of $MoSe_2$ has been fabricated and a gate tunable PV effect was obtained due to different gate modulation levels of the carrier densities in these flakes. The PV effect as a function of gate voltage ($V_g$) showed a maximum $V_{oc}$ ≈ -0.24 V and PCE ≈1.9 % [Figure 6(a) and 6(b)].[106] Similarly, gate-tunable PV effect has also been demonstrated in a vertical homojunction of n-type $MoS_2$:Fe and p-type $MoS_2$:Nb few-layered flakes by changing the charge carrier density through the Si-gate electrode.[107]

*Vertical heterostructures*

In contrast to homojunctions, the vdW heterojunctions, made by stacking 2D materials of different bandgaps, are expected to exhibit enhanced PV effect. A type-II vdW heterojunction composed of $MoS_2$ and $WSe_2$ monolayers has been fabricated, with gate tunability of the $J_{sc}$ and $V_{oc}$ resulted in a PCE of ≈ 0.2 %.[108] In another work, external quantum efficiency (EQE) ≈ 2.4 % has been achieved using a monolayer $MoS_2$/$WSe_2$ heterojunction sandwiched in between graphene electrodes, which can be further improved by increasing the number of atomic layers in the $MoS_2$/$WSe_2$ heterojunction.[39] A polymeric gate was also utilized on top of $MoS_2$/InP to obtain an electrically-tunable PV effect, leading to maximum PCE of ≈ 7.1 %.[109] Recently, a vertical vdW heterostructure consisting of multilayer InSe and Te has been fabricated, exhibiting a PV effect as well as a broadband photoresponse with an ultrahigh photo/dark current ratio exceeding $10^4$ and a high detectivity of ≈$10^{12}$ Jones under visible light illumination [Figure 6(c)][110]. In a vertical vdW p-n junction between few-layer p-BP and n-InSe, the suitable band alignment as well as the intrinsically high carrier mobilities in BP and InSe resulted in an EQE as high as 3%, indicating an efficient separation of the photogenerated charge carriers. The polarization-dependent photoresponse of the device was also investigated through scanning photocurrent microscopy which showed a substantially higher polarization sensitivity (photocurrent anisotropy ratio ≈ 0.83) than those of traditional BP photodetectors.[111]

To scale up the process, CVD-grown large area vertical heterostructures have also been employed. A gate-tunable PV effect has been obtained in CVD-grown $WS_2$/$MoS_2$ and $MoS_2$/$WS_2$ vertical heterostructures sandwiched between graphene electrodes.[112] The $I_{sc}$ and $V_{oc}$ increased monotonously from 30 nA and 0.7 mV to 120 nA and 2.4 mV, respectively, by varying the gate voltage. However, this enhancement is smaller than what has been reported for mechanically exfoliated samples.

In another recent report, high performance PV effect was demonstrated in WSe$_2$/WS$_2$ heterostructures due to effective modulation of the junction transport properties as a function of gate voltage, due to the ambipolar nature of WSe$_2$.[113] where a significant modulation of the output electrical power $P_{output}$ of the device with gate voltage was achieved [Figure 6(d)]. The heterostructure exhibited a maximum $V_{oc}$ of ≈ 0.58 V and PCE of ≈ 2.4%. The modulation of the PV properties was correlated to the combined effect of channel conductivity and quasi-Fermi level tuning at the interface. Theoretically, it has been further proposed that a high PCE ≈ 30.7% can be obtained using a dual-gated semiconducting 2H phase WTe$_2$/MoSe$_2$ vdW heterostructure.[114] By adjusting the dual-gate voltages, the photocurrents in the two subcells can be matched, leading to the tandem cell operation of the device.

Another research area is the study of the PV effect in 2D/3D heterostructures, which can theoretically absorb more of the incident light. By varying the gate voltage through an ionic polymer top gate in a MoS$_2$/GaAs heterostructure, the PCE was improved from 6.87 % to 9.03 %[115], where the improved barrier height $\Phi_{barrier}$ due to the shift in the Fermi level of MoS$_2$ under electrical gating led to the increase [Figure 6(e) and 6(f)]. Similarly, MoS$_2$/Si and MoS$_2$/GaN 2D/3D semiconductor heterojunctions have been fabricated for switching and rectification applications. By tuning the Fermi levels of MoS$_2$ via electrical gating, the devices exhibited over 7 orders of magnitude modulation in the rectification ratio and an ON/OFF ratio exceeding $10^7$.[54] In another recent study, a large lateral photovoltaic effect (LPVE) with ultrafast relaxation time (~2 μs) in a SnSe/p-Si junction has been reported. The diffusion of electrons laterally in the inversion layer formed at the SnSe/p-Si interface resulted in the large LPVE with ultrafast relaxation time.[116]

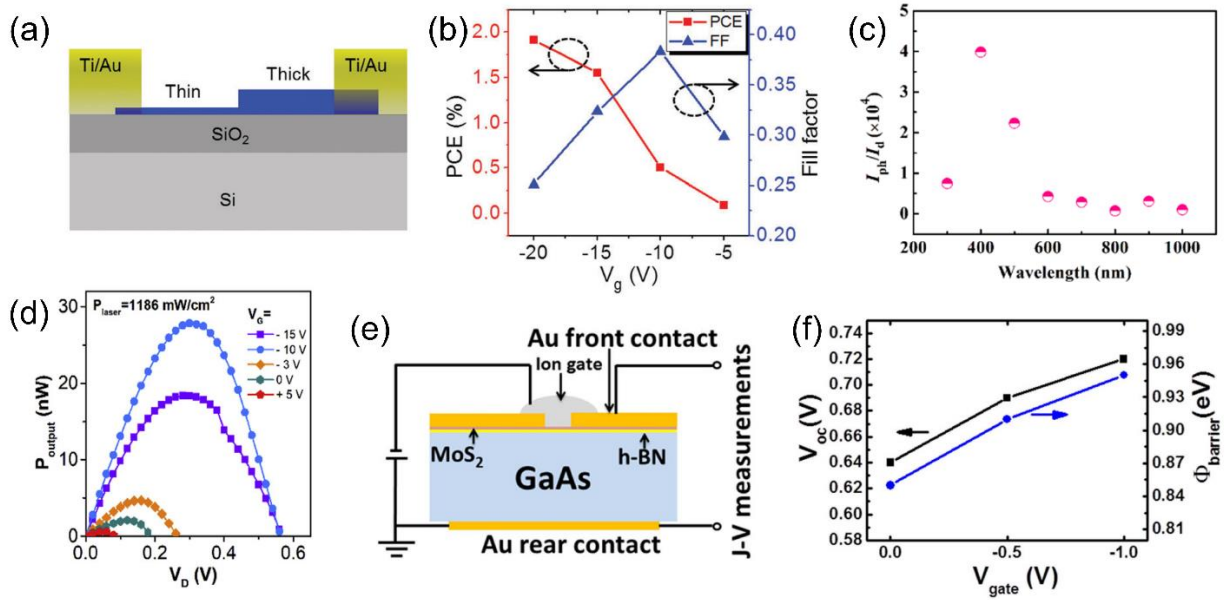

**FIG. 6.** (a) Schematic side view of the MoSe$_2$ homojunction device. (b) The back gate-dependent fill factor (FF) and power conversion efficiency (PCE). The PCE increases up to 1.9% with the increasing negative gate. Republished with permission of [106]. Permission conveyed through Copyright Clearance Center, Inc (c) Photo/dark current ratio under various wavelength light at V = −2 V in an InSe-Te vdW heterostructure.

### *3. Light emitting devices*

In general, 2D vdW materials exhibit layer dependent optical bandgaps, large exciton binding energies and high carrier mobilities, making them exciting candidates for designing novel optoelectronic devices such as phototransistors and LEDs.[99,117–119] Electrical tuning is the most preferred approach to tune optical excitations in 2D materials due to its fast tunability and easy integration of electrical tuning units on chip in nano/micrometer scale, and has been implemented in controlling the excitonic emission in 2D TMD-based LED devices.[120] Electroluminescence (EL) emerging from desired exciton species with unique emission characteristics is of fundamental importance for the practical implementation of the LEDs. In view of this, 2D TMDs represent a new class of functional materials for realizing electrically-tunable, exciton-mediated LEDs. In this section, we summarize the progress in realizing electrically controlled electroluminescence in 2D TMDs and their heterostructures using different device configurations.

*Lateral p-n junctions*

Gate-tunable EL has been reported in numerous TMD-based lateral p-n junctions, formed by either electrostatic gating or ionic liquid gating, where EL originating from different excitonic complexes is tuned by changing the carrier density.[100,101,120–123] The recent research efforts in this direction have focused on using CVD-grown TMDs for fabricating large area LED devices. Lateral p-n junctions using CVD-grown, polycrystalline WSe$_2$ and MoS$_2$ monolayers has been realized by placing an electrolyte on top of the TMD materials, where the electronic charges are induced by the formation of electric double layers.[124] The tunability of the EL intensity with the applied bias was demonstrated. However, EQE obtained in such devices was significantly low ($\approx 10^{-3}$ % for WSe$_2$ and $\approx 10^{-5}$ % for MoS$_2$), likely due to the poor quality of the CVD-grown samples.[124] Apart from the inorganic 2D-material LEDs, LEDs using 2D molecular semiconductors have also been fabricated, which hold great promise for ultrafast on-chip optical communication applications.[125] The lack of suitable bipolar ohmic contacts for achieving high injection levels of electrons and holes in the 2D material is a bottle neck for obtaining enhanced performance from the lateral p-n junction devices. In a recent study, EL from a dopant-free two-terminal device was observed by applying an AC voltage between the gate and the semiconductor [Figure 7(a)].[126] An excess of electron and hole populations simultaneously present in the monolayered TMD (MoS$_2$, WS$_2$, MoSe$_2$, and WSe$_2$) during the AC transient resulted in pulsed light emission at each $V_g$ transition [Figure 7(b)].

*Vertical heterojunctions*

In lateral p-n junctions, emission is spatially localized only near the contact regions due to difficulty in achieving uniform carrier injection throughout the monolayer, leading to low quantum efficiency. An improved performance of the LEDs can be realized using vertical vdW heterostructures due to the reduced contact resistance, large luminescence area and wider choice of TMDs, which can result in significantly improved EL efficiency. Light emission from p-n junctions composed of 2D/3D heterostructures such as n-type MoS$_2$/p-type silicon,[127] MoS$_2$/GaN[128] has been demonstrated, in which an electric field/current-driven tuning of the band alignment at the heterostructure interface or carrier redistribution was argued for the EL tuning.

Recent research work has focused on investigating light-emitting properties of interlayer excitons (X$_I$) in 2D/2D type-II heterostructures, which show great promise for novel excitonic device applications. Interlayer exciton emission in electrostatically-gated MoSe$_2$/WSe$_2$ heterobilayers has been investigated.[129] By biasing gate electrodes with opposite polarities ($V_{BG1}$ = -1 V and $V_{BG2}$ = 5 V), a p-n junction was realized and the EL dominated by interlayer exciton emission under forward bias condition was obtained. The EL peak was found to shift from 1.35 eV to 1.38 eV with a vertical electric field *via* the Stark effect. In another recent report, electrically-tunable interlayer exciton emission and upconverted EL *via* Auger scattering of interlayer excitons by injecting high carrier density into WSe$_2$/MoS$_2$ type-II heterostructure was demonstrated [Figure 7(c) and 7(d)].[130]

*Quantum well structures*

An extremely high EQE has been achieved in quantum well (QW) structures comprising of vertically stacked metallic graphene, insulating h-BN and semiconducting TMDs. An EQE as high as ≈ 8.4 % in a MoS$_2$-based, multiple QW structure has been obtained at low temperatures[131] and a high EQE (≈ 5%) at room temperature has been obtained using a WSe$_2$-based QW, in which a monotonic increase in the EQE as a function of bias voltage and injection current density was observed [Figure 7(e)].[132] In a recent study, a high-speed electrical modulation of light emission by integrating a photonic nanocavity (GaP) with a WSe$_2$ QW structure was reported.[133] The light emission intensity of the cavity coupled EL peak enhanced ~4 times compared to the cavity-decoupled peak as the bias voltage was increased. Furthermore, an electrical modulation of the EL revealed fast rise and decay times of ~320 ns and ~509 ns, respectively, by turning the voltage on and off, resulting in the fast-operational speed (~1 MHz) of the device [Figure 7(f)].

An ultralow turn-on current density of 4 pA·μm$^{-2}$, which is ~5 orders of magnitude lower than that of the best single QW device, has been obtained in a metal-insulator-semiconductor (MIS) vdW heterostack comprising of few-layered graphene (FLG), few-layered h-BN, and monolayered WS$_2$.[134] EL from the positively-charged *(X$^+$)* or negatively-charged *(X$^-$)* trions was reversibly controlled by electrostatic tuning of the TMD (WSe$_2$) into n- and p-type doping regimes under forward bias.[135] The EL intensity increased almost linearly with the tunnelling current and the lower-bound EQE was estimated to be ~0.1% for *X$^+$* and ~0.05% for *X$^-$*. Furthermore, electrically-driven light emission from multi-particle exciton complexes in TMDs (WSe$_2$ and WS$_2$) has been demonstrated using a MIS-type structure [Figure 7(g)].[136] By tailoring the parameters of the pulsed gate voltage, an electron-rich or a hole-rich environment can be created in the 2D semiconductors, where the emission intensity from different exciton species is tunable [Figure 7(h)].

*Valley polarized EL*

Owing to inherent broken inversion symmetry, monolayered 2D TMDs have been exploited to obtain valley-polarized EL. Electrically controlling circularly polarized EL was first demonstrated in monolayered and multilayer WSe$_2$ using ionic liquid gating.[137] Due to large carrier injection capability with ionic gating, inversion symmetry breaking and band structure modulation can be obtained in multilayer 2D TMDs. Consequently, circularly-polarized EL from monolayer and multilayer samples has been observed under forward bias. The EL intensity linearly increased with increasing bias voltage. Electrically tunable, chiral EL from large area CVD-grown monolayered WS$_2$ in a p$^+$-Si/i-WS$_2$/n-ITO heterojunction was found to be dominated by negatively charged excitons, with an increase in the injection current from 2.5 - 4.0 μA due to imbalanced carrier injection under forward bias resulting in an enhanced n-type doping. EL with a high degree of circular polarization ~81% at 0.5 μA was reported in the same device [Figure 7(i)].[138] Similarly, EL from MoSe$_2$ with circular polarization ~ 66% has also been obtained.[139]

Another popular approach for obtaining electrically controlled, valley-polarized EL is *via* spin-polarized charge carrier injection through a ferromagnetic semiconductor or electrode (Ni/Fe permalloy) into the TMD monolayers due to the spin-valley locking effect. Circularly-polarized EL at a (Ga,Mn)As/WS$_2$ p-n heterojunction and a WSe$_2$/MoS$_2$ heterojunction under forward bias has been obtained due to the imbalance of spin-injected carrier population at K and K′ valleys.[140,141]

*Electrically controlled single quantum emitters*

In addition to the direct gap excitons, luminescence from localized defect states in 2D TMDs can be used to obtain single photon quantum emitters, which are the fundamental building blocks for quantum photonics and quantum information technologies. The emission from such localized defect states can be controlled electrically, resulting in the realization of electrically controlled quantum emitters. Electrically controlled EL from the localized states has been demonstrated in graphene/h-BN/WSe$_2$/h-BN/graphene vertical heterostructures, where[142] a spectrally-sharp emission (at ~1.607eV, $V_b$ ~ -2.15 V) corresponding to a single defect state (SDE7) was obtained. As the electrical bias was increased, additional broad features corresponding to emission from other localized states also emerged [Figure 7(j)]. The full-width-at-half-maximum (FWHM) of the localized EL peak increased from 0.6 meV to 1.4 meV as the bias was increased above -2.3 V. In another report, electrically-driven light emission from defects in WSe$_2$ using both vertical and lateral vdW heterostructure devices was obtained.[143] The vertical structure exhibited a broad peak due to defect-bound exciton states, including a narrow emission peak (~ 1.705 eV) that was referred to as emission from a single defect. Electrical tuning of the emission from a single defect was demonstrated, where the emission was repeatedly switched on and off by sweeping external bias $V_b$ from 1.9 to 2.1 V. In contrast, the lateral heterojunction device exhibited EL originating from several single defects having line widths <300 μeV and a doublet structure with ∼0.7 meV energy splitting. Similarly, electrically driven single-photon emission from localized states in mono and bi-layers of WSe$_2$ and WS$_2$ has also been reported.[144] The electrical current dependence of the EL emission from the quantum emitter showed clear saturation, whereas the emission from the unbound monolayered WSe$_2$ excitons exhibited a linear relation between emission intensity and injected current.

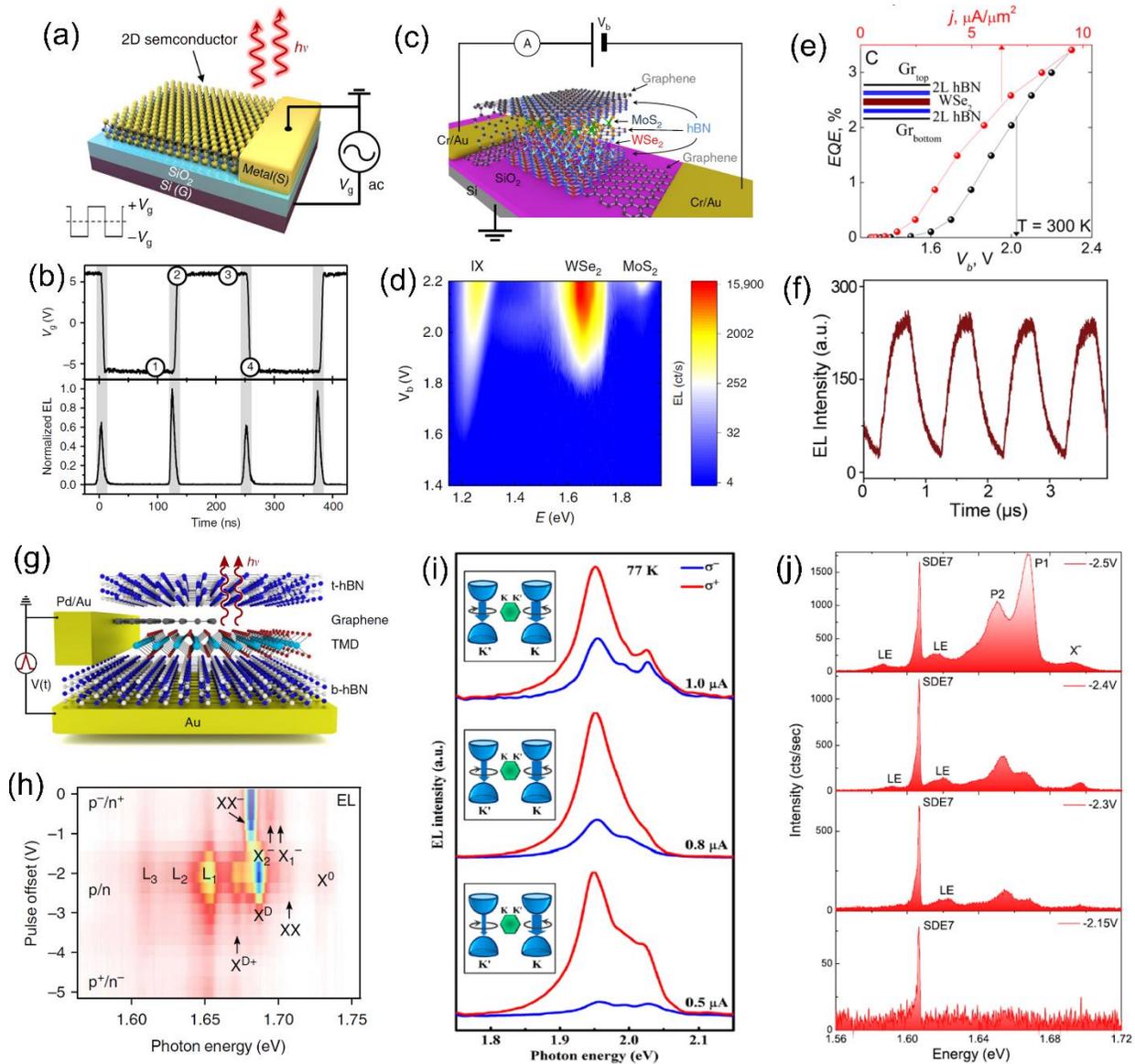

**FIG. 7.** (a) Schematic of the t-EL device. An AC voltage is applied between the gate and source electrodes and emission occurs near the source contact edge. (b) Time-resolved EL and the corresponding $V_g$, showing that EL occurs at the $V_g$ transients Reproduced with permission from [126] CC BY (c) Schematic drawing of the $MoS_2$/$WSe_2$ heterostructures with a middle monolayered h-BN spacer. (d) False color heatmap of the EL spectra as a function of bias voltage for sample without a monolayer h-BN spacer. Reproduced with permission from [130] CC BY (e) The external quantum efficiency for a $WSe_2$ quantum well plotted against bias voltage and injection current density at T = 300 K. The EQE monotonically increases even up to current densities of 1000 A/$cm^2$. Adapted with permission from [132]. Copyright © 2015, American Chemical Society (f) Electrical modulation of EL demonstrating 1 MHz electrical switch of the light emitter. Adapted with permission from [133]. Copyright © 2017 American Chemical Society (g) Schematic of the vdW heterostructure. The monolayered TMD and graphene are sandwiched between two multilayer h-BN flakes.[136] (h) Colormap of EL spectra at 5 K plotted as function of pulse offset $V_0$ in $WSe_2$. Reproduced with permission from [136] CC BY (i) Circularly polarized EL of the p+-Si/i-$WS_2$/n-ITO heterojunction

LED device at 77 K. The current-dependent circularly polarized EL spectra is normalized to the right-handed circular ($\square^+$) component at each injection current. Adapted with permission from [138]. Copyright © 2016, American Chemical Society (j) Evolution of EL spectra as function of bias. SDE7 and LE correspond to the EL peaks of the single-defect emitters and other localized states, respectively. Reproduced with permission from [142] CC BY

### *4. Optical modulators, mirrors etc.*

Extensive research efforts have been focused on designing ultrafast, low power and compact optical modulators for a variety of applications, such as optical interconnects, environmental monitoring, security and biosensing.[145–147] For designing high-performance optical modulators, materials with low optical-loss, large non-linear optical constants, broad wavelength operation, large tunability of optical constants and ease of integration with various optical components are desired. In this category, 2D materials have attracted significant attention, owing to their atomic layer thickness, strong light-matter interaction, broadband optical response and stimuli tunable opto-electronic.[147] 2D materials also offer advantages for low cost and large-scale integration to the well-developed silica fiber and silicon-based technology. Thus far, graphene has been the leading candidate for obtaining high speed optical modulation in an extremely broad spectral range, extending from ultraviolet to microwave regions, primarily due to its unique linear energy-momentum dispersion relation and high mobility.[148,149] However, other 2D materials such as monolayered TMDs and black phosphorus, which offer properties complementary to graphene, also exhibit similar potential and have been explored recently. The presence of a bandgap and parabolic band structure in 2D semiconductors can allow much higher tunability of the optical dielectric functions. Different tuning methods have been demonstrated to modulate the optical response of the 2D materials that can be characterized by the change in the dielectric constants or the complex refractive index of the materials.[150] The electrical control of optical response is particularly desired for data communication link applications. Herein, we present state-of-the-art progress in electro-optic modulators based on 2D semiconductors beyond graphene.

For the electro-optic modulators, the application of an external electric field is desired to tune both the real and imaginary components of the refractive index such that both the amplitude and phase of the optical field can be modulated. In a recent report, a modulation of the refractive index of monolayered TMDs, such as $WS_2$, $WSe_2$ and $MoS_2$, by more than 60% in the imaginary part and 20% in the real part around their excitonic resonance was demonstrated using electrical gating.[151] The giant tuning in refractive index was attained by changing the carrier density, which broadened the spectral width of excitonic transitions and facilitated the interconversion of neutral and charged excitons. The gate dependence of optical absorption and optical constants of TMDs was also investigated by several other groups [Figure 8(a) and 8(b)].[152,153] A modulation depth as high as ~6 dB for visible light (red light ~ 630 nm) has been reported.

The other popular approach that can be used for the manipulation of dielectric properties of TMDs is by the interaction of excitons with metallic plasmons.[154] In this work, narrow $MoS_2$ excitons coupled with broad Au plasmons, led to an asymmetric Fano resonance that was effectively tuned by the applied gate voltage to the $MoS_2$ monolayers. The gate-dependency of the Fano resonance is strongly sensitive to the modulation of the exciton-plasmon coupling strength and can be controlled by the $MoS_2$ exciton absorption at different external gate voltages. Active control of light-matter interactions is also critical for realizing plasmonic nanostructure-based electro-optic modulators. An electrical control of exciton-plasmon coupling strengths between strong and weak coupling limits in a 2D semiconductor has also been demonstrated with electrostatic doping [Figure 8(c) and 8(d)].[155] In addition to the coupling of the plasmonic modes to the neutral excitons, a strong coupling with negatively charged excitons was also obtained that can be switched back and forth with the gate voltage. In another recent study, electrical modulation of plasmon-induced exciton flux was demonstrated using Ag-nanowire waveguides overlapping with TMD transistors.[156] The laser-coupled-plasmon propagated through the Ag nanowire in the axial direction, which sequentially excited excitons of the TMD, and the exciton flux was modulated by the gate voltage. In most of the reports, electrical tuning of the optical properties of TMDs was achieved near their excitonic resonances only, where the refractive index and absorption can be modulated simultaneously at a maximum magnitude. In a recent report, electro-optic response of monolayered TMD ($WS_2$) at NIR wavelength regions was probed for integrated photonics applications.[157] Using an ionic liquid gate, high electron doping densities ($\approx 7.2 \pm 0.8 \times 10^{13}$ cm$^{-2}$ at 2 V) were induced in the monolayered $WS_2$ and a large change in the real part of the refractive index ($\approx$ 53 %), and a minimal change in the imaginary part $\approx$ 0.4 % was demonstrated [Figure 8(e) and 8(f)]. Also, a doping-induced efficient phase modulator with high $|\Delta n/\Delta k| \approx 125$ and low propagation losses was achieved.

Apart from TMDs, black phosphorus (BP) is also an emerging candidate for designing electro-optic modulators in the mid-infrared frequencies owing to its smaller bandgap. Several theoretical and experimental investigations have shown that an external electric field can result in a shift in BP's absorption edge due to the interplay between different electro-absorption mechanisms, mainly the field-induced, quantum-confined Franz-Keldysh effect and the Pauli-blocked Burstein-Moss shift.[158–160] These different mechanisms lead to distinct optical responses that are strongly dependent on the flake thickness, doping concentration and operating wavelength. To isolate and define the working mechanism of different electro-absorption effects, a recent report used two different field-effect device configurations with different gating schemes, wherein the BP either floats electrically in an applied field or is in direct contact [Figure 8(g) and 8(h)].[161] In the electrically floating case, the dominant tuning mechanism is the quantum-confined Stark effect while in the other case tunability is dominated by carrier concentrations effect e.g., by the Burstein-Moss shift. Near-unity tuning of the BP oscillator strength and electro-optic tuning of linear dichroismover a broad range of wavelengths, from the mid-infrared to the visible, by controlling the thickness of the BP was reported.

While significant progress has been achieved with optical modulation using 2D materials, it is safe to say that, with the exception of graphene, the field is in its infancy. This is evident from Figure 9, where we have compared depth of modulation with the wavelength of modulation. Graphene-based modulators not only show exceptional depth of modulation, but, due to its zero-gap nature, they can also operate well at telecom wavelengths, with performance comparable to Si, $LiNbO_3$ and III-V modulators. In contrast, TMDs are barely able to operate in the NIR range. The key issue at play for materials beyond graphene to serve in high-performance optical modulators is the lack of quality large-area material. Furthermore, most 2D semiconductors are in the 1.1 eV to 3 eV or < 0.5 eV range of band gap values. This makes it difficult to achieve modulation by means of a dominant photorefractive effect at a band or exciton edge in a 2D semiconductor in the telecom range ($\approx$ 0.8 eV). Finally, it is equally important to attain good metal contacts to inject and extract carriers at high speed with minimal resistance or other parasitic circuit elements, particularly at high operation speeds.

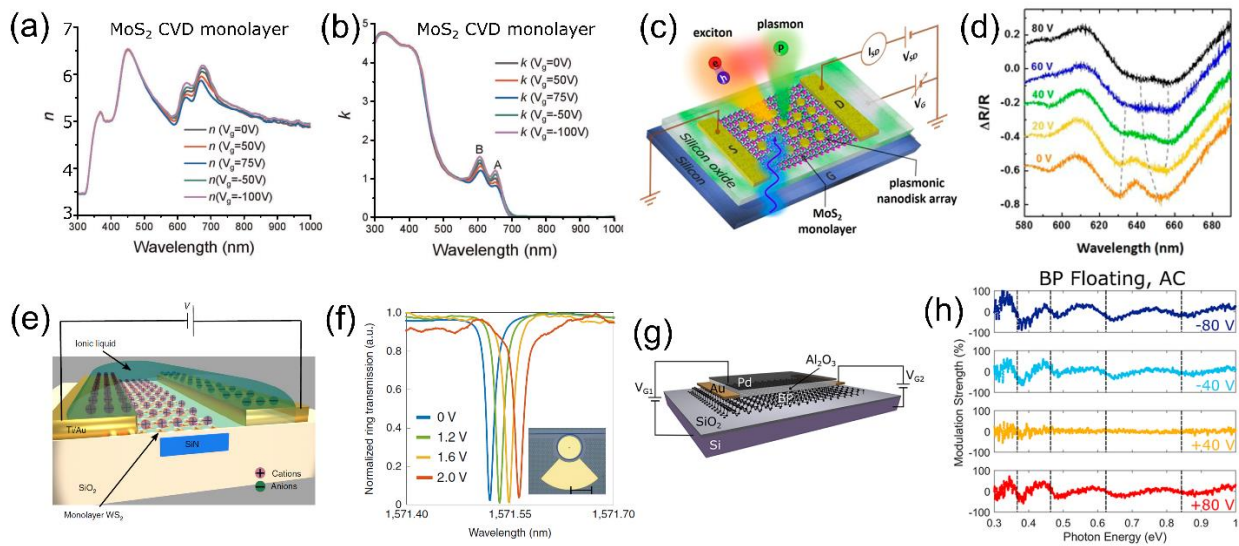

**FIG. 8.** Gate induced changes in the complex refractive index of MoS$_2$ monolayer: (a) the real part and (b) the imaginary part. Reproduced with permission from [152] CC BY (c) Schematic diagram of the MoS$_2$ monolayer-plasmonic lattice system integrated with an FET device. (d) Line cuts from the angle-resolved spectra at different gate voltage. Adapted with permission from [155]. Copyright © 2017 American Chemical Society (e) Schematic of the composite SiN-WS$_2$ waveguide with ionic liquid cladding. By applying a bias voltage across the two electrodes through the ionic liquid, charge carriers were induced in the monolayer. (f) Normalized transmission response of the microring resonator at different voltages applied across the two electrodes. Adapted with permission from [157]. Copyright © 2020, under exclusive licence to Springer Nature Limited (g) Schematic diagram of the device to obtain anisotropic electro-optical effect in few-layer black phosphorous (BP). (h) Tuning of BP oscillator strength with a field applied to the floating device, for light polarized along the AC axis. Adapted with permission from [161]. Copyright © 2017 American Chemical Society

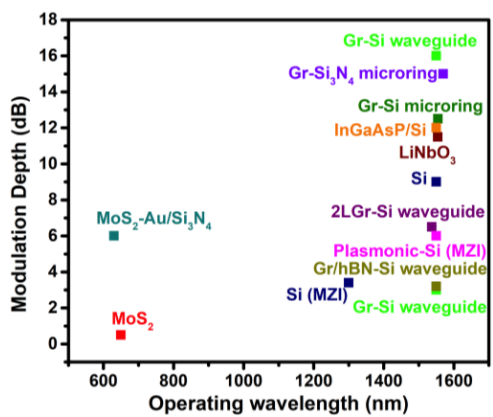

**FIG. 9.** Comparison of modulation depth vs. operating wavelength for 2D materials (graphene and MoS$_2$) and bulk materials including Si, LiNbO$_3$ and III-V (InGaAsP) on Si.[152,162–173] *Note: MZI, Mach-Zehnder interferometer.

## D. Sensors

Due to their unique physical properties, tunability and versatility, 2D materials are particularly attractive for multiple sensing applications. Here, we consider their use for gas (primarily environmental), humidity and biosensing as well as in strain and pressure sensing. The fact that their electronic structure is highly susceptive to and tunable by hybridization, chemical modification or doping makes them an ideal platform for regulation of catalytic properties. This is particularly important for their applications as chemical and biosensors, often allowing fine control of sensing by electrical means. We specifically focus our attention on the electrical means of read-out and control of such sensors. For another very powerful way to control such sensors, namely optical or optoelectronic detection, we address the readers to the other reviews in the area [174–176].

Based on the tremendous success of graphene-based sensors, efforts in constructing novel sensing platforms led to exploitation of other, more versatile and tunable 2D materials, e.g., TMDs, carbon nitride, boron nitride, phosphorene and MXenes, which will be discussed in this section of the review.

### 1. Gas sensors

In this review, we consider the recent developments in non-graphitic 2D material sensors that have followed in the wake of graphene. We discuss the different classes of 2D materials in application to sensing as well as their figures of merits, such as Level of Detection (LoD), selectivity, response/recovery time and detection range, and compare them with performance of graphene sensors.

When discussing the physical mechanisms of sensor operation, we only consider non-covalent (and thus reversible) interactions between the target analyte and 2D material, which include vdW forces, hydrogen bonding, coordination, and π-π interactions. The type of interaction for each 2D material will depend on its chemical and electronic structure, distinct structural features and surface chemistry. To be electrically detectable, any of these interactions should induce change in physical parameters such as conductivity, work function or permittivity. This is typically realized through the fundamental mechanisms such as modulation of doping level and/or Schottky barrier, as well as the formation of dipole and interfacial layer. For more details, see Meng et al.[174] and the references within.

In the majority of cases, the numerical output of a whole FET device is read out in the response to global gas exposure. Recently, Noyce et al.[177] provided an important insight into sensing mechanisms by realizing precise nanoscale control over the position and charge of an analyte using a charged scanning atomic force microscopy (AFM) tip acting as an effective analyte. The non-uniform sensitivity of the $MoS_2$ FET channel was demonstrated, showing time-stable, sensitive hotspots, where the signal-to-noise ratio was maximized at the center of the channel, and the response of the device is highly asymmetric with respect to the polarity of the analyte charge. The work reveals the important role of analyte position and coverage in determining the optimal operating bias conditions for maximal sensitivity in 2D FET-based sensors, which provides key insights for future sensor design and control.

Sensors serve an important role in the detection of common environmental gas pollutants including but not limited to nitrogen dioxide ($NO_2$ at 21 ppb), sulphur dioxide ($SO_2$ at 7.5 ppb), hydrogen sulphide ($H_2S$ at 5 ppm), ammonia ($NH_3$ at 20 ppm), and carbon monoxide (CO at 4 ppm). Each of these gases are considered to be toxic for human health, though the specific exposure limits vary significantly. The recommended exposure limits (as set by the Gothenburg Protocol[178] or the Paris Climate Agreement[179]) for each of these pollutants are shown in parenthesis above. See Buckley et al.[175] for more details. Additionally, carbon dioxide ($CO_2$) and methane ($CH_4$) are not toxic gases but are responsible for the greenhouse effect, leading to climate change. The exposure limits indicated in the parentheses above should be taken in account when performance of individual sensors is being evaluated, as in a number of cases the demonstrated LoD and detection range are far outside of specific environmental requirements.

*TMDs*

Among all 2D materials, TMDs and specifically $MoS_2$ have been the most intensely explored for gas sensing, where earlier works focused largely on mechanically exfoliated TMD materials of variable thickness [180,181] . $MoS_2$ FETs fabricated with 2-4 layers of $MoS_2$ exhibited better performance compared with monolayer $MoS_2$, which showed unstable response. For similar $MoS_2$ FETs, detection of both electron acceptors ($NO_2$) and electron donors ($NH_3$) under different conditions such as gate bias and light irradiation has been explored [182] [Figure 10(a)]. The results showed that a few-layer (5L) $MoS_2$ FET exhibited better sensitivity, recovery, and ability to be manipulated by gate bias and green light in comparison with its bilayer counterpart. It is noteworthy that the measurements were conducted in the high concentration range of $NO_2$, 10-1000 ppm. There have also been reports studying gas sensing of $MoS_2$ using more scalable material techniques, such as vapor phase growth of thin films [183] and flexible transistor arrays using all solution-processable materials [184].

Typically, the time-response of solid-state sensors is a challenging parameter, often showing unsatisfactory performance in ambient conditions. Poor response time and incomplete recovery at room temperature have often restricted the application of many 2D materials in high-performance practical gas sensors. Fast detection of $NO_2$ and reversibility have been demonstrated for $MoS_2$ gas sensors at room temperature[185]. While incomplete recovery and a high response time of ~249 s of the sensor were observed in ambient, $MoS_2$ exhibited an enhancement in sensitivity with a fast response time of ~29 s and excellent recovery to $NO_2$ (100 ppm) under photoexcitation at room temperature. The effect was attributed to the charge perturbation on the surface of the sensing layer under optical illumination. Driven by the same goal and using hybrid $MoS_2$ materials, Long et al.[186] presented fast detection and complete recovery (both at <1 min) to $NO_2$ at 200°C using $MoS_2$/Graphene/hybrid aerogels, while Kuru et al.[187] showed a response/recovery time of 40/83 seconds, respectively, to $H_2$ at room temperature using $MoS_2$ nanosheet−Pd nanoparticle composite. In a work by Park et al.[188], a highly porous h-$MoS_2$/Pt nanoparticle (NP) hybrid synthesized by pyrolis was used for $H_2$ detection, demonstrating 8/16 seconds for response/recovery time, respectively, for 1% of $H_2$. The response is among fastest for $H_2$ sensors based on 2D materials operated in standard ambient environment.

While experimental detection of non-polar gases (e.g., $CO_2$, $CH_4$ and etc.)) with 2D materials based sensors is generally problematic due to the lack of strongly pronounced mechanisms of adsorption, a modelling study using first-principles and Monte Carlo simulations revealed that single and double sulfur vacancies exhibit an excellent adsorption ability for both polar and non-polar gases[189]. The simulation results showed that $MoS_2$ with a single S vacancy could absorb 42.1 wt% of $CO_2$ and 37.6 wt% of $CH_4$ under a pressure of 80 bar at room temperature.

As mentioned above, heterostructures combining a 2D material and some other active component(s) often demonstrate an enhanced performance for gas sensing compared to their individual counterparts. For example, a $MoS_2$/$Co_3O_4$ thin film sensors were implemented for ultra-low-concentration detection of ammonia at room temperature[190], revealing high sensitivity, good repeatability, stability, selectivity and fast response/recovery characteristics. Zhang et al.[191] demonstrated a hydrogen gas sensor based on a complex Pd-$SnO_2$/$MoS_2$ ternary hybrid. The experimental results showed a response within 1-18% resistance change, swift response-recovery time (10 ~ 20 seconds), good repeatability and high selectivity toward hydrogen gas in the range of 30-5000 ppm at room temperature.

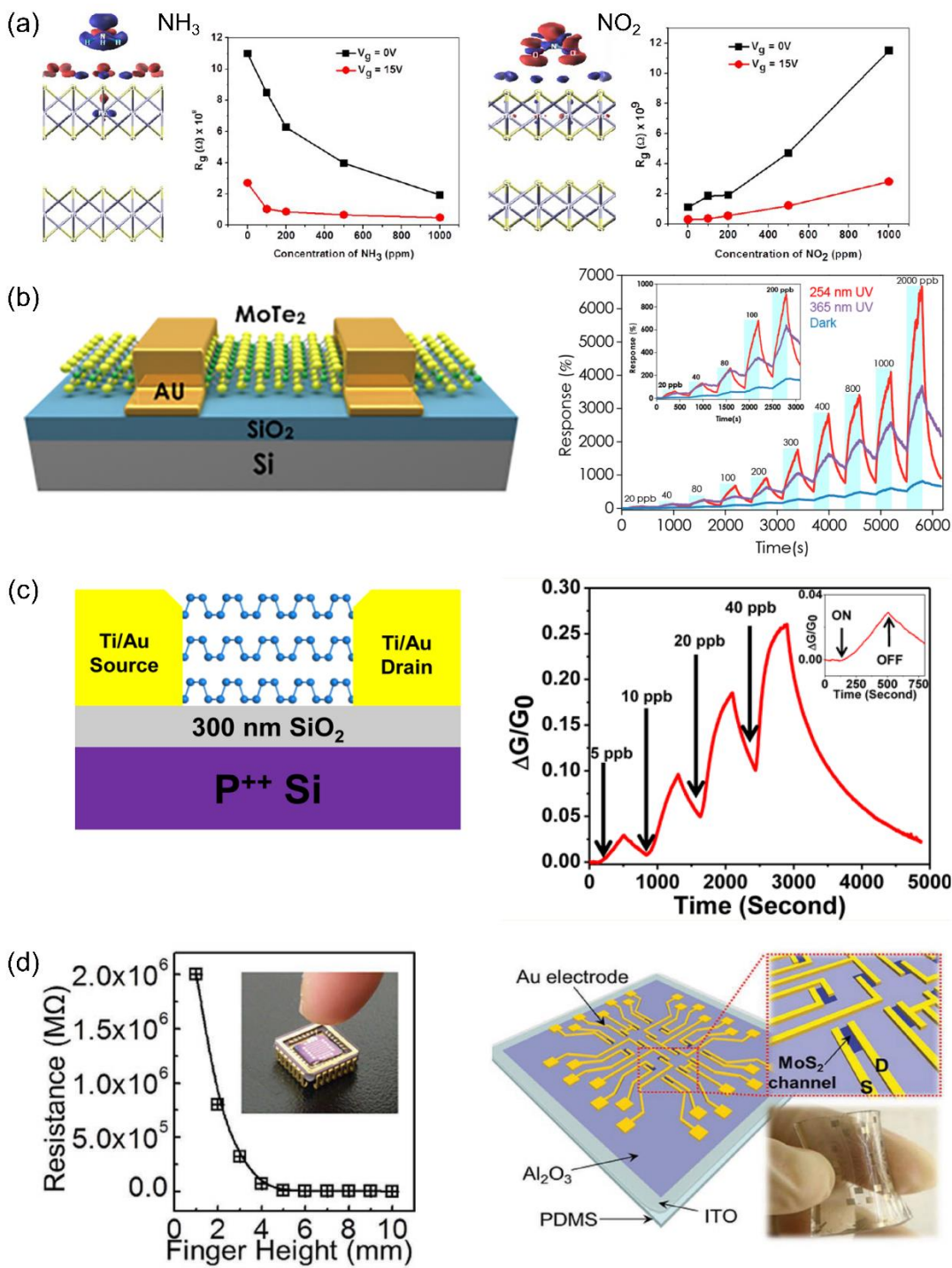

**FIG. 10.** (a) Left: Charge transfer in the case of $NH_3$ (top) and $NO_2$ (bottom) molecules on bilayer $MoS_2$ with $V_g = 0$ V. The red/blue isosurface indicates negative/positive charge. Right: Variation of resistance for $NH_3$ (top) and $NO_2$ (bottom) on bilayer $MoS_2$ at $V_g = 0$ and 15 V. Adapted with permission from [182]. Copyright 2013 American Chemical Society; (b) Top: Schematic diagram of $MoTe_2$ transistor. Bottom:

Dynamic sensing performance toward $NO_2$ in $N_2$ environment at concentration from 20 ppb to 2 ppm under different light illumination. Inset: dynamic sensing behaviors to $NO_2$ at concentration from 20 to 200 ppb. The light blue bars denote the gas concentration of each exposure. Adapted with permission of [192]. Copyright 2018 American Chemical Society. (c) Left: Schematics of a multilayer black phosphorus (BP) FET. Right: sensing performance of BP FET to $NO_2$ gas exposure (0.6–10 ppm). Adapted with permission from [193]. Copyright 2015 American Chemical Society; (d) Left: Humidity sensor: the exponential increase of the resistance with the finger moving closer to the $MoS_2$ device. Right: Schematic illustration of $MoS_2$ FETs array on soft PDMS substrate. Insets show the magnified $MoS_2$ channels and optical image of the sensor array, respectively. Adapted with permission from [194]. Copyright ©2017 Advanced Materials

In comparison to $MoS_2$, $WS_2$ has been significantly less explored for gas sensing applications, with the most notable works including complex 3D geometries assembled of 2D nanostructures. For example, a $WS_2$ nanoflake-based sensor showed a good sensitivity to ammonia (1-10 ppm) at room temperature with the response/recovery time of 120/150 seconds, respectively. Interestingly, the developed ammonia sensor also demonstrated excellent selectivity to formaldehyde, ethanol, benzene and acetone[195]. A sensor made of $WS_2$ nanosheets with size of 10 nm exhibited selectivity towards $NO_2$ and $H_2S$ [196], where the presence of oxygen and elevated temperatures (160°C) were necessary for $H_2S$ sensing. An interesting approach to increase the surface area (and thus number of active sites for molecular adsorption) of atomically thin TMDs was demonstrated by synthesizing 3D $WS_2$ nanowalls combined with CdSe – ZnS quantum dots (QDs), with a demonstrated detection limit of 50 ppb for $NO_2$ and a fast response time of ~ 26 seconds[197].

In general, many 2D disulfides have proven selectivity towards $NO_2$. A sensor based on $SnS_2$ platelets demonstrated high sensitivity (600 ppb measured) and superior selectivity to $NO_2$[198] due to the unique physical affinity and favorable electronic band positions of $SnS_2$ and $NO_2$. Density functional theory (DFT) studies of $NbS_2$ also suggest its high selectivity towards $NO_2$[199] depending on its edge configuration. In a recent work, a monolayered $Re_{0.5}Nb_{0.5}S_2$ sensor demonstrated high sensitivity, selectivity and stability towards $NO_2$ detection, accompanied by only minimal response to other gases, such as $NH_3$, $CH_2O$, and $CO_2$[200]. In the presence of humidity, the monolayer sensor showed complete reversibility with fast recovery at room temperature. Including other TMDs, the fast response/recovery rate accompanied by enhanced sensitivity under UV illumination was achieved in a p-type $MoTe_2$ gas sensor for $NO_2$ detection, demonstrating full recovery within 160 seconds after each sensing cycle at room temperature [Figure 10(b)][192]. The sensitivity of the sensor to $NO_2$ is significantly enhanced by an order of magnitude under 254 nm UV illumination as compared to that in the dark condition, leading to a remarkably low detection limit of 252 ppb (as derived from noise measurements).

*Other 2D Materials*

**Phosphorene and Blue Phosphorus** Following the initial prediction of superior gas sensing applications using first-principles calculations[201], semiconducting phosphorene (or 2D black phosphorus) has been proven to have great potential for gas sensing of CO, $CO_2$, $NH_3$, and NO (see e.g. Liu et al.[202]). Abbas et al.[193] experimentally demonstrated phosphorene-based FETs for $NO_2$ detection, achieving a lowest detectable concentration of 5 ppb [Figure 10(c)]. Phosphorene functionalized with gold has been shown to display selectivity to $NO_2$ (compared with $H_2$, acetone, acetaldehyde, ethanol, hexane and toluene) [203]. Although pristine phosphorene is typically insensitive to $H_2$, functionalization with Pt NPs led to improved $H_2$ sensing efficiency, as seen in the decreasing of the drain–source current and increase of the

ON/OFF ratio at low concentrations of $H_2$[203,204]. An impedance phosphorene-based sensor has demonstrated a strong response on the presence of a low concentration of methanol vapor (28 ppm) at ~1 kHz in impedance phase spectra, showing a high selectivity to methanol and absence of cross-selectivity from toluene, acetone, chloroform, dichloromethane, ethanol, isopropyl alcohol, and water due to different dielectric constants of these molecules[205].

Sensing properties of another 2D allotrope of phosphorus - blue phosphorus - were studied by first-principles calculations with respect to the adsorption behaviors of environmental gas molecules, including $O_2$, $NO$, $SO_2$, $NH_3$, $H_2O$, $NO_2$, $CO_2$, $H_2S$, $CO$, and $N_2$[206]. The calculations showed that $O_2$ tended to chemisorb, whereas the other gases were physiosorbed on monolayered blue phosphorus, showing different interaction strengths and distinct modifications to the band gap, carrier effective mass, and work function.

**h-BN** Applications of h-BN in gas sensing has remained challenging due to its inherently low chemical reactivity. Recently, a chemiresistor-type $NO_2$ gas sensor based on sulfate-modified h-BN has been investigated, demonstrating a linear response over a wide $NO_2$ concentration range and low LoD of 20 ppb [207]. Theoretical calculations predicted that the sulphate groups spontaneously grafted to the h-BN, effectively altering its electronic structure and enhancing the surface adsorption capability towards $NO_2$, in addition to a strong charge transfer between $NO_2$ and h-BN. Thus, applications of sulphate-modified h-BN in capturing environmentally hazardous exhaust from motor vehicles as well as combustion emissions monitoring have been proposed.

In a recent theoretical study, graphene/h-BN heterostructures were studied for detection of $NO$, $NO_2$, $NH_3$, and $CO_2$ gas molecules[208]. The strongest interaction was observed for the case of $NO_x$, where $NO$ and $NO_2$ molecules are more reactive at the interface regions of the heterostructure compared to the pristine ones, where large changes in the conductance with adsorption were seen. In addition, recent experimental work has successfully demonstrated $NH_3$ detection with graphene/h-BN devices[209]. The charge transfer from $NH_3$ to graphene strongly depends on the average distance between the graphene sheet and the substrate. Since the average distance between graphene and h-BN crystals is one of the smallest, the graphene/h-BN heterostructure exhibited the fastest recovery times for $NH_3$ exposure and revealed importance of substrate engineering for development of 2D gas sensors.

**MXene** Recently a new class of 2D materials - transition-metal carbides ($Ti_3C_2T_x$ MXene) – have received a great deal of attention for potential use in gas sensing, showing both high sensitivity and good gas selectivity. Integration of an effective superhydrophobic protection fluoroalkylsilane (FOTS) layer with MXenes has helped overcome one of the major problems typical for this class of material, e.g., environmental instability[210]. FOTS-functionalized $Ti_3C_2T_x$ displayed very good hydration stability in humid environment and showed good tolerance to strong acidic and basic solutions. The experiments also demonstrated very good sensing performance (e.g., sensitivity, repeatability, stability, selectivity and faster response/recovery time) to oxygen-containing volatile organic compounds (ethanol, acetone), as achieved in a broad relative humidity range of 5 - 80 %.

This review represents only a relatively small fraction of the research on the use of 2D materials sensors for gas detection. In a chart shown in Figure 11, we summarize the recent figures of merits: LoD (a), detection range (b) and response/recovery time (c), for graphene (as well as GO and rGO) and other layered 2D materials based on the results presented here as well as recent, more specialized reviews by Buckley et al.[175] and Meng et al.[174]. We limit this summary to $NO_2$ gas only, as being generally one of the easiest substances for detection and thus the most intensely studied in literature. It is important to emphasize that in many cases, the measurements were performed in very specific physical conditions (such as temperature, illumination, etc.), which are not reflected in the chart. The red dashed/dotted lines on the chart [Figure

11(a) and 11(b)] represent recommended annual mean limits of exposure as specified by the European Union (EU) and National Ambient Air Quality Standards (NAAQS), e.g., 21 and 58 ppb, respectively[211,212]. The area above the EU annual mean level (as highlighted by grey) demonstrates a potentially unsafe level of exposure for humans (note the exponential scale both for LoD and range). In terms of the response and recovery time, the results also vary significantly, with the recovery time being generally (in many cases, significantly) larger than the response one [Figure 11(c)]. Thus, the conclusions driven from these graphs are unfortunately not very positive. While each individual study often shows an interesting research breakthrough and outlines favorable detection conditions, the overall picture shows that, in the majority of cases, detection is achieved in conditions which are not suitable for environmental monitoring (e.g., falls significantly above red lines), yet may still be relevant for sensors employed in the chemical or defense industries. Graphene is generally more suitable for environmental diagnostics compared to other 2D materials, both in terms of higher sensitivity to $NO_2$ and generally a more favorable detection range, though this might be due to the larger number of studies conducted on graphene. Focusing on non-carbon-based 2D materials, some of the results obtained on BP and $MoS_2$ appear to be the most promising in terms of the LoD and detection range. The comparison between non-carbon 2D materials and graphene seems to be more encouraging in terms of the response and recovery times, where semiconducting 2D materials with a bandgap (BP, $MoS_2$, $SnS_2$) are characterized by generally faster times and would be more suitable for the immediate response to changing environmental conditions. This analysis demonstrates a crucial need not only to improve performance of the sensor but also to align it with the relevant targets dictated by the environmental detection needs, which have been unfortunately disregarded in many studies.

*Humidity sensors*

In addition to their wide-spread application in gas sensing, 2D materials, especially TMDs, have exhibited great potential for humidity sensing. For example, monolayered $MoS_2$ devices on a flexible PDMS substrate have demonstrated a high humidity sensitivity $> 10^4$, with their mobilities and ON/OFF ratios decreasing linearly at RH = 0~35%[194]. The authors showed an exponential increase of the resistance with a human finger moving closer to the $MoS_2$ device [Figure. 10(d)], with short response and decay times. In a similar approach, a hybrid ultrasensitive humidity sensor based on a $MoS_2$-$SnO_2$ nanocomposite revealed very good sensing parameters (e.g., response, fast response/recovery time and repeatability)[213] compared to the pure $MoS_2$ and $SnO_2$ counterparts, and graphene. In another work, a complex device, which incorporated large-area $WS_2$ as a sensing element, graphene as an electrode and thin flexible and stretchable PDMS as a substrate was utilized for humidity sensing (up to 90%) with fast response/recovery times (in a few seconds)[214]. The sensor was then laminated onto human skin and showed stable water moisture sensing behaviors, enabling real-time monitoring of human breath. Another flexible humidity sensor based on $VS_2$ has represented a new concept of a touchless positioning interface based on the spatial mapping of moisture[215]. In general, these flexible skin-attached chemical sensors (electronic skin or e-tattoo) are of great interest for many applications. Several interesting works have also emerged exploring applications of phosphorene in humidity sensing[202]. Yasaei et al.[216] realized a humidity sensor made of phosphorene nanoflakes. When the sensor was exposed to flows of $H_2$, $O_2$, $CO_2$, benzene, toluene, and ethanol, it resulted in enhancement of the electrical response only in presence of a water atmosphere.

Thus, the several 2D sensors summarized here have proved to be strong candidates for ultrahigh-performance humidity sensor toward various applications and have demonstrated a clear path towards developing low power consumption, wearable chemical sensors based on 2D semiconductors.

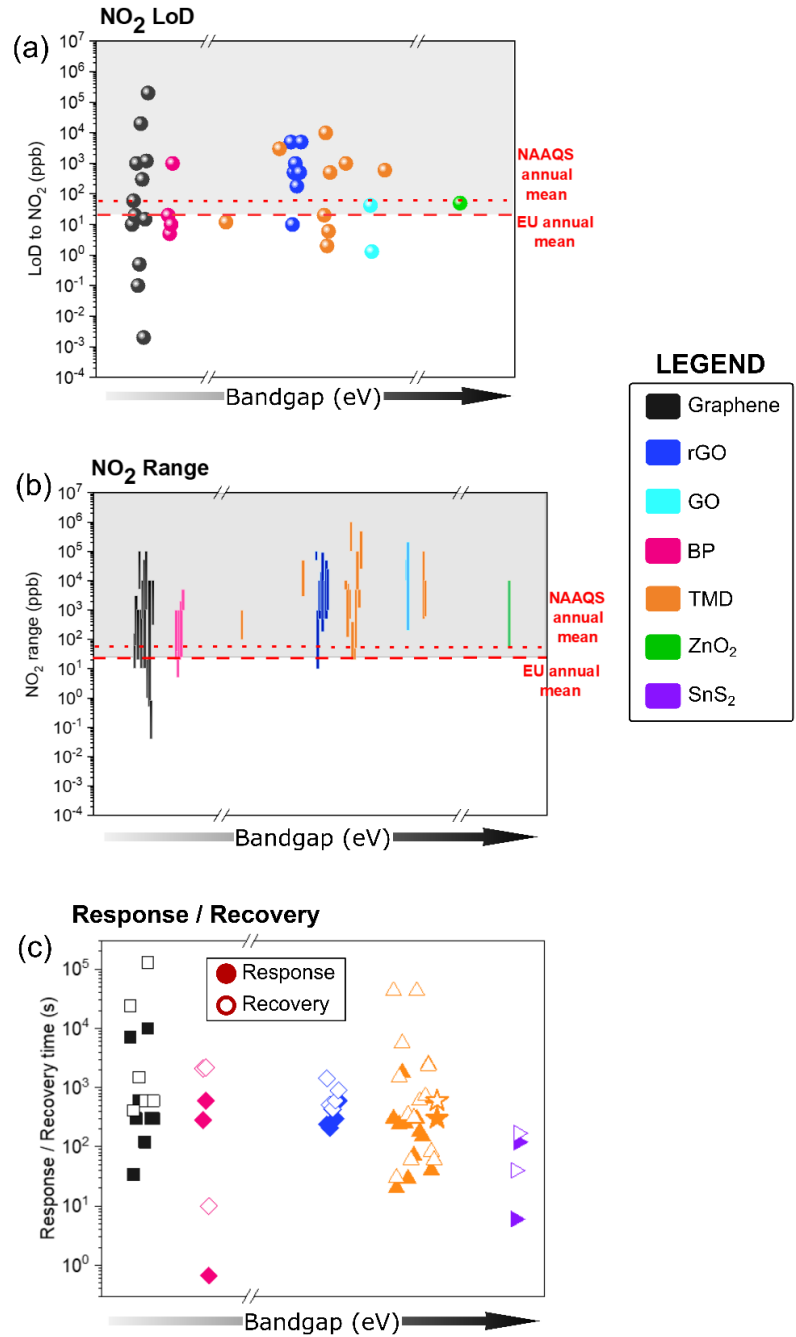

**FIG. 11.** Comparative figures of merit showing a combined performance of various 2D material sensors towards $NO_2$: (a) level of detection; (b) detection range and (c) response and recovery time. The red dashed/dotted lines in (a) and (b) represent recommended annual mean limits of $NO_2$ exposure as specified by EU and NAAQS, respectively. The grey panels in (a) and (b) show $NO_2$ concentrations above environmentally safe levels as specified in the text. The solid/open symbols in (c) represent the response/recovery time for each material. The offsets in the horizontal direction are introduced for clarity. The data is adapted from the recent reviews[174,175](see Tables within).

*Summary*

We have shown that, similarly to graphene, 2D material FET sensors bear the advantage of miniature size and good compatibility with CMOS technology. Moreover, the existence of the band gap allows for an additional degree of electrical control of adsorption/desorption of targeted molecules. The band gap becomes particularly useful for manipulation of the sensor's response/recovery time. It is important to note that, while individual comparison studies may confidently prove advantages of a certain 2D material to graphene for the main figures of merit, the cumulative comparison performed in this review for the case of $NO_2$ demonstrated an extremely broad spread of such parameters as LoD, detection range and response/recovery time, which overall is not dissimilar for graphene and non-carbon 2D materials. Thus, an overwhelming advantage of 2D materials over graphene has not been statistically demonstrated at the current technological level.

## 2. Biosensors

Success of graphene-based biosensors has stimulated a large research interest in the implementation of other 2D materials into biosensing platforms. Here, we present a short overview of the current state of the art and outline the most promising directions in the area of biosensors, focusing our attention on the electrical schemes of a sensor's control and optimization. For more detailed and insightful reviews in this area, we refer to the recent works by Bollella et al.[217], Zhu et al.[218], Hu et al.[219], Meng et al.[174], and Kou et al.[220].

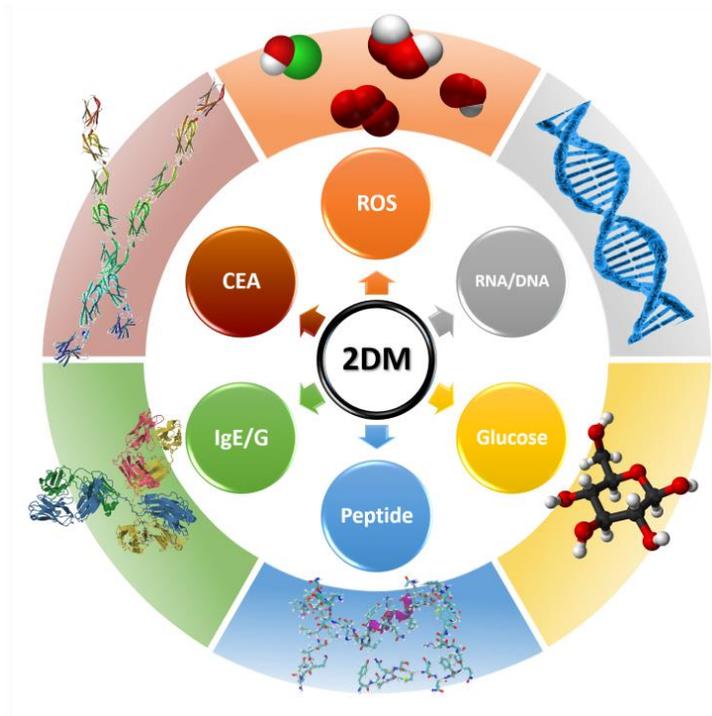

**FIG. 12.** Schematic diagram showing application of various 2D TMD materials in biosensing. Clockwise: in order of increasing complexity of molecules, starting from ROS (reactive oxygen species) to CEA (carcioembrionic antigen).

*TMDs*

Similar to gas sensing, among atomically thin 2D materials, TMDs have been the most intensely studied for electrochemical biosensing purposes[221]. Their high surface-to-volume ratio offers potential for the detection of large amounts of target analytes and low power consumptions. The current state of growth is such that 2D TMDs can be readily synthesized on a large scale and can be directly dispersed in aqueous solution without the aid of surfactants, providing environmentally friendly and even biocompatible and biodegradable solutions[174,222–224]. Figure 12 schematically summarizes the broad range of applications of 2D TMD biosensors for the detection of various molecules. Specifically owing to its high conductivity and a large number of active defects that provide sites for adsorption of biomolecules, $MoS_2$ is one of the most commonly used TMD materials for biosensing[174]. Either in its pristine form or as a part of hybrid structures/nanocomposites, $MoS_2$ has been used both as a platform for non-enzymatic sensing and as a biocompatible matrix for enzyme immobilization and development of both electrochemical sensors and biosensors [Figure 12 and Figure 13(a)][217,225,226].

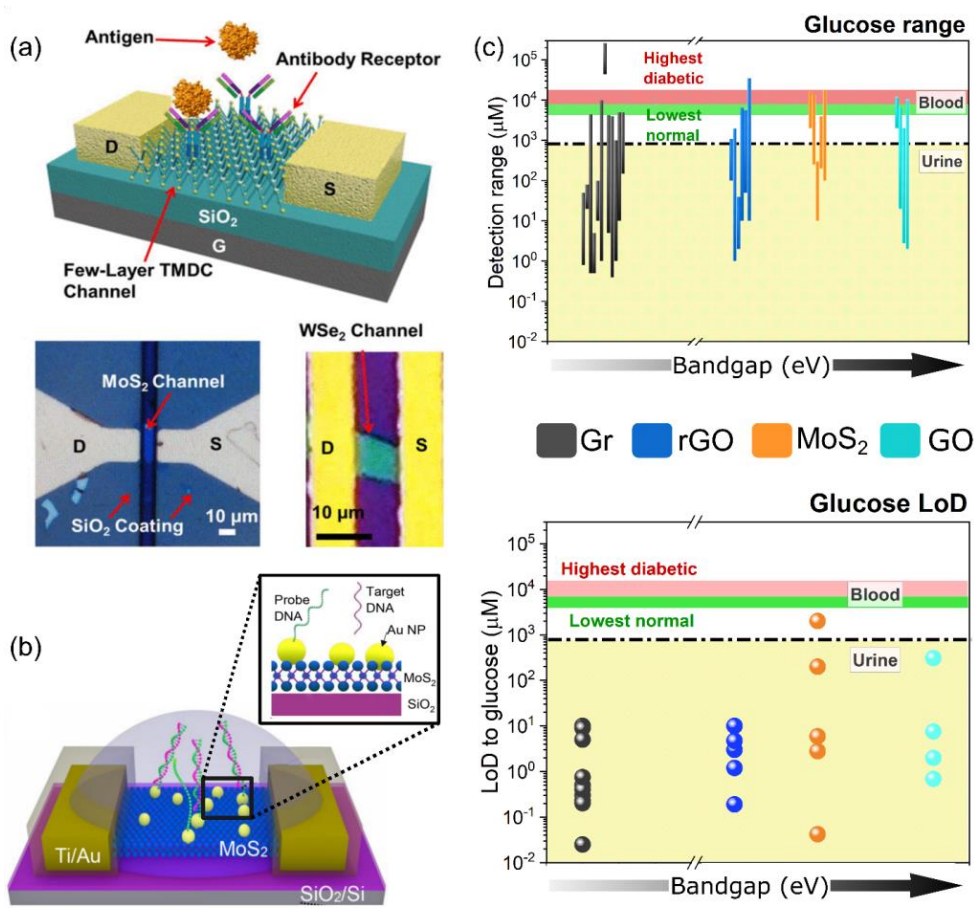

**FIG. 13.** (a) Top: The schematic diagram of TMD biosensors for detection of antibody – antigen reactions; Bottom: SEM and AFM images of $MoS_2$ and $WSe_2$ devices; Reprinted with permission from [225]. Copyright 2015, American Vacuum Society. (b) Schematic illustration of the $MoS_2$-Au NPs-DNA functionalized FET-based biosensor for screening of Down syndrome, adapted with permission from [227]; Copyright 2019 American Chemical Society; (c) Comparison of the main figures of merit: LoD (top) and MoS$_2$-based FETs have been successfully employed as a 2D platform for detections of various biomolecules, including streptavidin and biotin[228], ochratoxin[229–231], dopamine[232], anti-PSA[233,234], TNF-α[235,236], or bisphenol A[222]. We will further discuss these applications in order of increasing weight and complexity of biomolecules [Figure 12]. Specifically, for streptavidin and biotin detection (one of the strongest known binding reactions in biology), it has been demonstrated that MoS$_2$-based sensors provide specific protein sensing at concentrations as low as 100 fM. Superior performance of a MoS$_2$-based FET biosensor to graphene counterparts has been proven, e.g., ~ 70-fold better sensitivity of the MoS$_2$ biosensor was demonstrated[225,228] [Figure 13(a)]. The detection of glucose, one of the most important human biomarkers, using atomically thin 2D materials has been widely explored in recent years [237–239]. In general, two common routes have been developed, implementing both enzymatic [238,240] and non-enzymatic[239,241,242] sensors. MoS$_2$ has also often been used as a platform for the development of hydrogen peroxide, H$_2$O$_2$, an essential compound involved in many biological processes [243–247]. Similarly to glucose detection, enzymatic[244,245], typically involving an electrocatalytic reaction with haemoglobin[248], and non-enzymatic[249,250] routes have been explored.

TMDs also exhibit attractive properties for detection of small biomolecules (neurotransmitters, metabolites, vitamins, etc.). Detection and differentiation of DA (dopamine), AA (ascorbic acid) and UA (urine acid) have been demonstrated using MoS$_2$-sensors[240,251–255]. TMDs were also employed as a sensing platform for detection of biomarkers, such as carcinoembryonic antigen (CEA) [231]. Recently, MoS$_2$-based platforms were employed for detection of μ-opioid receptor, a synthetic opioid peptide and specific μ-opioid receptor agonist[256].

MoS$_2$-based sensors have also been utilized for detection of nucleic acids[257–259] and selective detection of dsDNA[260] and ssDNA[259] , where the detection mechanism was based on the different affinity of MoS$_2$ towards ssDNA and dsDNA[260]. Liu et al.[227] successfully used a MoS$_2$-Au NPs-DNA-functionalised FET-based biosensor for the screening of Down sindrome [Figure 13(b)]. The MoS$_2$ FET biosensors were able to reliably detect target DNA fragments (chromosome 21 or 13) with a detection limit below 100 aM, a high response up to 240%, and a high specificity, which satisfies the requirement for the screening of Down syndrome. In another series of works, MoS$_2$-based FET sensors have demonstrated high sequence selectivity capable of discriminating the complementary and noncomplementary DNA[261], RNA and ATP monitoring[262,263], and CEA detection[231,264].

A recent comprehensive study has highlighted important insights into the bioabsorption of CVD-grown monolayered MoS$_2$, including long-term cytotoxicity and immunological biocompatibility evaluations on live animal models [224]. The authors presented a MoS$_2$-based bioabsorbable and multi-functional sensors for intracranial monitoring of pressure, temperature, strain, and motion in animal models. A simple electrical implantable sensor based on monolayered MoS$_2$ was capable of monitoring intracranial temperature over a specified period before dissolving completely. They observed no adverse biological effects and verified that biodegradable MoS$_2$-based electronic systems offer specific, clinically relevant roles in diagnostic and therapeutic functions during recovery from traumatic brain injury [224].

Although currently significantly less explored than MoS$_2$ counterparts, WS$_2$ bioFETs have also been used for detection of glucose [237,265,266], IgE [267,268], steroid hormone (e.g., estradiol [268]) and nucleic acid aptamers (e.g., DNA [269,270] and micro-RNA [271]). Alternatively, working on a higher level of bio applications, wearable electronics or skin tattoos based on $PtSe_2$ and $PtTe_2$ with medical-grade Ag/AgCl gel electrodes have been demonstrated [272]. Specifically, in terms of sheet resistance, skin contact, and electrochemical impedance, $PtTe_2$ outperforms state-of-the-art gold and graphene electronic sensors. The $PtTe_2$ tattoos show 4 times lower impedance and almost 100 times lower sheet resistance compared to monolayered graphene tattoos, opening exciting applications in the development of advanced human–machine interfaces [272].

An interesting comparative analysis of TMD- and graphene-based biosensors has been performed recently by Bollella et al.[217]. It concluded that $MoS_2$-modified graphene platforms have shown the best results in terms of sensitivity[239,242]. In the case of $H_2O_2$ detection, the best electrochemical sensing electrode was realized with a $MoS_2$-CNT nanocomposite, which shows a wide linear range, the lowest LoD and the highest sensitivity [249]. The presence of incorporated metal NPs has been shown to improve the electrochemical performances of all sensors, where the $MoS_2$-based sensing platforms displayed the best results among the other 2D materials employed [217]. Regardless, a comparative analysis (LoD and detection range) of a large number of graphene, GO, rGO and $MoS_2$-based electrochemical glucose sensors (based on the results presented here as well as adapted from the comprehensive recent review by Meng et al.[174]) do not demonstrate an obvious advantage of one material against the other [Figure 13(c)]. The comparison shows that both the LoD and detection range vary significantly for each material. It is important to note that, although all compared materials demonstrated a generally low LoD, only a limited number of reported cases matched the clinically relevant detection range for glucose in blood [273]. Still, both carbon and $MoS_2$ based sensors could be a good platform for detection of glucose in urine, where significantly lower (only traces of the substance) levels are expected [274]. This example demonstrates a supreme importance of targeting development of 2DM sensors with a clear view on the application niche and on the underlying requirements.

*Other 2D materials*

Compared to TMDs, other 2D materials, such as black phosphorus (BP), boron and silicon nitrides, Mxenes, silicenes, etc., have been significantly less explored for biosensing applications. In many cases, the research is still limited to theoretical studies and predictions.

***Black phosphorus*** BP is a promising candidate for biosensing due to its inherent conductivity, biocompatibility and electrocatalytic properties. Recently, BP-based sensing platforms have been used for the detection of human immunoglobulin (IgG) and anti-IgG[271,275], $H_2O_2$ through immobilization of hemoglobin[276], myoglobin (iron- and oxygen-binding protein)[277] and leptin (a protein hormone, which is an important biomarker for liver diseases)[278]. However, the primary limitation for the use of BP in bio-applications is related to its relatively quick degradation due to moisture absorption and oxidation.

***h-BN*** Compared to BP, use of h-BN in biosensing provides the important advantage of chemical stability. Additionally, the good electrocatalytic performance of h-BN is highly beneficial for the development of electrochemical sensors. For example, h-BN sensors have been used for both enzymatic[279] and non-enzymatic[280] detection of $H_2O_2$ . h-BN electrodes support a high overpotential required for DA oxidation and have led to successful detection of DA in the presence of UA[281,282]. Furthermore, the simultaneous presence and differentiation between DA, AA and UA has been demonstrated[283,284] and complemented by the development of a non-enzymatic glucose sensor[285]. A h-BN-based biosensor has also been employed for detection of an important neurotransmitter called serotonin[286].

***MXenes*** have attracted significant research interest due to their metallic conductivity, hydrophilic surfaces, and good stability in aqueous environments. $Ti_3C_2T_x$-based FETs have been used for the detection of $H_2O_2$ (*via* immobilization of haemoglobin)[286–288], monitoring of hippocampal neurons (responsible for transmission of brain signals relevant to learning, emotions, and memory)[289], and for fabrication of an enzymatic glucose biosensor[290]. Moreover, Pt-doped $Ti_3C_2T_x$ biosensors have been demonstrated for detection of DA, AA, UA and acetaminophen[288]. Furthermore, MXene-$Ti_3C_2T_x$-based composites have been utilized for detection of nitrites in environmental water[291].

*Summary*

The current status of this rapidly developing field forebodes that 2D material-based bio- and electrochemical sensors can be widely used for detection of disease biomarkers in diagnostics and disease monitoring. Due to their advantageous physical and chemical properties (such as tunable conductivity, large surface area, biocompatibility, electronic anisotropy, etc.), 2D materials are well suited for continuous and real-time monitoring of specific molecules, even in complex environments such as the interior of living cells or blood serum [292]. Among other detection mechanisms, electrical schemes of detection remain the most robust and versatile for ON/OFF biosensing platforms. Most commonly they exploit FET-based platforms, which allow an additional degree of freedom (typically either through a back gate or top liquid gate) to tune the device sensitivity and response.

There are still a significant number of obstacles to overcome, biological specificity being one of the most crucial, where a large number of interfering substances in biological fluids may impact the accuracy and specificity of detection. The critical and reliable evaluation of the toxicity and biocompatibility of 2D materials is essential for *in vivo* applications. On a physical side, the limited understanding of the influence of the structural and compositional defects on sensing properties complicates reproducibility and device optimization [174]. Future developments should include novel synthetic methods, with a large degree of structural control. The development of novel hybrid materials and composites, e.g., by addition of other electroactive components, such as metal oxides, metals, graphene, or conductive polymers, is a huge advantage and has already been widely incorporated into the design of a new generation of biosensors. The improved catalytic activity and low-cost of such complex 2D hybrid materials make them a useful biocatalyst for multiple applications in environmental chemistry, biotechnology and clinical diagnostics.

## 5. Strain sensors

Recently, the application of 2D materials in micro/nanoelectromechanical systems (MEMS/NEMS) and energy conversion devices, active flexible sensors, actuators, etc., have relied on their inherit piezoelectric property in their atomically thin crystal structure. Here, we discuss several fascinating developments, mainly related to pressure, strain sensors and human-computer interfacing, where carrier generation, transport, recombination or separation is tuned through either electrical or mechanical stimuli. For comprehensive reviews in the area, we address the readers to recent publications[176,293], see also Section IV C: Piezo- and Ferro-electricity in 2D vdW materials for details.

*TMDs*

$MoS_2$ is probably the most explored piezoelectric semiconductor. The bandgap of $MoS_2$ is highly strain-tunable, which results in the modulation of its electrical conductivity and manifests itself as the piezoelectric effect. The piezoelectric effect is generally studied through application of the external strain to the devices, for example through the use of flexible substrates, where the electrons can be driven to external circuit when stretching the substrate or flowing back in the opposite direction when releasing the substrate[294]. To quantify the piezoelectric response experimentally and measure the piezoelectric coefficient ($e_{11}$), a free-standing monolayer of $MoS_2$ has been investigated using a combination of AFM probe-based nano-indentation and a laterally applied electric field with $e_{11} = 2.9 \times 10^{-10}$ C m$^{-1}$ [295]. The effect of the number of layers (e.g., odd vs. even) and orientation of crystals was also demonstrated. We further discuss applications of TMD materials in MEMS/NEMS, nanogenerators and strain-based humidity sensors.

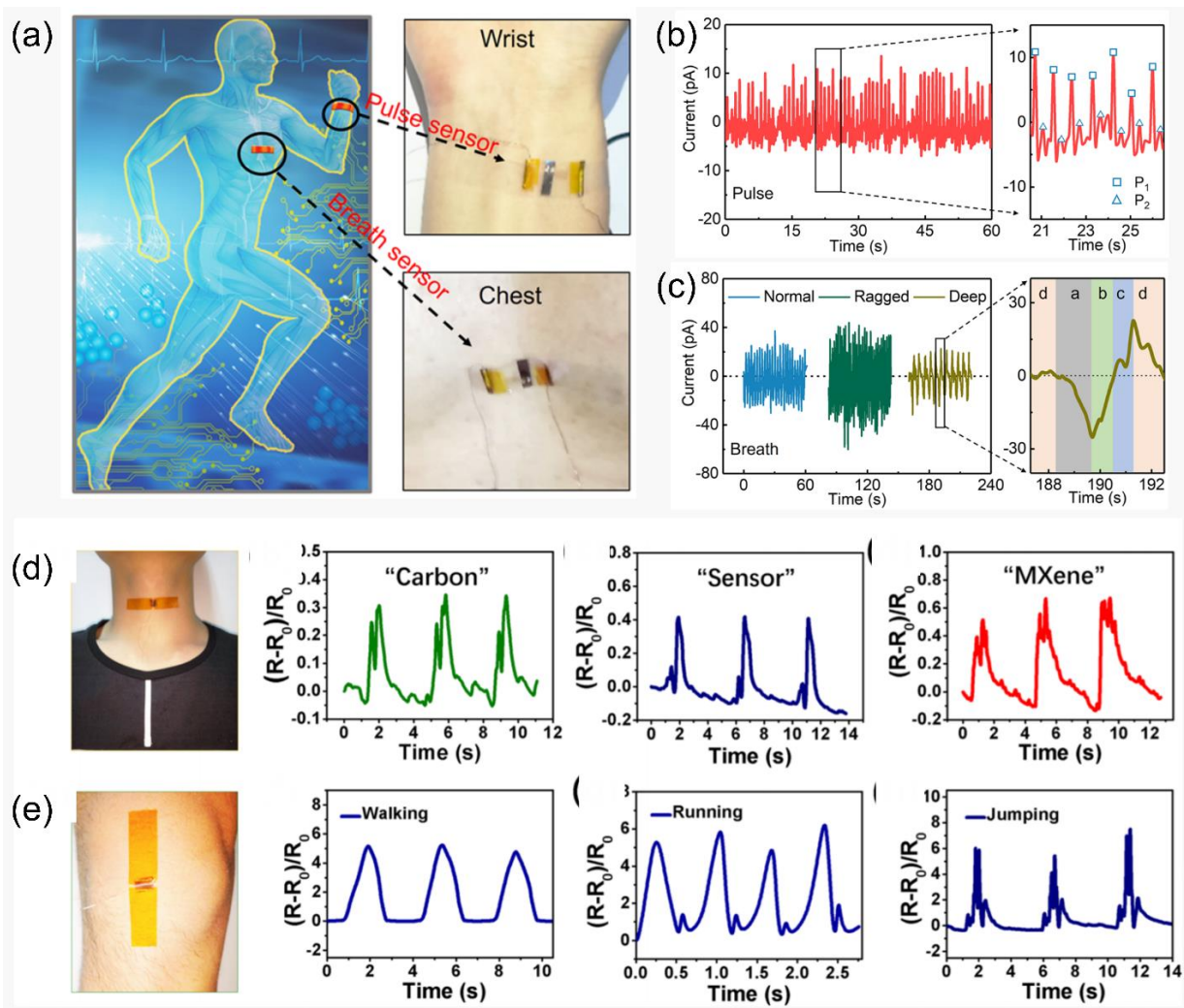

**FIG. 14. (a)** Realtime monitoring of physiological signals using a self-powered piezoelectric sensor based on a multilayer α-In$_2$Se$_3$ flake (left). The sensor is attached to the wrist (middle top) and chest (middle bottom) for monitoring arterial pulse and breath, respectively. Real-time monitoring of arterial pulse (b) and breath (c) signals. (c) Right: the time synchronization of current tracks to deep breath: a - breath in; b - breath gap; c - breath out; d - breath gap. Figures adapted with permission from [301]. Copyright 2019 American Chemical Society; **(d)-(e)** Left: photographs of a Ti$_3$C$_2$T$_x$ MXene/CNT/latex strain sensor attached to the throat and knee of a person. Right/top: relative resistance responses recorded during TMDs are model materials for MEMS/NEMS due to their atomically thin nature and coupling between electrical and mechanical properties (see e.g., Manzeli et al.[296] or Wagner et al.[297]). The calculated piezoresistive gauge factor was found comparable to state-of-the-art silicon strain sensors and higher than those based on suspended graphene. Electromechanical piezoresistive sensors were also realized in relatively-little explored 2D PtSe$_2$[297]. In this work, high negative gauge factors of up to −85 were achieved experimentally in PtSe$_2$ strain gauges. Integrated NEMS piezoelectric pressure sensors with freestanding PMMA/PtSe$_2$ membranes have been realized and exhibited very high sensitivity superior to previously reported devices. The low temperature growth makes PtSe$_2$ compatible with CMOS technology which is particularly attractive.

MoS$_2$ has also been widely studied for applications in nanogenerators [297,298], providing a new way to effectively harvest mechanical energy for low power-consuming electronics and realizing self-powered sensors. For example, in one of the earlier works it was demonstrated that a monolayered MoS$_2$ device on a flexible substrate under 0.53% strain produced a voltage of 15 mV and a current of 20 pA, corresponding to a power density of 2 mW m$^{-2}$ and a 5 % mechanical-to-electrical energy conversion efficiency[299]. In a separate study, it was shown that under applied strain of 0.48%, the output power of the MoS$_2$ nanogenerator in the armchair orientation was about twice higher than that in the zigzag orientation[300].

Realization of a novel type of MoS$_2$ humidity sensor was recently demonstrated, where exploitation of the piezoelectric effect allowed for a simple and stable way to enhance the sensor's sensitivity[194,303]. The authors showed that tensile strain generated in the sensor led to a larger current output and an enhanced sensitivity to humidity. The observed output current and humidity sensitivity were both enhanced when more electrons are moved to the conduction band under tensile strain at a positive gate bias[176]. The tunability of the sensor by strain was better achieved in a low humidity range, which was attributed to a better manifestation of piezoelectric effect when number of water molecules absorbed on the channel surface was small.

*Other 2D Materials*

**InSe** Although less explored than in TMDs, the piezoelectric effect in other 2D materials is an emerging field with many promising outcomes. Piezoelectric outputs up to 0.363 V for a few-layered α-In$_2$Se$_3$ device with a current responsivity of 598 pA for 1% strain were experimentally demonstrated, thus outperforming other 2D piezoelectrics by an order of magnitude. The self-powered piezoelectric sensors made of these 2D layered materials were successfully applied for real-time health monitoring [Figure 14(a), 14(b) and 14(c)][301]. In another work, strain sensors produced from large-scale CVD-grown In$_2$Se$_3$ exhibited two orders of magnitude higher sensitivity (gauge factor ≈ 237) than conventional metal-based (gauge factor ≈1-5) and graphene-based strain (gauge factor ≈ 2-4) sensors under similar uniaxial strain [304]. Additionally, the integrated strain sensor array, fabricated from the template-grown 2D In$_2$Se$_3$ films, displayed a high spatial resolution of ~500 μm in strain distribution, making this material platform highly attractive as e-skins for robotics and human body motion monitoring.

**MXenes** Emerging biowearables, various human-Artificial Intelligence (AI) interfaces and soft exoskeletons require urgent needs for high-performance strain sensors satisfying multiple sensing parameters, such as high sensitivity, reliable linearity and tunable strain ranges[305]. Recently, a number of fascinating studies have emerged exploring application of MXene hybrids in wearable electronics. Cai *et al.* demonstrated that a percolation network based on $Ti_3C_2T_x$ MXene/CNT composites could be designed and fabricated into versatile strain sensors[302]. The weaving architecture combined good electric properties and stretchability (attributed to the CNTs' network) and sensitivity of 2D $Ti_3C_2T_x$ MXene nanoplatelets. The resulting strain sensor was characterized by an ultralow detection limit of 0.1% strain, high stretchability (up to 130%), high sensitivity (gauge factor up to 772.6), tunable sensing range (30-130% strain) and excellent reliability and stability (>5000 cycles) [Figure 14(d) and 14(e)]. The versatile and scalable $Ti_3C_2T_x$ MXene/CNT strain sensors were proposed as a material platform for wearable AI, capable of tracking physiological signals for health and sporting applications in real-time. In a similar approach, wearable aerogel sensors that combined insulating 1D aramid nanofibers (ANFs) with conductive 2D MXene sheets demonstrated ultra-light weight, wide sensing range and good sensing ability. The resulting MXene/ANFs aerogel sensor showed a wide detection range (2.0~80.0% compression strain), sensitivity (128.0 $kPa^{-1}$) in the pressure range of 0 - 5 kPa and ultralow detection limit (0.1 kPa), opening applications in detecting human motions ranging from a light movement to vigorous loads in extreme sports. The MXene/ANFs aerogel with excellent integrated ability was proposed as a potential candidate for a human behavior monitoring sensor as well as for sensing under extreme conditions. In one of recent advanced approaches, Yang *et al.* realized wireless $Ti_3C_2T_x$ MXene strain sensing systems by developing hierarchical morphologies on piezoresistive layers and integrating the sensing circuit with near-field communication technology[305]. The wireless MXene sensor system could simultaneously achieve an ultrahigh sensitivity (gauge factor > 14,000) and reliable linearity ($\approx$ 0.99) within multiple user-designated high-strain working windows (130 - 900%). The wireless, battery-free MXene e-skin sensing system was able to collectively monitor the multisegmented exoskeleton actuations *via* a single database channel and was successfully used to assist limb rehabilitation.

*Graphene nitride* Anomalous piezoelectricity in 2D graphene nitride nanosheets (g-$C_3N_4$) has recently been demonstrated, where g-$C_3N_4$ was chosen because it naturally possesses uniform triangular nanopores and has advantageous piezoelectric properties (e.g., the linear relationship between piezoelectric response and applied voltage and the effective vertical piezoelectric coefficient of $\approx$ 1 $pmV^{-1}$)[306] .

*Germanene* By means of first-principles calculations, it was shown that in an AlAs/germanene heterostructure, both electric field and strain could be used to tailor its electronic band gap and dielectric function. Under a negative electric field and compressive strain, the material's bandgaps showed a near-linear decreasing behavior, whereas a dramatic and monotonous decrease of the bandgap as a response to a positive electric field and tensile strain was shown[307]. It was also predicted that the optical properties of the heterostructure could be improved by electric field and mechanical strain.

*Summary*

The experimental observations of piezoelectrical phenomena have yet to be fully demonstrated for all the materials predicted to be piezoelectric. In some advanced cases, such as 2D $MoS_2$ and MXenes, their piezoelectric properties have already enabled active sensing, actuating and new electronic components for nanoscale devices. Further applications of other 2D materials are expected to emerge in self-power nanodevices, adaptive biosensors, e-skins, tunable and stretchable electronics. Still, the remaining challenges remain substantial. Most importantly, the existence of interface states and disappearance of piezoelectric properties with thickness remain significant obstacles on the way to device realization. Another vital challenge is improving material stability, which should be addressed by further advances in materials processing and encapsulation technologies.

# IV. OUTLOOK ON EMERGING AREAS AND APPLICATION OPPORTUNITIES
## A. Tunable quasiparticle dynamics in 2D vdW materials

The presence of a band gap in many 2D materials lends them much more promise for optoelectronic applications than the gapless graphene. In contrast to bulk semiconductors, these materials feature very weak dielectric screening and strong spatial confinement of charge carriers. A consequence of this is the rich variety of excitonic quasiparticles which exist in these materials, particularly in 2D Mo and W chalcogenides and perovskites. The high binding energies of these quasiparticles means that they persist, even at room temperature dominating the optoelectronic behavior. This makes them exciting candidates for a range of applications, including light detection and emission, as well as spin- and valleytronic devices.

In this section of the review, we discuss the diverse range of excitonic quasiparticles that have been observed in these materials, with a particular focus on TMDs.

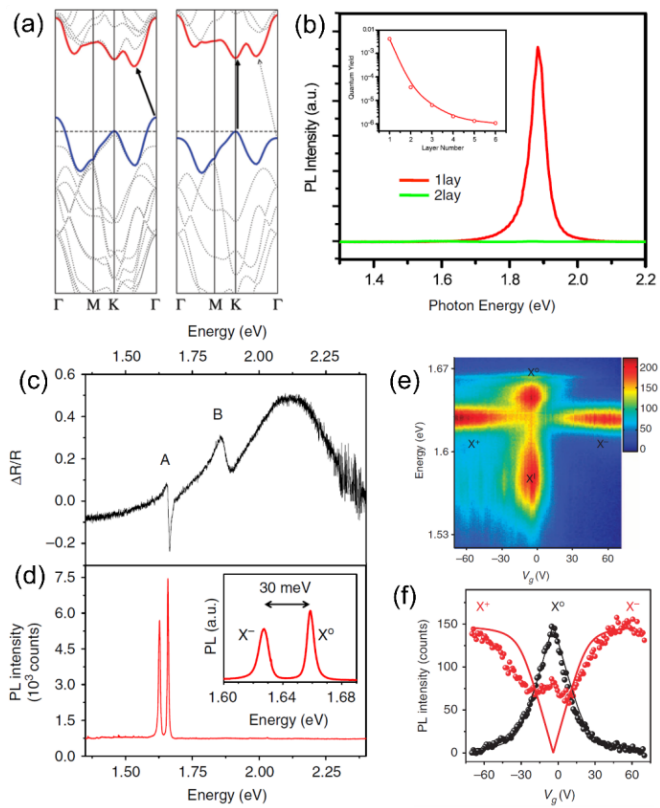

**FIG. 15.** The excitonic spectrum of transition metal dichalcogenides: (a) DFT calculated band structure, showing the indirect-direct transition in monolayered $MoS_2$. Adapted with permission from [308] Copyright © 2010, American Chemical Society. (b) Significantly enhanced photoluminescence for monolayered $MoS_2$ Reprinted figure with permission from [15] Copyright ©2010 by the American Physical Society. (c) Differential reflectance spectrum from monolayered $MoSe_2$, showing A and B excitons. (d) PL spectrum from monolayered $MoSe_2$ showing the neutral exciton ($X^0$) and negative trion ($X^-$). Inset shows the 30 meV binding energy difference in more detail. (e) PL spectra from an $MoSe_2$ FET, as a function of gate voltage, showing the change of the dominant excitonic species with carrier density. (f) Intensity of the neutral exciton and trion peaks from (e) as a function of gate voltage. Adapted with permission from [309]

### *The excitonic spectrum of TMDs*

The exciton is a quasiparticle formed of an electron and hole in a semiconductor, bound together by the Coulombic interaction. Being formed of two opposite charges, it has no net charge itself, so is often referred to as the neutral exciton, denoted by $X^0$, to distinguish it from its charged counterparts. Research into the excitonic properties of 2D TMDs was kickstarted with the demonstration that many of these materials undergo a transition from an indirect to a direct bandgap at the K and K' points, when reducing the number of layers down to a monolayer [Figure 15(a)][15,308]. This can be observed by a strong enhancement of photoluminescence (PL) caused by excitons recombining, emitting photons with energy approximately equal to the exciton binding energy [Figure 15(b)]. This indirect-direct bandgap transition has been previously reported for $MoS_2$, $WS_2$, $MoSe_2$ and $WSe_2$[310–313]. However, more recent studies have shown evidence that $WSe_2$ may remain indirect in monolayer, with a Q-valley minimum ~80 meV below the K-valley[314,315]. In addition, a very large spin-orbit coupling in TMDs has the effect of lifting the spin degeneracy of the band edges[311]. The resultant energy splitting, $\Delta_{SO}$, is particularly significant in the valence band, which leads to two pronounced resonances in the excitonic spectrum known as the A and B resonances [Figure 15(c)][309].

Beyond the neutral exciton, TMDs have also been shown to support quasiparticles called trions, often referred to as charged excitons[309,316,317]. These are composed of three bound charge carriers: either two electrons and a hole, for the negative trion, $X^-$; or two holes and an electron for the positive trion, $X^+$. These can be distinguished optically from the neutral exciton by their lower binding energies [Figure 15(d)]. For p- and n-type materials, excess charge carriers will tend to combine with neutral excitons to form positive and negative trions, respectively. This means that the excitonic makeup of the optoelectronic response, can be controlled effectively by gating 2D materials to control their carrier density [Figure 15(e) and 15(f)][309,318,319].

Furthermore, even higher order excitonic quasiparticles have been observed, including the neutral biexciton formed of two electrons and two holes, and the negative biexciton formed of three electrons and two holes. These quasiparticles are spectrally distinct from both the neutral exciton and trion, and their formation probability is enhanced by both carrier density and incident light power[320–324].

Lattice defects in TMDs may lead to Coulombic trapping potentials, resulting in the formation of spectrally narrow defect-bound excitons. These become significant at low temperatures, when the trapping potential is greater than the thermal energy of the system[325]. These defect-bound excitons have been shown to be effective single-photon sources, analogous to nitrogen vacancy centers in diamond, with emission properties tunable by magnetic field and gate voltage[326–330].

To gain a full understating of the excitonic behavior of TMDs, we also need to account for so called dark excitons, which are excitonic states that cannot typically be excited directly *via* photons due to optical selection rules, enforced by conservation of spin[331,332] or momentum[314,315,333]. Nevertheless, these states play an important role in determining the lifetime of bright excitonic states, which can relax into energetically more favorable dark states at high enough temperatures[334].

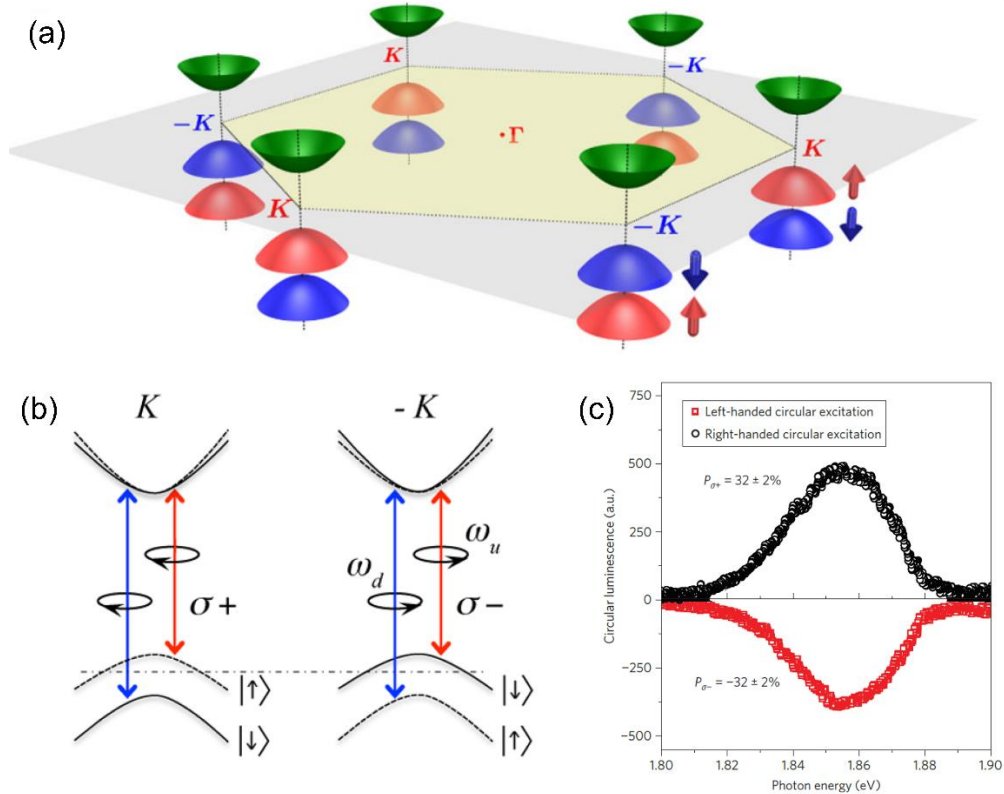

**FIG. 16.** (a) Band edge schematic showing different signs of the spin-orbit splitting for spin up and spin down carriers at the K and K' valleys. (b) Schematic showing the different optical selection rules for spin up and spin down carriers at the K and K' valleys. Adapted from [335]. Copyright ©2021 by the American Physical Society. (c) Polarization resolved PL spectra from $MoS_2$ under circularly polarized excitation. Positive (negative) values correspond to more right (left)-hand polarized emission. Adapted with permission from Springer nature: AIP Publishing, Applied physics Review [336] ©2012

*1. Spin and valley polarized excitons*

Another interesting consequence of the large spin-orbit coupling of TMDs arises thanks to circular dichroism, which these materials also possess. Time-reversal symmetry dictates that the spin splitting at the K and K' valleys must be of opposite sign[335], leading to a valley-dependent energy difference for carriers of a particular spin [Figure 16(a)]. This enables selective population of excitons within either the K or K' valley, by optically pumping with either right- or left-hand circularly polarized light [Figure 16(b)].

The degree of valley polarization can be experimentally determined by measuring the circularly polarized components of the emitted photoluminescence intensity [Figure 16(c)], which has been demonstrated for numerous monolayered TMDs[336–340].

These properties make valley polarization a controllable degree of freedom, comparable to charge and spin, that can be utilized in valleytronic devices. However, a major limitation is imposed by short, picosecond-scale lifetimes of valley polarized states, caused by strong electron–hole exchange interactions[341–344]. These lifetimes can be significantly enhanced by taking advantage of interlayer excitons in van der Waals heterostructures (see Interlayer excitons).

## 2. Interlayer excitons

A heterobilayer is formed of two distinct 2d materials, layered together in a vdW heterostructure. For TMDs, the range of different bandgaps and work functions means heterobilayers typically form vertical p-n junctions, with type II band alignment. This means that electrons and holes in the heterostructure have energy minima in different materials. Experimental measurements, with techniques including angle resolved photoemission spectroscopy and scanning tunneling spectroscopy, have confirmed that this is the case for a wide range of TMD heterobilayers [345–348]. The band structure of the heterobilayers is then further altered by hybridization of the band structures of its constituents, an effect which is governed by the momentum-varying interlayer hopping potential[349], and by the superposition of a moiré period for non-lattice matched and rotated crystals.

In these heterobilayers, an interlayer exciton (IX, often referred to as an indirect exciton) can be formed by the creation of a regular intralayer exciton (DX, or direct exciton) in either layer, followed by interlayer charge transfer, leaving a Coulomb-bound electron-hole pair with each carrier in a different material. The resulting exciton configuration is shown schematically in the insets to Figure 17(a). A great advantage of interlayer excitons is significantly enhanced lifetimes, due to the spatial separation of the carriers[350]. For charge transfer to be energetically favorable, the binding energy of the interlayer exciton must be lower than that of the intralayer exciton [Figure 17(b)]. This means that interlayer excitons can be observed in PL spectra as a lower energy peak, arising from the heterobilayer [Figure 17(a) and 17(c)][350–355].

Similar to the case of intralayer excitons, valley polarization has been observed in interlayer excitons[356–358]. It has also been demonstrated that the charge transfer process which converts intralayer to interlayer excitons, conserves the spin-valley polarization of the excited charge carriers[359,360]. However unlike for intralayer excitons, these valley-polarized states are orders of magnitude more persistent, with lifetimes on the order of microseconds. This makes interlayer excitons ripe for exploitation in valleytronic devices.

The properties of interlayer excitons can be further tuned by electrical gating. The broken inversion symmetry of a heterobilayer mean that the relative band offsets of the constituent 2D materials can be adjusted *via* the field effect. This, in combination with the out-of-plane electric dipole possessed by interlayer excitons, allows effective external control of the exciton energy. Gated $WSe_2/MoSe_2$ devices have shown gate-dependent energy shifts ~70 meV[350]. Additionally, time- and polarization-resolved PL revealed significant gate-dependent changes in the lifetimes of interlayer excitons and polarized states [Figure 17(d)][356].

Beyond electronic tunability, it has recently been demonstrated that the properties of interlayer excitons can also be altered by the twist angle between two layers, through modification of the period of the moiré pattern caused by the overlapping lattices[353,361–364].

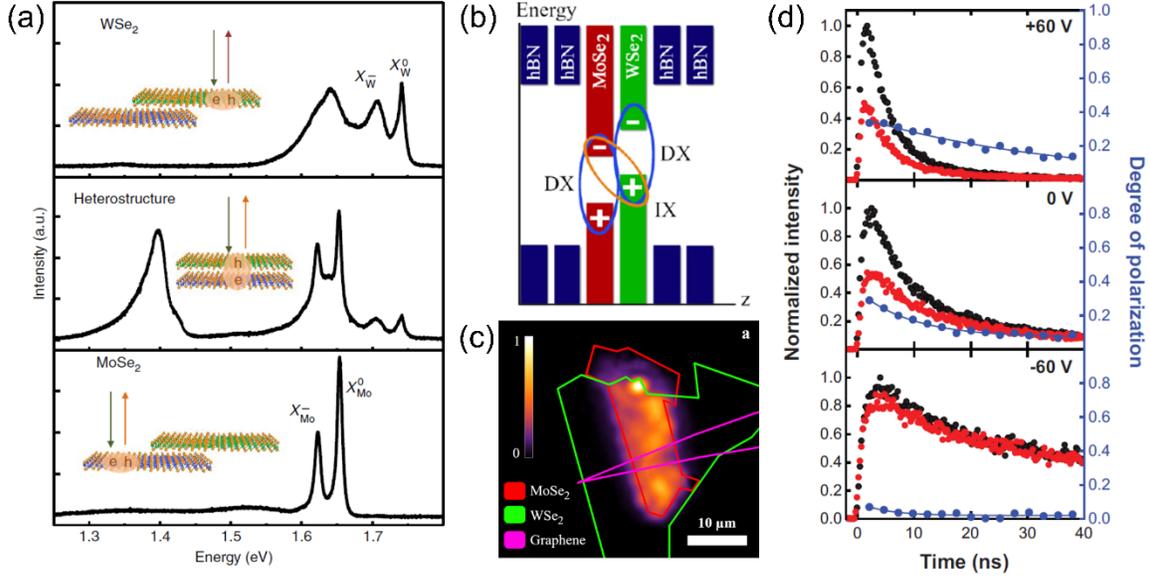

**FIG. 17.** (a) PL spectra from h-BN encapsulated WSe$_2$ and MoSe$_2$ monolayers, as well as a heterobilayer of both materials. A lower binding energy interlayer exciton can be seen in the spectrum from the heterobilayer. Figure adapted from [350]. (b) Simplified band edge alignment of an h-BN encapsulated MoSe$_2$/WSe$_2$ heterobilayer. The direct (DX) and interlayer (IX) excitons are marked with ellipses. (c) Spatial map of interlayer exciton PL intensity from a graphene-contacted MoSe$_2$/WSe$_2$ heterobilayer. The light emission is confined to the area of overlap between the two TMDs. Adapted with permission from [354]. Copyright © 2020 American Chemical Society. (d) Time- and polarization-resolved IX PL, showing right- and left-hand circularly polarized components (in black and red, respectively) and total degree of polarization (in blue) at three different gate voltages. Adapted with permission from [356].

*3. Exciton-polaritons*

The family of 2D-material quasiparticles is further expanded with the addition of exciton-polaritons. These are hybrid light-matter quasiparticles, resulting from strong coupling between an electromagnetic wave and the electric dipole associated with an exciton. This strong-coupling regime is enhanced by the large exciton-binding energies and sharp resonances in TMDs, meaning these exciton-polaritons persist even at room temperature. These materials are therefore interesting platforms to study strong light-matter interactions.

Strong coupling is typically achieved by placing 2D materials in a photonic microcavity between two mirrors [Figure 18(a)], which concentrates the local light intensity by exciting a cavity resonance[365–368]. In this geometry, exciton-polaritons in monolayered WSe$_2$ have shown room temperature valley coherence, whose phase can be effectively controlled *via* the Zeeman effect, through application of a magnetic field[369]. Additionally, in a cavity-embedded field-effect device combining WS$_2$ and MoS$_2$, gate-controlled, polariton-mediated energy exchange was demonstrated, between the excitons originating from each 2D material [Figure 18(b)][370].

An alternative method of achieving strong light-matter coupling is through the use of scanning near-field optical microscopy (SNOM). This uses tightly focused near-field light focused at the apex of an atomic force microscope probe to both excite and detect exciton-polaritons, enabling high-resolution spatial mapping [Figure 18(c)]. In multilayer slabs of TMDs, thickness-dependent internal waveguide resonances can be excited, which can interact strongly with exciton-polaritons. SNOM studies of WSe$_2$[371] and MoSe$_2$[372] have shown this results in exceptionally long polariton propagation lengths of over 12 μm [Figure 18(d)], as well as thickness-tunable polariton wavelengths.

Various further methods have been demonstrated to induce exciton-polaritons, without the need for a cavity or SNOM tip. WS$_2$ nanodiscs have been shown to support internal Mie resonances and novel anapole states, which couple strongly with excitons, forming polaritons whose energy can be tuned *via* the disc radii[373]. In monolayered WS$_2$, strong coupling has been demonstrated between excitons and waveguide modes, formed by pattering the TMD into a photonic crystal[374]. Additionally, WS$_2$ nanogratings fabricated on gold have shown strong coupling between excitons, cavity modes and plasmon polaritons[375].

For further detail on the physics behind exciton-polaritons, we point readers to previous reviews on 2D material polaritonics[376,377].

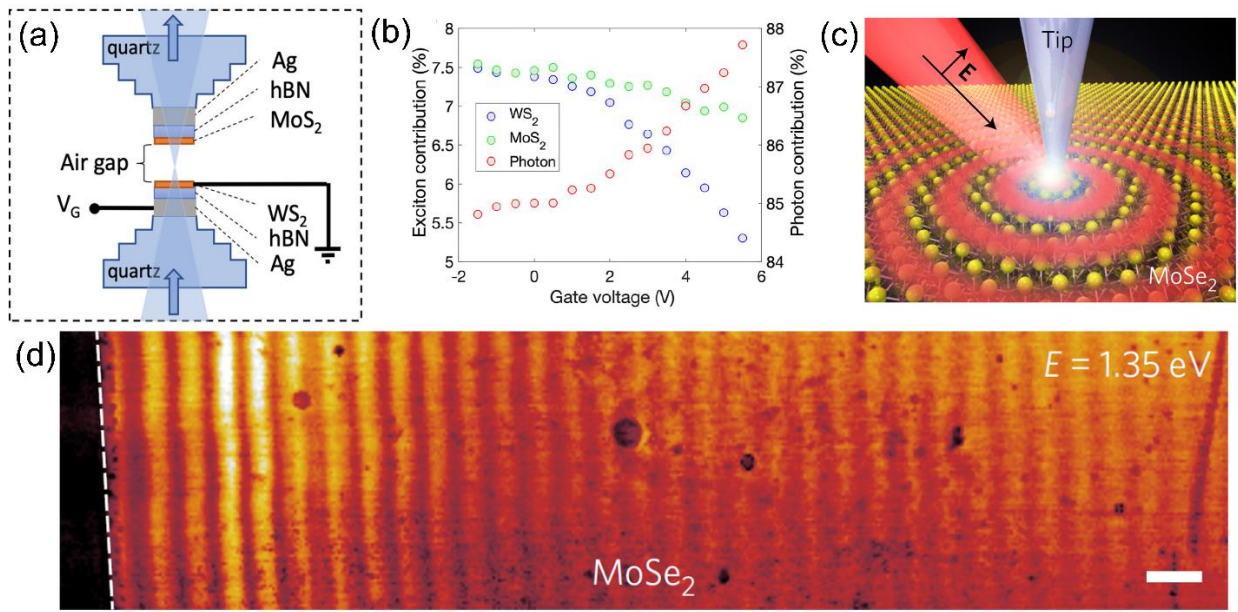

**FIG. 18.** (a) Schematic figure of a photonic microcavity incorporating MoS$_2$ and a WS$_2$-based FET. (b) Contributions to the emission spectrum from WS$_2$ MoS$_2$ and cavity modes as a function of gate voltage. Adapted with permission from [370]. (c) Diagram of a SNOM experiment, near-field light is focused at the apex of a sharp metallized tip, exciting and scattering exciton-polaritons in a 2D material. (d) SNOM image, showing long propagation lengths of waveguide-coupled exciton polaritons in MoSe$_2$ (scale bar 1 μm). Adapted with permission from [372]

## B. Electrically controlled magnetism in 2D vdW materials
### 1. *Magnetism in 2D van der Waals materials and opportunities*

Magnetism in 2D vdW materials has been sought after for engineering ultra-scaled magnetic devices and tunable magnetic phenomena in low dimensions. Theoretically, intrinsic magnetism in atomically thin materials was believed to be prohibited due to enhanced thermal fluctuations, as per the Mermin-Wagner theorem.[378] Therefore, previous efforts were focused on extrinsically inducing magnetism in 2D vdW materials through defect engineering, doping, intercalation and/or band structure engineering.[379–384]

Recently, however, it was discovered that several atomically-thin vdW materials do, in fact, sustain long-range magnetic order through the inclusion of magnetic anisotropy that opens up a spin wave excitation gap, thereby suppressing the thermal agitations and resulting in finite Curie/Neél temperatures.[18,385–387] The discovery of intrinsic ferromagnetism (FM) and antiferromagnetism (AFM) in 2D vdW materials provides unprecedented opportunities for studying various exotic properties such as spin fluctuation-driven generation of new quantum phases and topological orders, etc. In addition, it provides an ideal material platform for experimentally realizing ultrathin 2D spintronic devices for sensing, quantum information and memory applications.[388–390] Magnetic 2D vdW materials are easily integrable into heterostructures without the need of lattice matching, making them suitable for exploring emergent interfacial phenomena such as multiferroicity, quantum anomalous Hall effect and unconventional superconductivity.[391–393] The 2D heterostructures also provide access to various external stimuli such as optical, electrical and mechanical tuning of their physical properties. In particular, electrical manipulation of magnetism in 2D vdW materials provides an exciting opportunity for realizing low-power and high-speed spintronic devices compatible with existing semiconductor technology. In this section, we particularly focus on the recent advances and current understanding of electrically manipulating magnetic order of 2D vdW materials and realizing new functional devices.

Electrical control of magnetism through electrostatic gating has been demonstrated in several recently discovered 2D vdW magnets. The carrier density or the Fermi level position of a 2D magnet can be modulated, which in turn affects the magnetic exchange interactions and magnetic anisotropy. However, this state-of-the-art approach is volatile, as it requires persistent electrical control. Ferroelectric switching in magnetoelectric-multiferroic systems, where an applied electric field modifies the magnetization in 2D magnets by coupling through electrical polarization, is considered to be an effective approach for achieving non-volatile electrical control of magnetism for practical applications.

## 2. *Gate controlled magnetism in 2D vdW materials*
*Intrinsic ferromagnetic semiconductors/insulators*

Manipulation of magnetism through electrostatic gating was first demonstrated in few-layer, insulating 2D magnets such as $CrI_3$[394] and $Cr_2Ge_2Te_6$[395]. In particular, atomically-thin $CrI_3$ has been shown to exhibit intriguing layer-dependent magnetic order: each monolayer of $CrI_3$ has a FM ordering while the stacking between the layers is AFM. The weak interlayer exchange interactions are easily susceptible to external electrical perturbation, thus enabling a unique route for controlling magnetic order in $CrI_3$.[18] Both the linear magnetoelectric effect (ME) and electrostatic doping have been reported for tuning magnetism using single-gated or dual-gated field-effect devices [Figure 19(a)]. The magnetism was probed by the magnetic circular dichroism (MCD) or magneto-optical Kerr effect (MOKE) microscopy. For the linear ME effect, bilayered $CrI_3$ with graphene and h-BN as gate electrodes and dielectric, respectively, in a dual gated device was used [394]. An increase in both the magnetization in the AFM stacking state and the critical magnetic field for a spin-flip transition ($H_c$) were observed with increasing the gate voltage due to the linear ME coupling effect. A remarkable phenomenon of complete switching of the magnetic order from FM stacking to AFM stacking has also been demonstrated near $H_c$ through the application of an external electric field [Figure 19(b)]. Tuning of the magnetic order by the linear ME effect is only possible in samples with even number of layers that exhibit both broken time reversal as well as spatial inversion symmetries. In monolayered $CrI_3$, spatial inversion symmetry is still present, therefore no tuning of the magnetism through linear ME effect can be observed [394].

In contrast to linear ME effect, electrostatic doping can control magnetism in both even and odd number of layers of CrI$_3$ by controlling the doping density with gate voltage [396]. In monolayered CrI$_3$, the hole (electron) doping was found to strengthen (weaken) the magnetic order, where significant tuning of the coercive force ($H_c$) up to ≈ 75 %, saturation magnetization ($M_s$) ≈ 40 % and Curie temperature ($T_c$) ≈ 20 % was achieved [Figure 19(c)]. In bilayered CrI$_3$, $H_c$ could be continuously decreased with increasing electron doping density (*n*) where, above a certain critical density $n ≈ 2.5×10^{13}$ cm$^{-2}$, a complete transition from an AFM stacking to FM stacking state was attained due to a significant decrease in the interlayer exchange coupling with electron doping [Figure 19(d)]. It can be inferred that electrostatic gating induces both linear ME and doping effects, simultaneously. Thus, the exact mechanism for tuning magnetism through electrostatic gating remained elusive until a recent study investigated both the effects independently.[397] They found that electrostatic doping played a major role in controlling the magnetic order in bilayered CrI$_3$ [Figure 19(e)]. Thus, doping could be an efficient and more general approach in controlling the magnetism in 2D vdW magnets.

Furthermore, both ionic liquid and solid gate-dielectric gating, which can induce different levels of doping densities, have been used for tuning magnetism in vdW Cr$_2$Ge$_2$Te$_6$ [398]. The doping density with ionic liquid gating is usually ≈ 100 times higher than what can be achieved with solid state gate dielectrics. A large tuning of the magnetism with ionic liquid gating was attained for thicker (≈ 19 nm) Cr$_2$Ge$_2$Te$_6$ samples that was unlikely with solid state gate dielectrics. An ultrahigh-sensitive MOKE set-up was used to investigate the magnetization and the saturation field ($H_{sf}$) was found to reduce by a factor of two at gate voltage ($V_g$) = -4 V compared to that measured at $V_g$ = 0 V [Figure 19(f)]. However, for thinner (≈ 3.5 nm) Cr$_2$Ge$_2$Te$_6$, solid Si gating was used. Both electron and hole doping resulted in enhanced $M_s$ that was also consistent with the first principle calculations [Figure 19(g)]. Moreover, electron doping resulted in relatively larger magnetization compared to hole doping due to shifting of the Fermi level into the conduction band by filling Cr-*d* orbitals that gives rise to larger magnetizations than the *p* orbital for the valence band from Te atoms [398].

### *Layered metallic ferromagnets*

Compared with CrI$_3$ and Cr$_2$Ge$_2$Te$_6$ ferromagnetic insulators, 2D magnetic metals such as MnSe$_2$,[387] VSe$_2$[399] and Fe$_3$GeTe$_2$[385] exhibit high Curie temperatures (T$_c$ ≈130 K - 300 K). This provides an ideal material platform wherein coexisting itinerant electrons and local magnetic moments can enable interplay of both spin and charge degree of freedoms. As of yet, there are only few reports that demonstrate gate tunability of the magnetic properties of 2D metallic magnets. Recently, it has been demonstrated that doping induced by ionic gating can elevate the Curie temperature from 100 K to room temperature in Fe$_3$GeTe$_2$ thin flakes [Figure 19(h)] [400]. Under gate voltage, a high electron doping density $n ≈ 10^{14}$ cm$^{-2}$ was induced, leading to substantial shifts of the electronic bands and a large variation in the DOS at the Fermi level that profoundly modulated the Curie temperature and coercivity of the Fe$_3$GeTe$_2$ thin flakes.

### *Dilute ferromagnetic semiconductors/insulators*

Although MnSe$_2$, VSe$_2$ and Fe3GeTe$_2$ have been shown to be room temperature FM, as of yet, there has not been a report of a 2D intrinsic FM at room temperature that simultaneously possesses semiconducting nature. Thus, there is a significant effort to impart magnetism to non-magnetic materials through the creation of dilute magnetic semiconductors. There are numerous theoretical papers focused on introducing magnetic dopants into various 2D materials. Some examples include predicted ferromagnetism in Mn-, Fe-, Co-, V-, Cr- and Zn-doped MoS$_2$,[401–403] V-doped WSe$_2$,[404] and Co-doped phosphorene.[405] Specifically, for Co-doped phosphorene, the authors demonstrated that the exchange interaction and magnetic ordering could be tuned by adding holes or electrons to the system (e.g., gating), which may result in a FM to AFM transition. Experimentally, however, the challenges in synthesizing dilute 2D magnetic semiconductors lie in preventing interstitial substitutions, clusters, or alloy formations during the synthesis that would result in drastically different electronic properties of the host material and magnetism not inherent to the material. Recent reports have successfully demonstrated room temperature FM and confirmed uniform alloying through electron microscopy in Mn-doped MoS$_2$ (≈ 3 at%), Fe-doped MoS$_2$ (≈ 0.5 at%), V-doped WS$_2$ (up to 12 at%), and V-doped WSe$_2$ (≈ 0.1 at% to ≈ 4 at%).[406–410] Of these, however, the tuning of magnetism *via* electric fields has only been demonstrated thus far for V-doped (0.1 at%) WSe$_2$ by Seok Yun et al.[409] They studied the phase contrast of ferromagnetic domains under back-gate biases from -10 V to 20 V using magnetic force microscopy, observing non-monotonic variations in the phase contrast between domains at different gate biases. Although growth of these dilute magnetic semiconductors based on TMDs are currently limited to micrometer-sized individual domains, they provide the foundation to explore applications in room temperature spintronic devices.

### *Other layered 2D magnets*

AFM semiconductors such as MnPSe$_3$, FePS$_3$, etc. are theoretically predicted to exhibit transition from AFM semiconductor to FM half-metal with both electron and hole doping.[411,412] The spin polarization direction doping that can be controlled through an external gate voltage is opposite for electron and hole [Figure 19(i)]. Such an AFM to FM transition may provide a new means of magnetization switching for memory devices.

Furthermore, to give an overview of possible magneto-optical applications of so far explored 2D magnetic materials in comparison to other bulk magnetic materials a comparison between the Curie/Neel temperature vs. bandgap of 2D materials and other bulk ferromagnetic metals and insulators is plotted in Figure 20.

### *Ferroelectric switching controlled magnetism*

Ferroelectric switching in magnetoelectric-multiferroic devices is an efficient approach in controlling magnetism due to inherent coupling of magnetic and ferroelectric orders. Heterostructures combining different ferromagnetic and ferroelectric materials can exhibit strain-mediated magnetoelectric coupling which holds a great advantage in designing next generation spintronic devices [413]. To date, no experimental work on designing magnetoelectric 2D vdW heterostructures has been reported. This is most likely because most of the recently investigated 2D vdW ferroelectric materials lack out-of-plane spontaneous polarization, which is a prerequisite for magnetic manipulation. Recently, a few theoretical works have been reported with the aim to probe mechanisms for controlling magnetism in 2D vdW materials by the manipulation of ferroelectric polarization that can lead to experimental efforts in the near future.[414–418] Here, we briefly summarize some of these representative works.

Using first-principles calculations, a recent study predicted a robust control over magnetism in 2D FeI$_2$/In$_2$Se$_3$ bilayered vdW heterostructures by changing the direction of the ferroelectric polarization.[414] The magnetic order of FeI$_2$ was found to change from FM to AFM by reversing the polarization direction from +*P* to -*P* state. In -*P* polarization state, the Fe-*3d* orbital exhibited a half-filled, high-spin $d^5$ state and the direct exchange leads to a strong AFM ordering whereas, in +*P* polarization state, the Fe-3*d* orbital consisted over a half-filled, high-spin $d^6$ state that preferred the FM ordering *via* the super-exchange interaction. In another work, density functional theory was used to demonstrate a non-volatile electrical control of 2D ferromagnets in a multiferroic heterostructure consisting of monolayered $CrI_3$ and ferroelectric $Sc_2CO_2$.[415] By changing the polarization state of $Sc_2CO_2$, the $CrI_3$ 2D FM semiconductor was transformed into a half metal that was attributed to the charge transfer at the heterointerface and the broken time reversal symmetry in the FM state. Similarly, in another report the electrical control of magnetic behavior of AFM $MnPSe_3$ by changing the direction of the ferroelectric polarization in $YMnO_3$ substrate was demonstrated. A weak magnetism was introduced owing to exchange interaction between $MnPSe_3$ and surface $O_2$ atoms.[416]

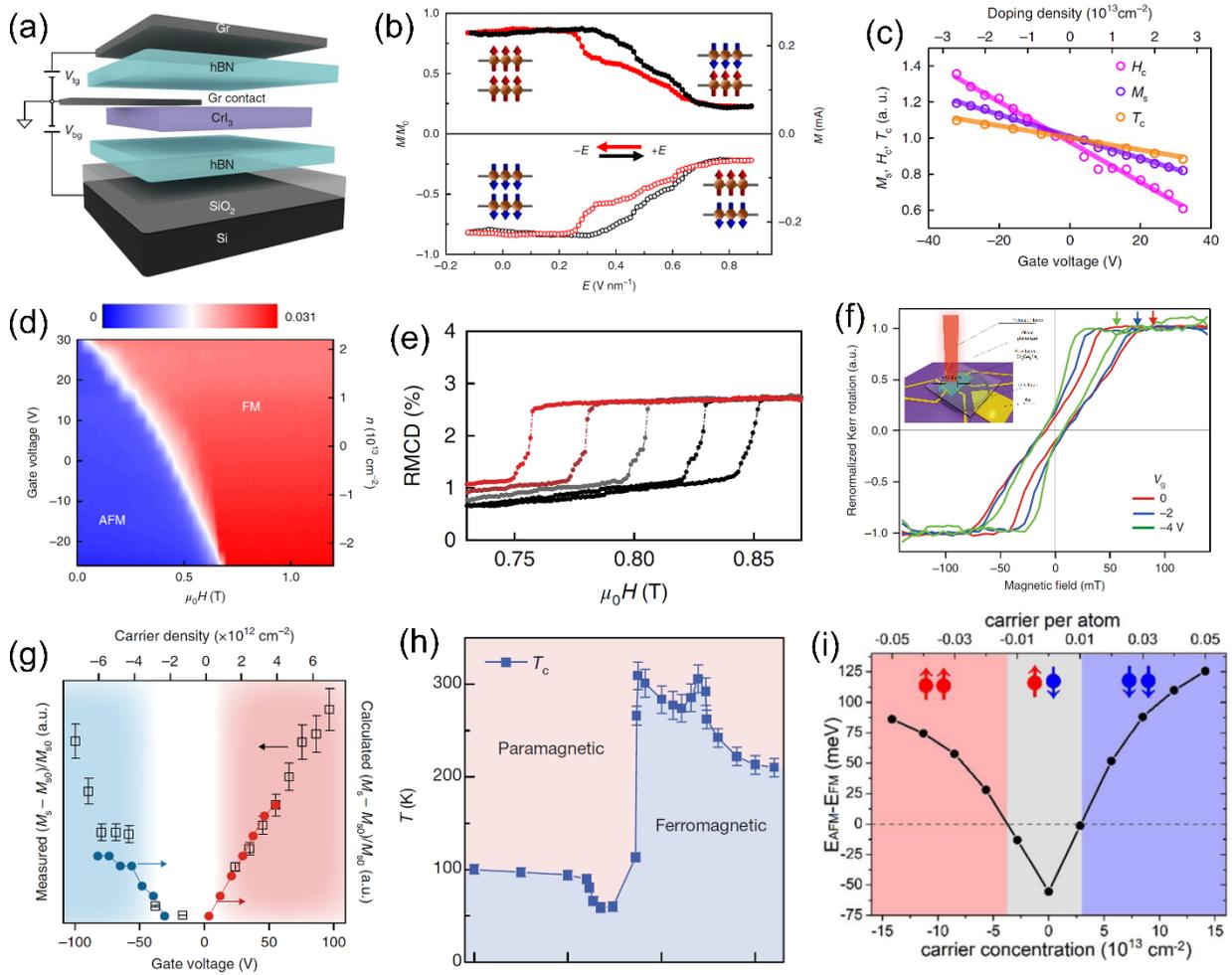

**FIG. 19**. (a) Schematic of a dual-gated bilayered $CrI_3$ device Adapted from Springer nature: Springer, Nature Nanotechnology .[397]© 2018 (b) Switching of the magnetization as a function of an applied electric field *E*. The top and bottom panels are measurements at 0.44 T and -0.44 T, respectively. Adapted from Springer nature: Springer, Nature Nanotechnology [394]© 2018 (c) Magnetic properties of monolayered $CrI_3$ as a function of gate voltage $V_g$ and induced doping density *n*. (d) Doping density-magnetic field phase diagram at 4 K for bilayered $CrI_3$ device. Adapted from Springer nature: Springer, Nature Nanotechnology [396]© 2018 (e) RMCD signal of a dual gated bilayered $CrI_3$ device as a function of applied magnetic field

at different doping levels from 0 cm$^{-2}$ (in black) to 4.4 × 10$^{-12}$ cm$^{-2}$ (in red). Adapted from Springer nature: Springer, Nature Nanotechnology .[397]© 2018 (f) Renormalized Kerr angle measured at 20 K at different ionic gate voltages of 0, -2 and -4 V. Arrows indicate saturation field for the loops measured at each $V_g$. Inset shows a schematic of the experimental set-up for Kerr measurement using ionic liquid gating. (g) Normalized spin magnetization as a function of hole and electron carrier density, respectively. Open squares are the experimental data points and solid circles with solid lines are the simulated data. Adapted from Springer nature: Springer, Nature Nanotechnology [398]© 2018 (h) Phase diagram of the trilayered Fe$_2$GeTe$_3$ sample as $V_g$ and temperature are varied. Adapted from Springer nature: Springer, Nature Nanotechnology [400]© 2018 (i) Relative energy of AFM and FM states under the variation of $n$ for 2D MnPSe$_3$ calculated with HSE06 functional. The positive and negative values are for electron and hole doping, respectively. The up and down arrows indicate up-spin and down-spin, respectively. Adapted with permission from [411]. Copyright ©2014 American Chemical Society

### *3. Gate control of critical spin fluctuations*

When a material goes through a continuous magnetic phase transition, strong and highly correlated spin fluctuations are expected near the critical point. These spin fluctuations are dependent on the dimensionality of the magnetism and dictate resulting critical phenomena. For a 1D chain, strong fluctuations typically destroy any long-range magnetic ordering whereas in 3D, there is a limited range of phase space where fluctuations become critical.[419,420] For 2D vdW magnets, on the other hand, there is a balance between fluctuations and long-range order such that they provide ideal platforms to access and possibly tune critical spin fluctuations. Recently, researchers have demonstrated real-time imaging of critical spin fluctuations in CrBr$_3$ encapsulated in h-BN *via* a custom-engineered magnetic circular dichroism (MCD) imaging technique.[421] As the temperature increased towards $T_c$, they observed spatially resolved fluctuations in the magnetization of monolayered CrBr$_3$ under zero applied magnetic field, whereas no such fluctuations were observed in a 3D bulk sample. Furthermore, an applied gate voltage using graphene as the gate material could be used to tune $T_c$, *i.e.,* how far away the system is from the critical point, thus enabling control of the critical spin fluctuations at a fixed temperature. By turning critical fluctuations on and off with a gate voltage ($V_g$ < 0.5 V) and using real-time feedback control, the magnetization state in the monolayered CrBr$_3$ can be switched between the fully spin-up state (state "1") and the fully spin-down state (state "0") by purely electrical means. This concept of switching states with critical fluctuations could potentially lead to opportunities in low-power magnetic processing and storage since, in principle, the energy cost of the switching comes from the feedback measurement.

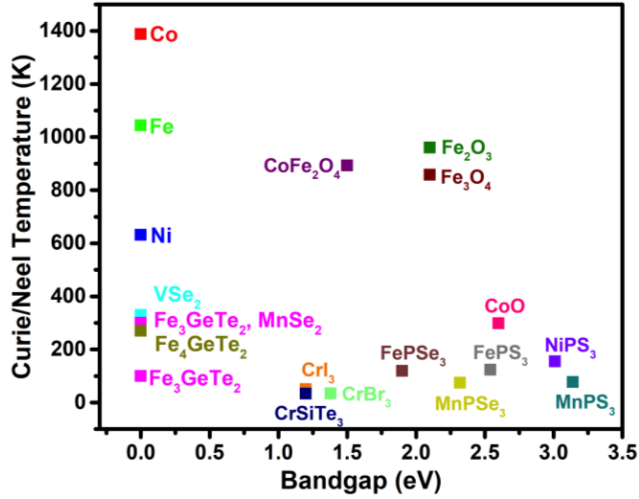

**FIG. 20.** Comparison of Curie/Neel temperature vs. bandgap of 2D materials and other bulk ferromagnetic metals and insulators. The data is adapted from [387,396,398–400,422–437].

## 4. Emerging device applications
### Magnetic tunnel junctions

Magnetic tunnel junctions (MTJs) that exploit magnetoresistance are the fundamental building block of spintronic devices. MTJs based on vdW materials offer a clear advantage of facilitating all area tunneling due to uniform barrier thickness. Recently, several research groups have successfully fabricated vdW MTJs using graphite/CrI$_3$/graphite heterostructures, taking advantage of the unique AFM stacking of few-layered CrI$_3$ [Figure 21(a)].[438–440] The CrI$_3$ functions as a spin filter tunnel barrier, whereby switching the magnetic order in CrI$_3$ with a magnetic field leads to a giant tunneling magnetoresistance (TMR). With its AFM stacking state, bilayered CrI$_3$ exhibited low current due to hinderance in the spin-conserving tunneling of electrons through the adjacent layers. However, the application of a magnetic field transformed the bilayered CrI$_3$ into a fully spin polarized state, resulting in a significant increase in current. Electrical tuning of the TMR in bilayered CrI$_3$ has also been demonstrated, where a maximum ratio ≈ 530% at 290 mV bias voltage was achieved [Figure 21(b)].[439] For trilayered and four-layered CrI$_3$ devices, the TMR peaked at ≈ 3200% and 19,000%, respectively.[439] In a separate study, electrically-controlled TMR was demonstrated using a four-layered CrI$_3$ tunnel barrier sandwiched by monolayer graphene contacts in a dual-gated structure.[441] Under fixed magnetic field, bistable magnetic states of the CrI$_3$ tunnel barrier can be switched reversibly by sweeping $V_g$ between -2.4 V to +2.4 V, resulting in different tunneling currents. Further, the $V_g$-controlled TMR was demonstrated in fully antiparallel or parallel spin configurations, which could be modulated between 57,000% and 17,000%. A combination of electrically tunable and spin-dependent tunnel barrier, Fermi level variation and magnetic proximity effects in the graphene contacts induced by CrI$_3$ were suggested to be possible explanations for the TMR modulation.

### Spin valves

Tunnelling spin valves using vdW Fe$_3$GeTe$_2$ magnetic electrodes separated by a thin h-BN layer have also been fabricated [Figure 21(c)].[442] A relative change in the orientation of magnetization in the Fe$_3$GeTe$_2$ electrodes led to a change in the tunneling resistance due to spin valve effect and a TMR as large as 160% at 4.2K, corresponding to a spin polarization of 66%, was achieved. The TMR was further tuned by applying a voltage bias between two $Fe_3GeTe_2$ electrodes. A bias dependent non-linearity in the differential tunneling conductance for antiparallel configuration and a steep decrease in TMR upon increasing voltage bias was observed, which originated due to bias dependent opening of inelastic tunnelling channels and enhanced spin relaxation rates [Figure 21(d)].[442]

*Spin field-effect transistors*

Spin FETs hold great promise for non-volatile and low power operation of the data storage devices. Spin FETs based on vdW heterostructures have been previously reported, however, their performance suffers significant challenges due to inefficient spin injection and spin relaxation in the semiconductor channels through the FM contacts.[443,444] The 2D half metals hold great promise for high-efficiency spin FETs. An electric field-induced half metallicity in $2H-VSe_2$, which is an A-type AFM vdW material, was recently investigated for applications in spin FETs using DFT calculations.[445] After applying an electric field above a critical value (e.g., E > 0.4 V/Å), the energy levels of the constituent layers were lifted in opposite directions, leading to the gradual closure of the gap of singular spin-polarized states and the opening of the gap of the opposite spin channel, thereby inducing half metallicity. The interlayer charge transfer under electric field also transformed the interlayer exchange coupling of bilayered $2H-VSe_2$ from AFM stacking to FM stacking. A switching in the polarity of the half metal was also realized by reversing the direction of the electric field. Therefore, the vertical electric field can act as a "gate" to switch the half metallicity as well as control its polarity, in analogy to conventional semiconductor transistors.[445] Recently, intrinsic 2D vdW magnetic materials, such as dual gated graphene/$CrI_3$/graphene tunnel junctions, have been exploited for designing tunnel field effect transistors with spin-dependent output [Figure 21(e)].[446] Due to spin filtering effect of few-layered $CrI_3$, which can be easily controlled *via* $V_g$ under a constant magnetic bias near the spin flip transition, the device was efficiently switched between a low and high conductance state, resulting in a high-low conductance ratio approaching 400% for a tunnel FET with a four-layered $CrI_3$ tunnel barrier [Figure 21(f)].

*Magnonic and Skyrmionic Devices*
*Electrical Control of Magnons & Magnonic Devices*

The rich physics of spin waves in magnetic materials has been attractive from both a fundamentals and potential applications standpoint.[447] Spin waves, which have quanta called magnons, are collective spin excitations in a magnetically ordered material; in a semi-classical sense, they can be understood as precessions of the spins as a wave. Magnons are able to transfer angular momentum, and thus researchers have been considering their use in information transport and processing for next generation "magnonic" devices. The major advantage of using magnons instead of spin diffusive currents to transport information is that, with magnons, the angular momentum transfer occurs without the actual motion of the electrons, thus drastically reducing losses due to Joule heating. In addition, magnons can have frequencies in the low-THz range for fast switching speeds and allow access to wave-based computing concepts.[447] Researchers have utilized optical techniques including Raman spectroscopy and magneto-optical Kerr effect, to probe magnons in $FePS_3$[448], $NiPS_3$[449], and $CrI_3$[450–453]. In addition, long-range magnon transport over several micrometers has been measured in vdW antiferromagnet $MnPS_3$, demonstrating these materials as possible channel materials for magnonic applications.[454]

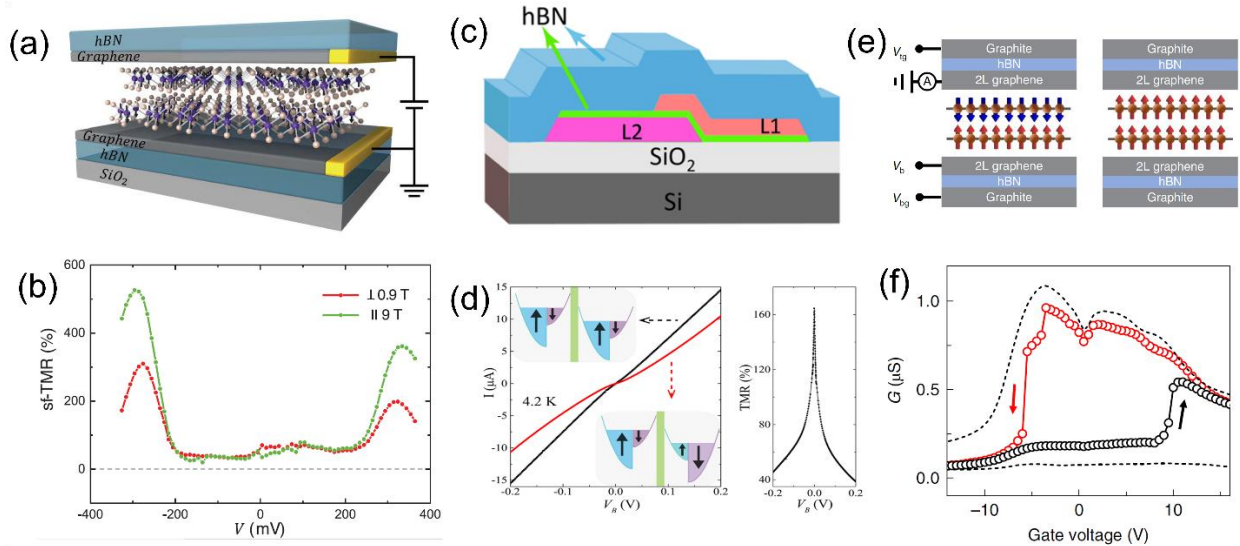

**FIG. 21.** (a) Schematic of 2D spin-filter magnetic tunnel junction with bilayered CrI3 functioning as the spin-filter encapsulated between few-layer graphene contacts.[439] (b) Spin filter-tunnelling magnetic resistance ratio as a function of bias. Adapted with permission from [439] (c) Schematic representation of the spin valve device structure Fe2GeTe3 FM electrodes (L1 and L2). (d) Bias dependence of I-V curves measured with the magnetization in the two Fe2GeTe3 electrodes pointing parallel (in black, B = 0 T) and antiparallel (in red, B = -0.68 T) to each other, and corresponding bias dependent tunnelling magnetoresistance is shown in the right panel. Adapted with permission from [442]. Copyright ©2018American Chemical Society (e) Schematic representation of the operating principles of a spin-T-FET based on a gate-controlled, bilayered CrI3 spin filter in the tunnel junction. Arrows indicate the spin orientation in the CrI3 layers. The left and right panels correspond to a low and a high tunnel conductance state, respectively. (f) Tunnel conductance of a T-FET with a four-layered CrI3 tunnel barrier repeatedly switched by gating under a constant magnetic bias of 1.77 T. Adapted from Springer nature: Springer, Nature Nanotechnology[446]© 2019

Recently, gate tunability of a 2D magnon was demonstrated by using an ultrafast optical pump/magneto-optical Kerr probe technique.[451] The sample included a bilayered CrI3- monolayered WSe2 heterostructure encapsulated in layers of h-BN, where the WSe2 served to significantly enhance the optical absorption of the pump pulse and thus increase the magnon excitation in the CrI3 [Figure 22(a)]. At a constant applied magnetic field value, the magnon frequency could be tuned continuously from 80 GHz to 55 GHz by varying the voltage from -13 V to +13 V. In addition, the saturation field $H_S$, e.g., the applied magnetic field needed to align the spins in both layers with the field, could be tuned by as much as 1 T with applied gating. Once the magnetic field exceeds $H_S$ and the spins were in the FM stacking configuration, however, the gate voltage had a negligible effect. The change in magnon frequency and $H_S$ as the gate voltage was varied were both the result of the interlayer exchange field $H_E$ and anisotropy field $H_A$ decreasing linearly with increasing electron doping [Figure 22(b)]. Being able to locally control magnons *via* gate control may open avenues for reconfigurable spin-based devices.

Magnon-assisted tunnelling in atomically thin magnet has also been studied by using a 2D FM as a barrier between two layers of graphene.[438,455,456] In contrast to non-magnetic barriers where conservation of momentum for tunnelling electrons is satisfied *via* phonons, tunneling mechanisms for FM barriers mostly consist of the emission of magnons at low temperatures. Applying a bias voltage $V_g$ across the two graphene layers resulted in a tunneling current $I_t$ through the CrX3 material (X = I, Br, Cl), where the differential conductance $G = dI_t/dV_g$ displayed step-like increases at values of $V_b$ that related to the magnon energies in the material [Figure 22(c) and 22(d)]. In addition, the tunneling conductance peaks dispersed in $V_g$ under an applied magnetic field, as expected for magnon-assisted tunneling. Tunneling through 2D FMs with magnon emission could lead to novel methods to inject spin-polarized currents into other 2D materials for spintronics, since one-magnon processes only allow one type of spin polarization to tunnel.[455]

*Skyrmions*

Recent studies have also focused on stabilizing magnetic skyrmions, e.g., nanoscale vortex-like spin-textures in 2D magnetic materials. Depending upon the type of interaction between the spins, these skyrmions can be categorized as either Néel-type or Bloch-type and are stabilized by Dzyaloshinskii-Moriya interactions (DMIs). Several research groups have demonstrated the stabilization of skyrmions in 2D magnetic materials, such as $Fe_3GeTe_2$[457,458] and $Cr_2Ge_2Te_6$[459], that were imaged using *in situ* Lorentz transmission electron microscopy (L-TEM). However, these skyrmions could be stabilized in the presence of external magnetic field applied along the out-of-plane direction [Figure 22(e) and 22(f)]. A recent study demonstrated that Néel-type magnetic skyrmions in $Fe_3GeTe_2$/[Co/Pd]$_{10}$ heterostructure could be stabilized without the need of an external magnetic field whereby, the magnetic interlayer coupling between $Fe_3GeTe_2$ and out-of-plane magnetized Co/Pd multilayers played a crucial role in stabilizing the skyrmions.[460]. The magnetic field was also not required for stabilizing skyrmions in twisted moiré superlattices consisting of 2D ferromagnetic layer twisted on top of an AFM substrate in which the coupling between the FM layer and the AFM substrate produces an effective exchange field.[461] The skyrmion properties can also be tuned with the moiré periodicity and anisotropy. For practical applications of skyrmions-based spintronic devices an electrical control of the 2D skyrmions is desired. It was demonstrated initially for bulk $BaTiO_3$/$SrRuO_3$ perovskite heterostructures that the electric field generated through ferroelectric polarization can lead to nonvolatile control of skyrmions *via* the magnetoelectric coupling effect.[462] A recent experiment demonstrated that the magnetic skyrmions can also exist in the form of bimerons in the vdW LaCl/$In_2Se_3$ heterostructure and can be manipulated by switching the polarization in $In_2Se_3$ using electric fields.[463] Owing to the low Curie temperatures of the so far investigated 2D magnetic materials, skyrmion systems can only be stabilized at low temperatures thus far, limiting the practical applications of these systems. Therefore, further efforts to stabilize skyrmions in 2D magnetic materials at room temperature without an external applied magnetic field and their efficient manipulation under external electric fields are desired. Spin-texture, including skyrmion engineering *via* materials composition, interfaces or external stimuli is therefore expected to be a major area of future research in these classes of materials.

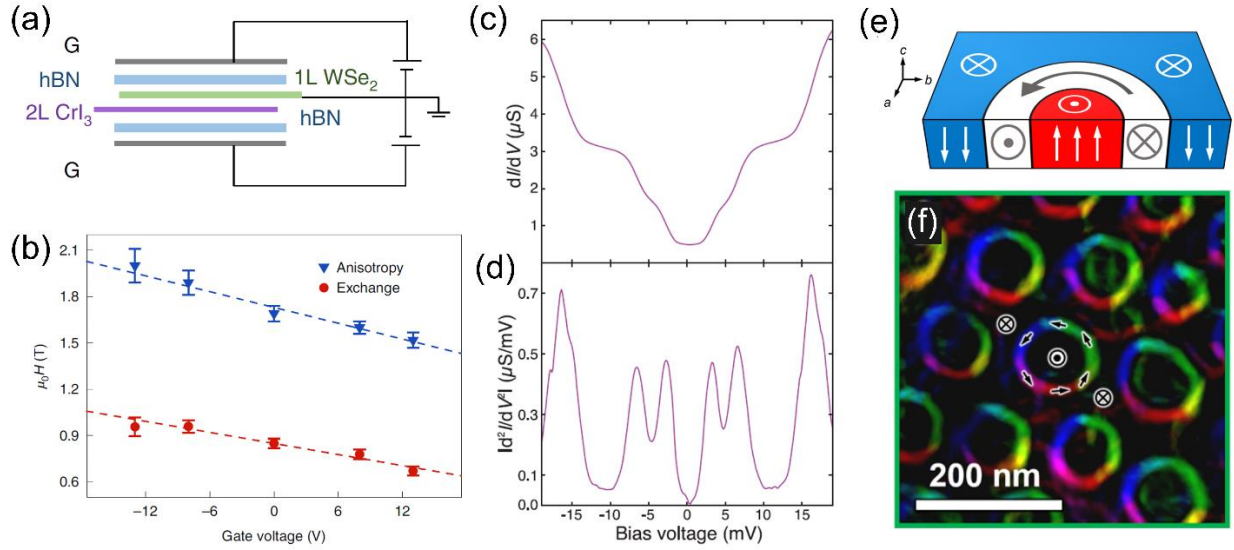

**FIG. 22.** (a) Schematic of the dual-gated device consisting of $CrI_3$-$WSe_2$ heterostructure encapsulated by h-BN. The graphite layers were used as electrical contacts. (b) Anisotropy and exchange fields extracted at different gate voltages. Adapted from Springer nature: Springer, Nature Nanotechnology.[451] © 2020 (c) Differential conductance vs. bias voltage for a bilayer $CrI_3$ barrier device at zero applied magnetic field. (d) Absolute value of $d^2I/dV^2$ vs. bias voltage extracted from differential conductance data in (c). Adapted with permission from [438] The peaks in $d^2I/dV^2$ correspond to phonon or magnon excitations of the barrier or electrodes. (e) Schematic of the skyrmionic bubble spin texture. (f) Magnetization maps of the Lorentz transmission electron microscopy data of the skyrmionic bubbles. Adapted with permission from [459]. Copyright © 2019 American Chemical Society

### C. Piezo- and Ferro- electricity in 2D vdW materials
#### 1. Piezoelectricity in 2D vdW materials

Piezoelectricity refers to the polarization in response to the applied mechanical stress originating from the non-centrosymmetry in the structure. When the size of a piezoelectrical material changes due to applied mechanical stress, the mass centers of the negative and positive charges shift and no longer coincide with each other, leading to the formation of a dipole. Dipoles inside the material cancel out with each other but those on the surface do not, which results in the polarization. Although non-centrosymmetry is generally regarded as a straightforward and effective method to describe whether a material is potentially piezoelectric [464,465], it is worth mentioning that some materials with non-centrosymmetry fail to show piezoelectricity. For example, those belonging to 432 (*O*) point group, because the piezoelectric charges developed along the <111> polar axes cancel out with each other. A significant amount of research interest has been dedicated to 2D vdW piezoelectric materials due to their ultrathin geometry, weak interlayer interaction and outstanding piezoelectric response, which makes them promising candidates for next-generation flexible and nanoelectronics. While some 2D materials do not show any piezoelectricity in their bulk form due to their centrosymmetric crystals, this inversion symmetry can be broken once they are thinned down to few layers. In fact, the percentage of non-centrosymmetric structures increases from 18 % in bulk crystals to 43 % in monolayers [4]. In this section, we will review theoretical and experimental reports on the in-plane, out-of-plane and intercorrelated piezoelectricity within 2D vdW materials.

**In-plane piezoelectricity**

*Transition metal dichalcogenides (TMDs)* Monolayers of 2H-phase TMDs materials have a honeycomb structure of which lattice points are occupied by transition metal and chalcogenide atoms alternatively, which breaks the in-plane centrosymmetry. Thus, monolayered TMDs have been long predicted as potential piezoelectric materials. In 2014, Wu *et al.* [299] reported the first experimental study on piezoelectricity in single layer MoS$_2$. Meanwhile, an alternative piezoelectric response between finite and zero was observed in few-layered MoS$_2$, due to the breaking and recovering of inversion symmetry in odd and even number of layers respectively [Figure 23(a)][299,466]. Similar layer-number-dependent piezoelectric response phenomenon was also observed in other TMDs [467].
Additionally, modulation of the bond charges by electric field have experimentally succeeded in inducing a piezoelectric response in TMD materials with even layer number [Figure 23(b)] [467]. A non-zero second harmonic generation (SHG) signal was observed in WSe$_2$, which is indicative of inversion symmetry breaking and the indication of the existence of piezoelectricity. The application of a gate voltage between -20 to 40 V led to the bond charges ($q_n$) on all six bonds to be shifted upwards. Here the average polarization remained zero and therefore, no electric-field-induced second-harmonic generation was expected. However, when the gate voltage exceeded the threshold value, a hole accumulation layer formed on the W $5d_{x^2-y^2, xy}$ orbitals led to screening of the field. Therefore, the charge on the W-Se bond was compelled to shift further toward the W atoms in the bottom half-monolayer while not affecting the top half-monolayer [Figure 23(c)]. Consequently, the polarization changes and electric-field-induced second-harmonic generation is observed. Furthermore, electric fields can also be applied in bilayer MoS$_2$ to break inversion symmetry. In this case, the nonlinear polarization led to a more hole-like character in the layer in direction of the DC field and vice versa. This resulted in a net polarization from the bilayer and therefore a piezoelectric response [468].

*Hexagonal boron nitride (h-BN) and related materials* Similar to the 2H-phase TMD monolayers, the boron and nitrogen atoms in monolayer h-BN alternately occupy the hexagonal lattice points, also leading to a piezoelectric response. Its in-plane piezoelectricity was predicted by Michel *et al.*[469] based on Born's long-wavelength theory. Later, the non-centrosymmetric structure for monolayer h-BN was probed using SHG and a similar layer dependent SHG response was also observed in multilayered h-BN films [299,466]. In addition, boron-V group binary and ternary monolayers BX (X=N, P, As, Sb or mixed) were also investigated by first principle calculations. This family of 2D materials show excellent in-plane piezoelectricity compared to conventional piezoelectric materials such as α-quartz, α-SiO$_2$ and GaN[470].

*Intercorrelated piezoelectricity*

*α-In$_2$Se$_3$* α-In$_2$Se$_3$ is a layered semiconductor and regarded as the most thermodynamically stable form among the five known phases of In$_2$Se$_3$ (α, β, γ, δ, κ respectively) at room temperature. Although the bulk structure of α-In$_2$Se$_3$ has been widely studied experimentally, the precise alignments of the atomic layers remain controversial. Possible alignments of the atomic layers for α-In$_2$Se$_3$ have been theoretically investigated by Ding *et al.*[471] using DFT. Based on the alignment of the layers, Ding predicted the existence of both in-plane and out-of-plane piezoelectricity in α-In$_2$Se$_3$. Unlike in the TMDs or h-BN, the interlayer distances in the Se(1)–In(2)–Se(3)–In(4)–Se(5) quintuple layer are not the same. The Se(3) is slightly shifted off-center toward the neighboring In(2) atom[472]. Thus, the inversion symmetry along the c-axis for each quintuple layer is broken, giving rise to an appreciable out-of-plane piezoelectricity. In-plane piezoelectricity is due to the in-plane centrosymmetry breaking. The observation of multi-directional piezoelectricity in mono- and multilayered α-In$_2$Se$_3$ was reported by Xue *et al.*[473] With a thicker α-In$_2$Se$_3$ sample, both *d$_{33}$* piezoelectrical coefficient (out-of-plane direction) and lateral piezo-response increased and presented a saturated trend up to ~ 90 nm in thickness [Figure 23(d) and 23(e)]. Multi-direction piezoelectricity in α-In$_2$Se$_3$ offers an opportunity to enable piezoelectric devices responding to strain from all directions.

*Janus structures of TMDs.* Unlike the conventional TMDs, the transition metal atomic layer in a Janus structure is sandwiched between two different chalcogen atomic layers. Therefore, the inversion symmetry along the c-axis in typical TMDs no longer exists here in Janus structures, giving rise to the multi-directional piezoelectricity. The first Janus of TMDs, MoSSe, was first discovered in 2017 and was synthesized by artificially replacing one S-atom layer with one Se-atom layer[473]. The Janus structure has a great potential in piezoelectric applications due to its outstanding piezoelectric performance, as supported by theoretical calculations. The strongest out-of-plane piezoelectric response has been theoretically predicted in multilayers of MoSTe. Its piezoelectric coefficient was around 5.7-13.5 pm/V depending on the stacking sequence, which was larger than the commonly used bulk piezoelectric material AlN (~5.6 pm/V) [474]. Additionally, 2D Janus Te$_2$Se monolayers displays a large in-plane piezoelectric coefficient ($d_{11}$) of 16.28 pm/V and out-of-plane piezoelectric coefficient ($d_{31}$) of 0.24 pm/V given its structural asymmetry and flexible mechanical properties[475].

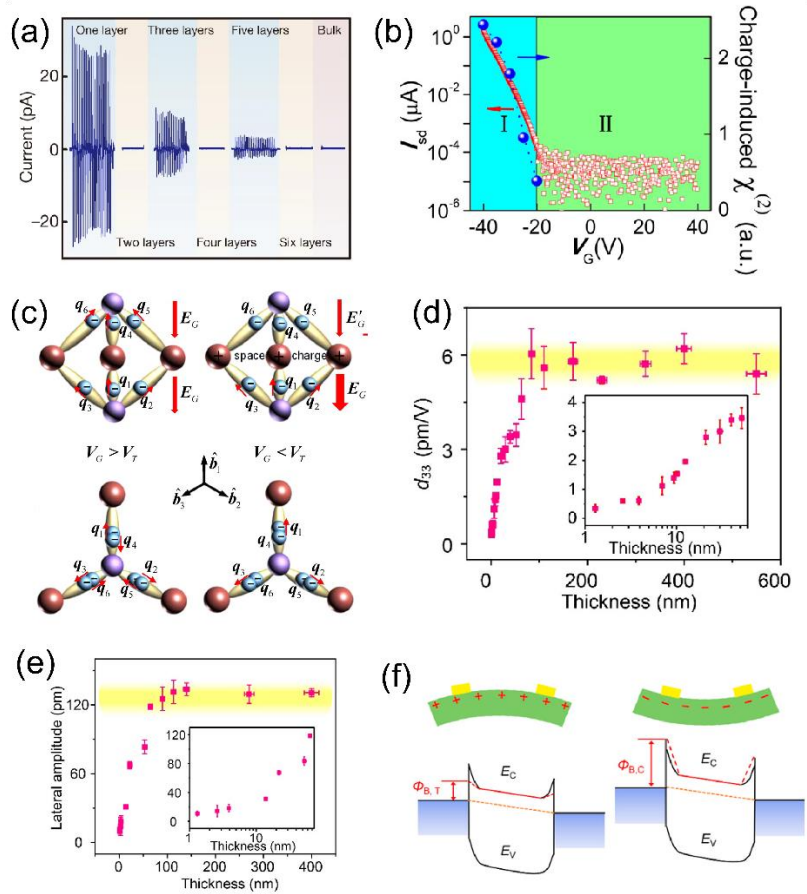

**FIG. 23.** (a) Evolution of the piezoelectric outputs with increasing number of atomic layers (n) in MoS$_2$ flakes. [299] (b) $I_{sd}$ vs $V_G$ curve of the bilayered WSe$_2$ device. The blue region (I) represents $V_G < V_T$ (hole

accumulation) while the green region (II) represents $V_G > V_T$. Here charge-induced second-order susceptibility $\chi^{(2)}$ is related to the magnitude of the square root of SHG intensity. (c) (Top) side-view and (bottom) top-view for the schematic diagram of the bond-charge distributions within one WSe$_2$ unit cell for $V_G > V_T$ and $V_G < V_T$, respectively. [467] (d) Thickness dependence of the $d_{33}$ piezoelectric coefficient of α-In$_2$Se$_3$. (e) Thickness dependence of the lateral piezo-response of α-In$_2$Se$_3$. [473] (f) Band diagrams under of (left) tensile strain with nonzero source–drain bias; (right) compressive strain with nonzero source–drain bias for In$_2$Se$_3$ thin film. [472]

*Applications utilizing 2D piezoelectric materials*

Due to their ultrathin geometry, weak interlayer interaction and outstanding piezoelectric response, 2D piezoelectrics are promising platforms for the design of complex heterostructures with desired functions, which will be needed to meet the urgent requirements of next-generation flexible and nano-scaled devices, such as electronic skin. In this session, we will briefly introduce the application of two-dimensional piezoelectric materials in actuators and strained-tuned electronics.

Actuators are an important tool in a wide variety of equipment including atomic force microscopy (AFM) and scanning tunneling microscopy (STM) for positioning objects with high accuracy. Actuators based on ultrathin piezoelectric materials may pave the way to positioning an object with extreme precision. Two-dimensional prototypical piezo-actuators based on CdS film have been proposed by Wang et al.[476]. High vertical piezoelectricity ($d_{33}$ = 32.8 pm/V) in CdS thin film was predicted using finite element calculation, where a 3-nm-thick CdS thin film was used as the actuating material and DC voltages from -1 V to -5 V were applied on its surface as driven voltages. With increasingly negative driven voltage, the deformation of the sample was linearly increased from ~30 to ~150 pm, which could be used for subatomic deformation actuators.

Another area where 2D piezoelectrics can be taken advantage of is through mechanical modulation of charge transport on flexible substrates in nanoscale electromechanical devices. It has been demonstrated that the source–drain current increases (decreases) considerably in an In$_2$Se$_3$-Pd Schottky junction device when the In$_2$Se$_3$ flake was under a small tensile (compressive) strain of ± 0.1%. The mechanism was similar to a previous report utilizing ZnO thin layers[477]. Under compression, the Schottky barrier height was lowered, giving rise to an increase in current. Otherwise, the Schottky barrier became higher and current decreased[472] [Figure 23(f)].

## 2. Ferroelectricity in 2D vdW materials

In contrast to piezoelectric materials, which only possess electric polarization upon applied external stresses, ferroelectric materials exhibit spontaneous polarization in the absence of the electric field, which can be reversed by the application of external electric field, making it possible to achieve a switching in the polarization states. Meanwhile, the spontaneous polarization represents the lack of centrosymmetry in the materials and thus all ferroelectric materials have piezoelectricity. Typically, these features only exist below the Curie temperature ($T_c$) and above the critical thickness. Below the critical thickness, materials become paraelectric because of depolarizing field cancels out with the spontaneous polarization.[478]. Recent 2D vdW ferroelectric materials have become highly attractive for device applications due to their flexible bandgap tunability and smaller critical thickness in out-of-plane direction. Thus, they are of growing importance in a variety of applications such as ferroelectric random-access memory (FeRAM), ferroelectric field effect transistors (FeFETs) and ferroelectric tunnel junctions (FTJ). In this section, we will review both theoretical and experimental reports on ferroelectricity in various 2D vdW materials with various polarization directions.

*In-plane Ferroelectricity*

*Group-IV monochalcogenides* Group-IV monochalcogenides, with formula MX (M=Si/Ge/Sn and X=S/Se/Te) have been predicted to possess large spontaneous polarization based on their hinge-like crystal structures. Interestingly, stable in-plane spontaneous polarization was discovered in ultrathin SnTe films with only one unit-cell thickness and the corresponding Curie temperature was around 270K, as reported by Chang *et al*.[479]. Strong evidence of the existence of ferroelectricity in these 2D layers were observed at liquid helium temperatures. This was evident in the stripe domain formation, the slight lattice distortion and band-bending as well as the ability to manipulate the polarization through the application of an electric field all under STM [Figure 24] [479]. Further experiments revealed that the physical mechanism behind the Curie temperature enhancement in atomically thin SnTe compared to it bulk counterpart arises form a phase transition that occurs as the 2D layers are thin down from the bulk α-phase of SnTe to the γ-phase at the atomic limit, which is an antipolar orthorhombic structure [480]. In addition, ultrathin SnTe layers with different thickness (from two unit-cell to four unit-cell) exhibited in-plane ferroelectricity with Curie temperatures above room temperature [Figure 24(d)] [479].

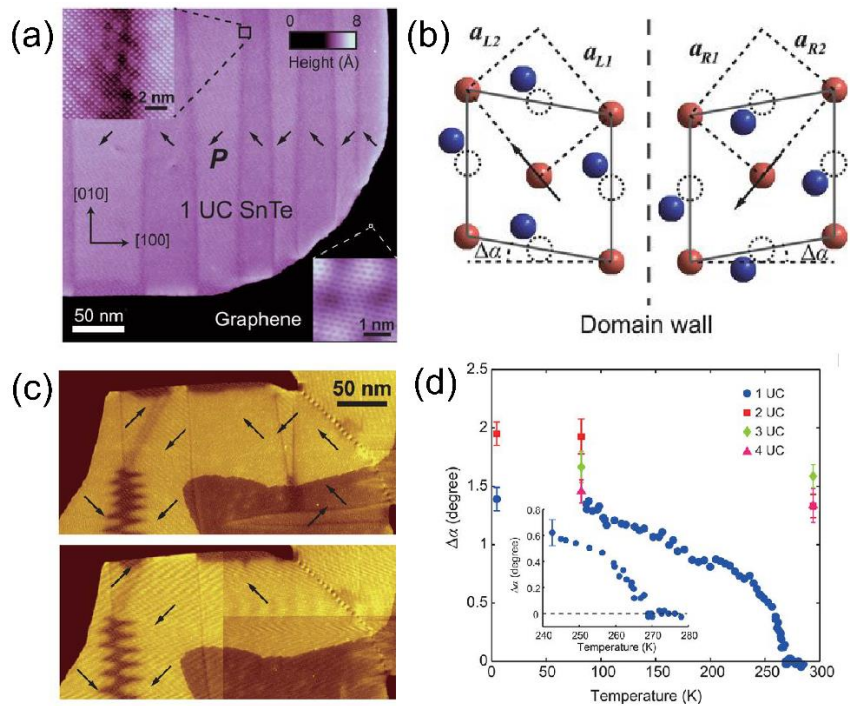

**FIG. 24.** (a) The stripe domain of a one unit-cell SnTe film. (b) Schematic of the lattice distortion and atom displacement in the ferroelectric phase. (c) Topography images of the same area before (top) and after (bottom) a 5 V voltage pulse is applied. (d) Temperature dependence of the distortion angle for the 1- to 4 unit-cell SnTe films. [479]

Based on the discovery of stable in-plane ferroelectricity in monolayer SnTe, a growing number of first-principles calculations have been carried out on other group-IV monochalcogenides to investigate the existence of ferroelectricity. 2D GeTe was predicted to possess a stable in-plane spontaneous polarization. The Curie temperature of monolayer GeTe was up to 570K and could be further enhanced by external tensile strain[481]. 2D SnS has also been proven to have in-plane ferroelectricity in experiments. Recently, Higashitarumizu *et al.*[482] put forward the ferroelectric limit of SnS towards monolayer thickness. The polarization was along the armchair direction and robust ferroelectricity was identified in SnS below the critical thickness of 15 monolayers. Due to the inversion symmetry in the c-direction of SnS, the zero net-polarization is expected in even-layer SnS. However, the inversion symmetry can be lifted by an external perturbation such as a substrate or under an electric field[482].

*Other phosphorene analogues* Both monolayered SbN and BiP possess a phosphorene-like structure, which can be regarded as a distorted NaCl-like structure where one type of ion displaces along the armchair direction with respect to the zigzag direction, leading to the spontaneous in-plane polarization along the armchair direction. Because of the relatively large electronegativity difference and displacement between Sb and N, SbN was predicted to have a much higher spontaneous polarization ($\sim 7.81\times10^{-10}$ C/m) compared to many other 2D ferroelectric materials. BiP also had a sizeable polarization ($\sim 5.35\times10^{-10}$ C/m) close to that of other strong 2D ferroelectric materials such as GeS. From the *ab initio* molecular dynamics simulation, SbN and BiP have relatively high Curie temperature (1700K and 800~900K respectively), showing the existence of stable ferroelectricity [483].

*Out-of-plane Ferroelectricity*

*TMDs* In contrast to in-plane ferroelectricity, out-of-plane ferroelectricity is fairly challenging for ultrathin materials because of the depolarization field along the c-direction. When the thickness of the material approaches a critical thickness, the depolarization becomes stronger due to the uncompensated charge at the surface/interfaces[484]. For traditional ferroelectric materials like ultrathin perovskite films, this critical thickness is about 3-unit cells[485]. In 2018, Fei *et al.*[486] reported that two or three-layered 1T' $WTe_2$ exhibited spontaneous out-of-plane electric polarization that could be switched using gate electrodes. Bi-stability near $E_\perp = 0$ was observed in the conductance G of undoped tri-layered and bi-layered devices [Figure 25(a) and 25(b)], which was characteristic of ferroelectric switching. The corresponding Curie temperature was up to 350K. However, a similar phenomenon wasn't observed in the monolayer device [Figure 25(c)]. First-principles calculations proposed that the polarization of $WTe_2$ bilayer stemmed from uncompensated interlayer vertical charge transfer depending on in-plane slippage[487]. On the other hand, Yuan *et al.*[484] reported the discovery of room-temperature out-of-plane ferroelectricity in $d$1T-$MoTe_2$ (t-$MoTe_2$) down to the monolayer limit. Local piezo-response force microscopy (PFM) hysteretic loops were recorded and the PFM contrast revealed that the polarization was anti-parallel in the two domains, confirming the existence of out-of-plane ferroelectricity [Figure 25(d)]. From cross-sectional STEM images, a few Te atoms were found to move towards the Mo plane in the out-of-plane direction by around 0.6Å while the others largely remained still. This, plus the trimerized structure, both contribute together to the rise of spontaneous polarization perpendicular to the lattice plane.

***AgBiP$_2$S$_6$*** Ferroelectricity in $AgBiP_2S_6$ comes from the off-centering $Ag^+$ and $Bi^{3+}$ ions with opposite directions. Xu *et al.*[488] predicted $AgBiP_2S_6$ exhibited stable out-of-plane ferroelectricity with a thickness of 0.6 nm, with spontaneous polarization of 1.2 pC/m. The depolarization field was overcome by the compensated ferrielectric ordering, which was similar to the previous report on $AgBiP_2S_6$[89] .

*Intercorrelated Ferroelectricity*

***α-In₂Se₃*** In 2017, α-In$_2$Se$_3$ was first predicted to possess stable room-temperature in-plane and out-of-plane ferroelectricity simultaneously based on first principle calculation by Ding et al.[471] and proven by experiments one year later. In rhombohedral (3R) α-In$_2$Se$_3$ with thickness from 1 to 6 layers, the phase of in-plane and out-of-plane ferroelectricity changed simultaneously, indicating the existence of intercorrelated ferroelectricity [Figure 25(e) and 25(f)][471]. There was a 180-degree difference between the phase contrast of the even and odd layers of α-In$_2$Se$_3$ as shown in Figure 25(g). In the even-layer, the polarization canceled out due to the interlayer antiparallel-polarization phenomenon in α-In$_2$Se$_3$. Also, polarization manipulation by electric fields was observed using PFM. For non-volatile memory devices where good stability at high temperature are required, 3R α-In$_2$Se$_3$ is attractive in that it exhibits robust and stable out-of-plane ferroelectricity down to 10 nm from room-temperature to ~200°C[489]. Intercorrelated ferroelectricity has also been found in hexagonal (2H) α-In$_2$Se$_3$ at room temperature. Such intercorrelated behavior in 2H α-In$_2$Se$_3$ is akin to that in 3R α-In$_2$Se$_3$ due to its analog structure. Most strikingly, the electric-field-induced polarization switching and hysteresis loop were, respectively, observed down to the bilayer and monolayer (~1.2 nm) thicknesses, which designated it as the thinnest layered ferroelectric and verified the corresponding theoretical calculations[490].

***CuInP₂S₆ (CIPS)*** Room-temperature out-of-plane ferroelectricity has been discovered in 2D CIPS materials dating back to 2015 and the corresponding critical thickness was reported as 50 nm[491]. Its spontaneous polarization arises from the off-center ordering in the Cu sublattice and the displacement of cations from the centrosymmetric positions in the In sublattice. Later, ferroelectricity in an ultrathin CIPS film was discovered. Switchable polarization remains as the film is thinned down to 4 nm (~5 layers) with Curie temperature around 320K, as confirmed by PFM [Figure 25(h)][89]. Meanwhile, in-plane polarization in ultrathin CIPS was not realized until 2019 by Deng *et al.*[492]. In-plane polarization did not diminish until it reached its critical thickness, which was around 90 to 100 nm. More importantly, above 90 nm, the in-plane phase changed simultaneously with the out-of-plane phase, indicating that the polarizations were intercorrelated. When the thickness was above 90 nm, Cu atoms moved toward the sulfur plane, inducing out-of-plane and in-plane displacement simultaneously. However, below 90 nm, the monoclinic structure was transformed into the more stable trigonal structure. In this case, the Cu atomic displacement only led to out-of-plane displacement and thus in-plane ferroelectricity disappeared.

***MXene*** MXenes have a layer structure resulting from the selective etching of the A atom layer of the parent phase MAX (Section II). The out-of-plane ferroelectricity in MXene originates from the asymmetric surface functionalization. Recently, Zhang *et al.* predicted three types of ferroelectric MXene phases, type-I: Nb$_2$CS$_2$ and Ta$_2$CS$_2$; type-II: Sc$_2$CO$_2$ and Y$_2$CO$_2$ ; and type-III: Sc$_2$Cs$_2$ and Y$_2$Cs$_2$ [493]. All these phases were predicted to exhibit out-of-plane and in-plane polarization along the c-axis and arm-chair direction, respectively. More importantly, robust out-of-plane and in-plane ferroelectricity have been found in the ferroelectric MXene phases, which was comparable or even stronger than the reported strong 2D ferroelectric materials such as In$_2$Se$_3$[471], GeSe and GeS [486].

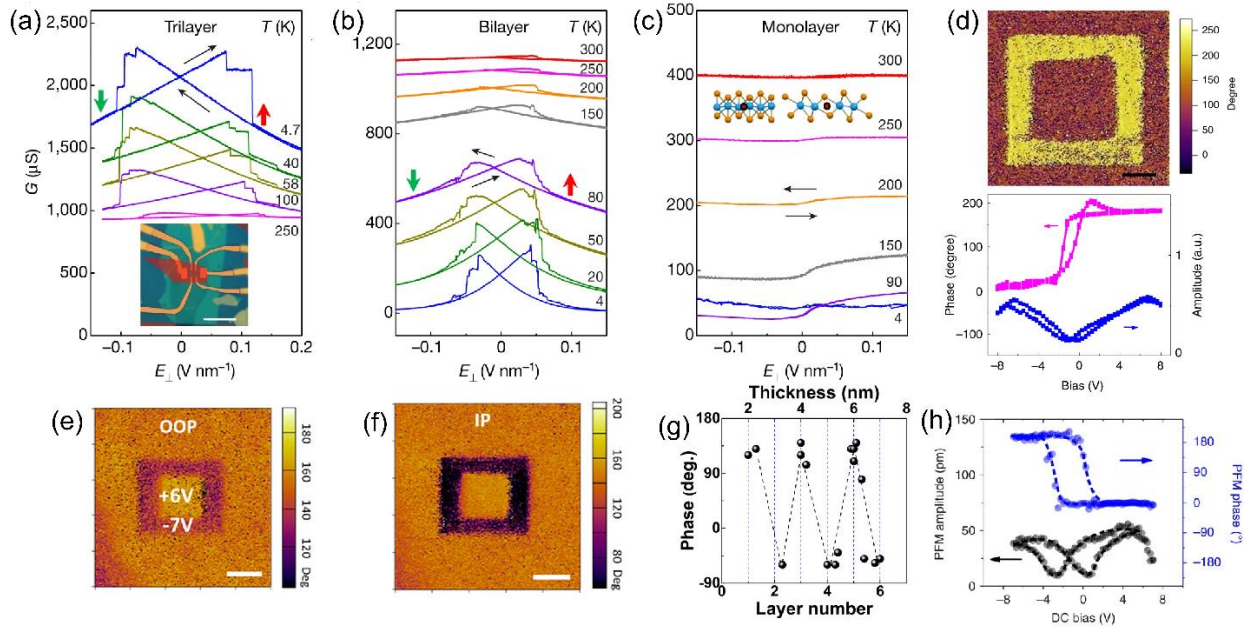

**FIG. 25.** (a)-(c) Conductance G for undoped tri-layered, bi-layered and monolayered 1T'-WTe$_2$ devices. [486] (d) Top: PFM phase image of monolayered $d$1T-MoTe$_2$ - scale bar is 1 μm. Bottom: PFM phase hysteretic and butterfly loops of monolayered $d$1T-MoTe$_2$. [484] out-of-plane phase image (e) and the corresponding in-plane phase image (f) of a 6 nm thick In$_2$Se$_3$ flake acquired immediately after writing two square patterns with a size of 2 and 1 μm by applying −7 and +6 V voltages consecutively. (g) Thickness-dependent in-plane phase (1L–6L) of In$_2$Se$_3$ flakes. [88] (h) PFM amplitude (black) and phase (blue) hysteresis loops of 4 nm thick CIPS flakes during the switching process. [89]

### 3. 2D Multiferroic Materials

Multiferroics refers to a special kind of material that exhibits more than one of the primary ferroic properties simultaneously. Among them, magnetoelectric multiferroic and ferroelastoelectric multiferroic are the most widely reported. Currently, a significant amount of research interest has been dedicated to 2D magnetoelectric multiferroic materials due to their potential applications in ultrathin ferroelectric and magnetic storage. Additionally, magneto-electric coupling provides a new path to control magnetism using an electric field and vice versa. However, it is difficult for ferroelectricity and ferromagnetism to coexist naturally in one phase. In contrast, many ferromagnets tend to be metallic, and thus screen out electric polarization[494]. Different mechanisms including lone-pair mechanism, charge ordering and more have been proposed to circumvent this anticorrelation.

Monolayer transition metal phosphorus chalcogenides (TMPCs) CuMP$_2$X$_6$ (M = Cr, V; X = S, Se) have been predicted to be a promising magnetoelectric multiferroics. Their ferroelectricity originates in the Cu atomic displacement while the magnetism arises from the indirect exchange interaction between the Cr/V atoms[494]. Moreover, Li et al.[495] reported that the introduction of a small twisted angle in vdW bilayers could result in the formation of different ferroelectric domains where some of them possess non-zero polarization. In this way, ferroelectricity could be introduced into 2D ferromagnetic materials, such as MXenes, VS$_2$, MoN$_2$, and so on. From first principle calculations, multiferroicity could also be induced by the combination of charge and orbital order.[495] . In the case of CrBr$_3$, this can be achieved by doping one electron int the primitive cell. This gives rise to an anomalous asymmetric Jahn-Teller distortion of two neighboring Cr-Br$_6$ units which brings simultaneous charge and orbital order. This resulting spatial electron-hole separation and spontaneous symmetry breaking leads to 2D multiferroicity in the CrBr$_3^{0.5-}$ system[496].

The coupling of ferroelasticity and ferroelectricity is also intensively studied for its significant potential in future ultrathin mechano-opto-electronic applications. Monolayered group IV monochalcogenides (MXs) were predicted to have ferroelastoelectric multiferroicity by Wang *et al*.[497]. Similar to monolayer black phosphorus, a well-studied ferroelastic material, MX has a hinge-like structure and thus also exhibits the ferroelastic order. The two distinct chemical elements in monolayer group IV monochalcogenides give rise to appreciable difference in the electronegativity and a large relative atomic displacement. This represents a new class of 2D multiferroic semiconductors with large in-plane spontaneous polarization, spontaneous ferroelasitc lattice strain, and small domain wall energy which are important additions to the existing realm of multiferroic bulk materials, interface structures, and thin films.

*Application for 2D ferroelectric materials*

Owing to the tunable bandgap and reduced critical thickness in the out-of-plane direction, 2D vdW ferroelectric materials are widely investigated for application to memory and logic devices, including ferroelectric random-access memory (FeRAM), ferroelectric field effect transistors (FeFETs) and ferroelectric tunnel junctions (FTJ). Among them, $d$1T- MoTe$_2$ is a promising candidate for the construction of non-volatile FTJ devices at the nanoscale limit because of its small critical thickness in out-of-plane direction ($\approx$1 layer) and a $T_C$ above-room-temperature. The corresponding *I-V* curve of monolayer $d$1T-MoTe$_2$ on Pt was measured by conductive atomic force microscopy (C-AFM) and the ON/OFF resistance ratio of the FTJ device was roughly 1000[484]. Liu *et al*.[89] also designed a Au/CuInP$_2$S$_6$/Si ferroelectric diode structure with polarization switching in the out-of-plane direction in the CuInP$_2$S$_6$ ferroelectric thin layer. The ON/OFF resistance ratio was about 100, which was comparable to tunnel junctions based on conventional ferroelectric oxides. Moreover, a lateral Au/α-In$_2$Se$_3$/Au resistive memory device with in-plane polarization switching was demonstrated. The ON/OFF resistance ratio was about 10. Also, due to the semiconductor nature of α-In$_2$Se$_3$, the device was also sensitive to visible light. In this case, a multifunctional memory device with four resistive memory states could be tuned by both electric field and light, showing potential application in advanced data storage technologies[88].

Benefiting from the weak vdW interaction between layers, complex heterojunctions could be designed with different functionalities. Si *et al*.[87] demonstrated a room-temperature FeFET device with MoS$_2$ and CuInP$_2$S$_6$ 2D vdW heterostructure, where the 4-μm-thick CuInP$_2$S$_6$ layers served as a ferroelectric gate insulator to a 7 nm thick MoS$_2$ channel. The corresponding ON/OFF resistive ratio was significantly enhanced by back-gating. Furthermore, Huang et al.[498] inserted an extra layer of h-BN between MoS$_2$ and CuInP$_2$S$_6$, which effectively stabilized the polarization of CuInP$_2$S$_6$ and further improved the device performance. An ultrahigh ON/OFF ratio up to $10^7$ as well as a large memory window was observed under $V_{GS} = \pm80$V and ultralow programming state current.

**D. Phase Modulation in 2D Materials**

At the monolayer limit, TMDs can be stabilized into two prominent crystal phases. These are the H and T phases. In both phases, the transition metal layer is sandwiched between two chalcogen layers, however, these transition metals are trigonally coordinated in the 1H-phase and octahedrally coordinated in the 1T-phase with respect to their chalcogen atoms. These two primary phases can be further developed into more phases due to different stacking sequences of layers (such as 2H- and 3R-phases) and in-plane distortion (such as 1T' and $T_d$ phases). Charge injection has been proposed as one of the major mechanisms to induce phase transformation in these 2D TMD materials. From the interaction of the transition metals with their chalcogen ligands, crystal field theory suggests that the d-orbitals of the transition metal in the H-phase splits into three energy levels ($d_{xz,yz}, d_{x^2-y^2,xy}$ and $d_{z^2}$), while in the T phase, the d-orbitals splits into two energy levels ($d_{x^2-y^2,z^2}$ and $d_{xz,yz,xy}$). The free carrier population in the 2D crystal dictates phase stability and therefore the stability can switch between the two phases depending on the number of electrons filled in those orbitals based on Hund's rule. Other pathways to induce phase transformation in these 2D TMD materials have been widely reported. [505-514] Here we review phase transformation induced through carrier density modification that will serve useful in electrically active, tunable and reconfigurable devices.

*1. Charge induced phase transformations through chemical means*

One of the most prominent examples to induce phase transformations in TMDs is through the use of n-butyllithium, which has been shown to be an excellent reagent for intercalation of lithium between the 2D layers. The corresponding mechanism is proposed to occur in the following sequence: a) generation of active species in the form of monomer, dimer or tetramer; b) surface adsorption and activation of the adsorbent; c) electron transfer from the lithium-carbon bond to the solid; d) diffusion of the alkali into the lattice and dimerization and diffusion of the alkane into the solution[501]. To further understand the transformation of the phase during this process, an *in situ* electrochemical TEM study was carried out on $MoS_2$ 2D flakes[502]. The study revealed that the intercalated lithium-ion triggered the sliding of layers in $MoS_2$ and thus inducing a transformation of the 2H phase to the 1T. The lithium ion occupied the interlayer S-S tetrahedron site in the 1T-$LiMoS_2$ structure and can be removed using mild annealing treatment, resulting in the gradual restoration of the semiconducting 2H phase[502,503]. Based on this approach, phase-engineered low-resistance contacts for ultrathin $MoS_2$ transistors were designed. Through localized conversion of the semiconducting 2H phase to the metallic 1T, abrupt lateral 2H/1T interfaces can be formed, yielding a contact resistance ($R_c$) as low as 0.24 kΩ-μm [Figure 26(a) and 26(b)] [504].

*2. Charge induced phase transformations through non-chemical means*

Using *in* situ aberration-corrected scanning transmission electron microscopy (STEM), Lin *et al.*[505] investigated phase transformations in $MoS_2$ monolayers. The electron-beam irradiation induced a variety of phase transformations, including the 2H-to-1T, 1T-to-2H and 1T-to-1T' phase transitions as illustrated in Figure 26(c), where the electron beam introduced vacancies into the lattice, inducing the phase transformation. Although joule heating is inevitable under the electron beam, these results shed light into the great possibility of achieving dynamic control over structural phase transformations in TMD monolayer devices. In fact, modulation of the phase in TMD monolayers through electrostatic gating in devices was suggested by Li *et al.*[506] using density functional theory. Excess charge introduced into the structure led to switching of the ground state from 2H to the 1T' phase. For an undoped monolayer of $MoTe_2$ under constant-stress, a semiconductor-to-semimetal phase transition was predicted to occurs as the surface charge density was within the range of less than -0.04 e or greater than 0.09 e per formula unit. Consequently, a small negative gate voltage was adequate to trigger the phase transition in monolayer $MoTe_2$. Interestingly, for $MoTe_2$, the 2H phase is the energetically favorable structure while for $WTe_2$, it is the 1T phase. Therefore, in an alloy of $Mo_xW_{1-x}Te_2$, the energy difference between the 2H and 1T' phase per formula unit can be further tuned, opening up the possibility of reducing the total energy requirement for phase transformation. [506]

In 2017, Wang et al.[507] experimentally realized the phase transformation between 2H and 1T' phases of monolayer MoTe$_2$ using ionic liquid (DEME-TFSI) gate in a device structure. Their results were supported by the hysteretic loop of gate-dependent Raman measurements, as the top gate voltage cycled between 0 and 4.4 V as show in Figure 27(a). Here, $A_1'$ (171.5 cm$^{-1}$) and $A_g$ (167.5 cm$^{-1}$) are Raman characteristic modes for 2H and 1T' phases, respectively. In addition, the orientation of the crystal was preserved before and after gating as highlighted by polarization-resolved Raman measurements in Figure 27(b) and SHG measurements in Figure 27(c). Furthermore, a 2H-1T' phase transformation can also occur in MoTe$_2$ through the introduction of tellurium vacancies, when the vacancy concentration exceeds >2%[508]. This opens up the possibility of engineering defects into the monolayer crystal, and with the aid of alloying MoTe$_2$ with tungsten, one could envision the engineering of low energy barriers for phase transformation for low voltage switching devices.

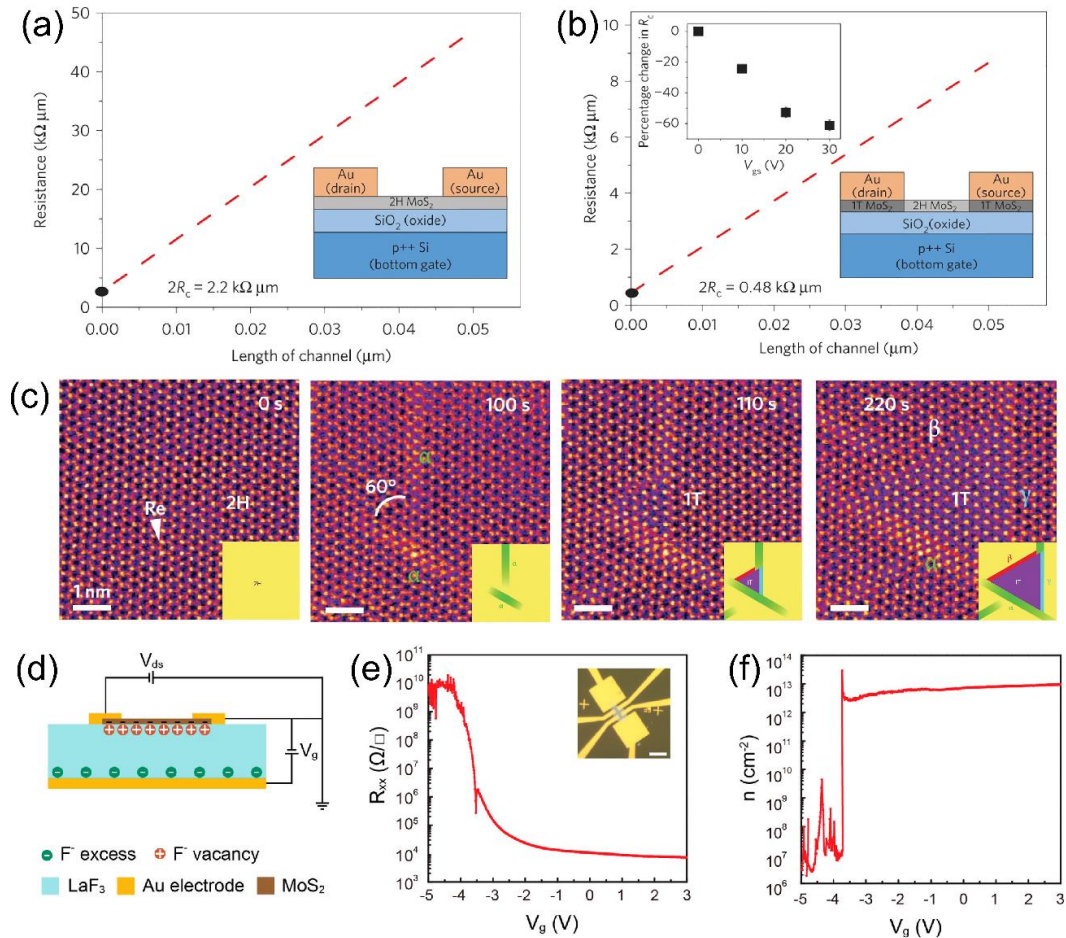

**FIG. 26.** Resistance versus 2H channel lengths for Au deposited directly on: (a) the 2H phase and (b) the 1T phase. [504] (c) Atomic movements during 2H→1T phase transformation in the monolayer MoS$_2$ under electron beam irradiation at an elevated temperature of 600°C. [505] (d) Schematic diagram of a MoS$_2$/LaF$_3$ electric-double-layer transistor when $V_g > 0$. Electronic property modulation LaF$_3$ gating: (e) gate bias-dependent sheet resistance, (f) and corresponding carrier density measured at 220K. [509]

Further optimization of the ionic liquid gating method was proposed to address some of its technical issues including possible side reactions/intercalation, challenges for characterization due to limited access to the active device surface, and strain at the interface[509]. Back-gating with solid-state $LaF_3$ substrates are promising alternatives to ionic liquid gates due to its insulating property, mechanical robustness and their affective ability to achieve high charge densities through the formation of an electric-double-layer. The fast ion conduction enables the realization of solid-state devices exhibiting comparable performance to ionic-liquid and electrolyte-based devices. To examine the tunability of the electrical properties on $MoS_2$ with $LaF_3$ gating, transfer characteristics and the corresponding sheet carrier density were measured at 220 K. [Figure 26(d)] Within a gate voltage from −5 to 3 V, the channel resistance was tuned from more than 1 GΩ/□ to several kΩ/□ [Figure 26(e)], crossing the critical value of resistance for the insulator−metal transition in $MoS_2$. The measured sheet carrier density, on the order of $10^{12} \sim 10^{13}\,cm^{-2}$ [Figure 26(f)] in the high conductance regime, hinted the possibility of phase transformation from the semiconducting to a possible semi-metallic phase.

## 3. *Field induced phase transformations*

Phase transformation in 2D materials using electric fields was first realized in STM. As reported by Zhang et al.[510], a localized 2H to 1T phase transformation in $TaSe_2$ was achieved with an electric field developed at the apex an STM tip. The high field induced collective motion of Se atoms by $\frac{\sqrt{3}}{3}a$ along one of its energetically favored directions. The transformed area of the T-phase increased with increase bias voltage at the tip. Furthermore, the changes in the Ta coordination resulted in a distinct charge density wave (CDW) state in the phases. A similar phase transformation was also observed in $TaS_2$[511]. Later on, Zhang et al.[512] reported phase transformations in $MoTe_2$ and $Mo_xW_{1-x}Te_2$ from 2H to $2H_d$ and $T_d$ phases in resistive random access memory (RRAM) devices. The $2H_d$ phase was suggested to be a transient state between the semiconducting 2H and the metallic 1T' or $T_d$ phases, which was formed upon the application of a set voltage [Figure 27(d)]. Pulse measurements also showed that the phase transition time was smaller than 10 ns, opening up the possibility of realizing ultrafast resistive switching in electrically tunable devices. Moreover, the programming voltages were tunable by varying the thickness of the TMD layers, where the ON/OFF ratio went up to $10^6$, which is promising for RRAM. Furthermore, terahertz pulsed light field excitation has also been elucidated as a means to induce phase transformations in TMD layers[513]. The terahertz pulse was shown to induce a terahertz-frequency interlayer stain shearing with a large strain amplitude in the $WTe_2$, leading to a topologically distinct metastable phase transformation. A relativistic ultrafast electron diffraction (UED) technique was utilized to reconstruct the shear motion and crystallographically quantify the corresponding atomic displacements by measuring more than 200 Bragg peaks [Figure 27(e)]. The measured diffraction pattern in equilibrium was consistent with the orthorhombic phase of $WTe_2$. The modulation of the intensities of many Bragg peaks, plus the changes observed in time-resolved SHG measurements [Figure 27(f)], indicated a structural transition between a topological phase and a trivial phase in the $WTe_2$ lattice. It was suggested that the phase transition from the orthorhombic ($T_d$) to a monoclinic (1T') phase in $WTe_2$ occurred *via* hole doping at a density of about $10^{20}\,cm^{-3}$. Microscopically, the field accelerated the electron population away from the topmost valence band, which constituted an interlayer antibonding orbital. This destabilized the interlayer coupling strength and launched a shear motion along the in-plane transition pathway from the $T_d$ to the 1T' phase with a new equilibrium position. A similar ultrafast THz-induced phase transformation was discovered in $MoTe_2$ by Shi et al.[514] with a timescale within 10 ns. An Irreversible 2H to 1T' phase transformation in monolayer and bilayer

MoTe$_2$ occurred owing to a high carrier density generated by the THz electric field through Poole-Frenkel ionization and subsequent cascading process.

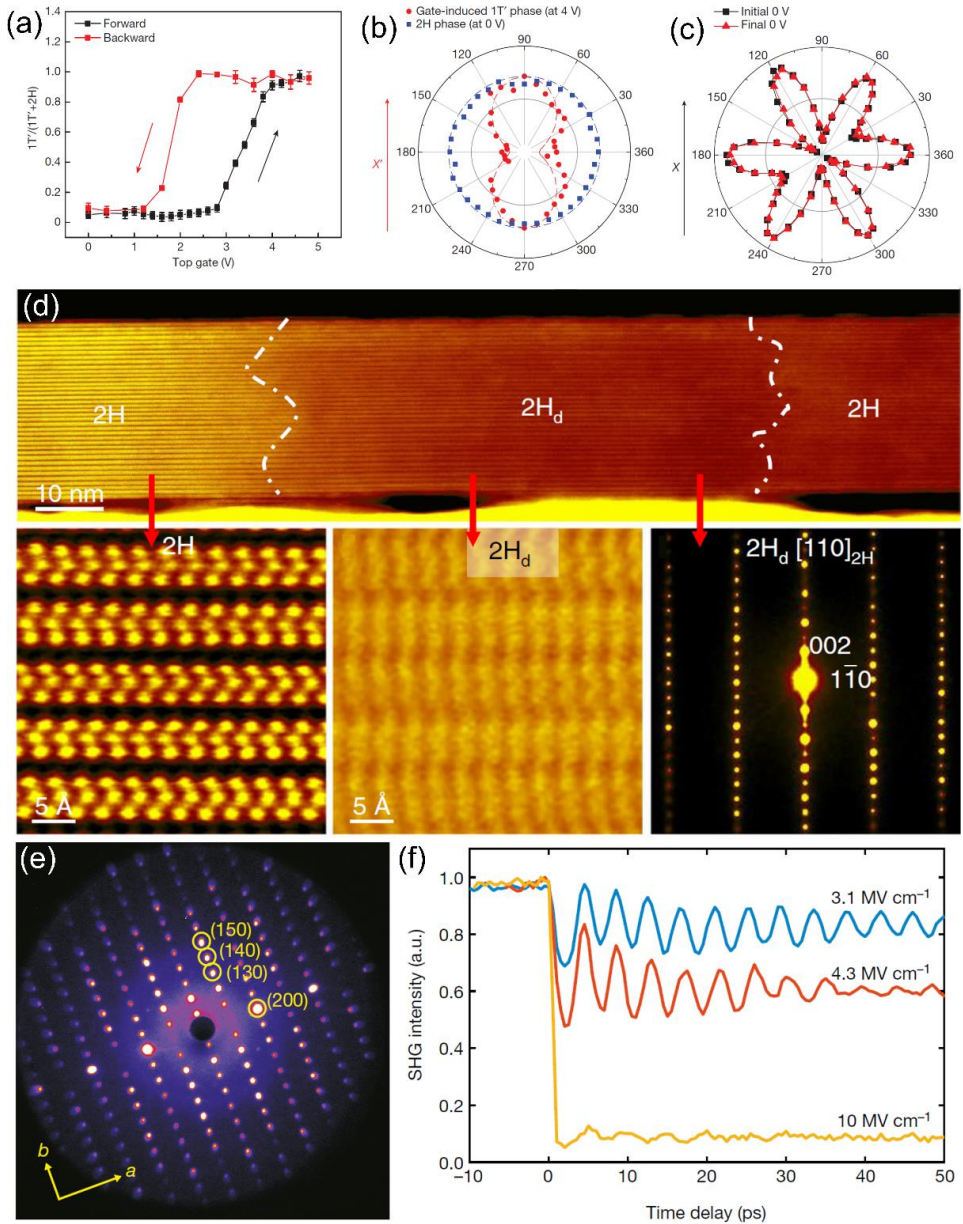

**FIG. 27.** (a) Gate-dependent Raman intensity ratios. The ratio F = 1T'($A_g$)/[2H($A'_1$)+1T'($A_g$)] (y-axis) shows hysteresis under an electrical field scan, with a loop width as large as 1.8 V, (b) Raman intensity from monolayered MoTe$_2$ as a function of crystal angle, (c) SHG intensity from the same monolayered sample as a function of crystal angle. [507] (d) Higher-magnification HAADF image of Mo$_{0.96}$W$_{0.04}$Te$_2$. bottom images: Atomic-resolution HAADF images taken along the [110]$_{2H}$ zone axis (left and middle) and corresponding nanobeam diffraction pattern for the distorted 2H$_d$ area. (right) [512] (e) Measured diffraction pattern of WTe$_2$ at equilibrium and (f) Pump-induced SHG time traces of WTe$_2$ at various pump field strengths. [513]

**E. Controlling Correlated States with Electric Fields in Twisted Heterostructures**

The field of twistronics, which investigates how the properties of a heterostructure can be tuned by the angle (the twist) between layers, has been a rapidly expanding field in the 2D vdW materials community. The aligned stacking of individual 2D materials is a well-developed technique at this time, where complex heterostructures can be created with sub-degree control over the rotational alignment of two layers. Stacking layers this way creates moiré patterns, where the wavelength of the superlattice can be tuned either by lattice mismatch or rotational alignment. This type of precise control on the resulting band structure is truly novel to 2D vdW materials, where other subsets of materials are limited by epitaxy. In the following section, we discuss how electric fields, especially used to change the carrier density in an FET, can be used to access correlated states in twisted 2D materials.

In a general sense, among some of the most fascinating states of matter are those where the interaction between particles is very strong, e.g., strongly correlated states of matter. Examples include fractional quantum Hall states and strongly correlated quantum materials, such as heavy fermions, quantum spin liquids, ultra-cold atom lattices, Mott insulators, and high $T_c$ superconductors. In condensed matter systems, strong correlations often occur when the system has a high density of states, such as in electronic "flat bands" where the vanishing Fermi velocity results in Coulomb interactions dominating over kinetic energy, amplifying the effects of electron-electron interactions.

The electronic bands of a single sheet of graphene, with their Dirac cone nature, are nowhere near flat. In 2011, however, theorists predicted that in twisted bilayer graphene (tBLG) with a small relative twist angle between the layers, exotic phenomena could occur[515]. While twist angles of $>3°$ generally resulted in the tBLG displaying the properties of two isolated graphene sheets, small twist angles resulted in moiré bands appearing in the band structure due to strong interlayer coupling. At the "magic angle" of approximately 1.1°, the moiré wavelength was extremely long and the Fermi velocity went to zero, leading to flat moiré bands and thus strong correlations. These results also indicated that the magic-angle flat bands would only contain electron densities of $\approx 10^{12}$ cm$^{-2}$, suggesting that, for the first time, back-gating could be used to significantly tune the electron density in a correlated flat band material without the need for chemical doping.

The first graphene moiré superlattices with small twist angles studied showed the presence of moiré bands, with dips in the conductance observed when the carrier density was tuned *via* a backgate to be at integer filling factors of $n_s$, with $n_s$ and $-n_s$ corresponding to +4 and -4 electrons per moiré unit cell, respectively[516,517]. Then in 2018, researchers were able to gain enough control over the twist angle to fabricate devices with magic-angle tBLG (MAtBLG), with a twist angle of $\approx 1.1°$[5,518]. By having dual-gated structures, both the carrier density and the displacement field could be tuned by gating. In addition to the insulating states at integer fillings, which could be explained by single particle band gaps, MAtBLG devices displayed a plethora of phases associated with strongly correlated electrons, including correlated insulating states at $n_s/2$ filling (where it is expected to be metallic)[518] and superconductivity[5]. The full phase diagram as a function of temperature and carrier density strongly resembled many of the fascinating properties of the cuprate superconductors. Unlike in the cuprates, where mapping out the phase diagram for different electron densities requires synthesizing a new material with different chemical doping, the full phase diagram as a function of temperature and carrier density for MAtBLG could be easily mapped out on the same sample using modest values (minimal carrier density of $1.2\times10^{12}$ cm$^{-2}$ from charge neutrality) of gate electric field to tune the Fermi energy through the flat bands. The superconductivity in MAtBLG occurs at extremely low carrier densities ($\approx 10^{11}$ cm$^{-2}$), orders of magnitude lower than carrier densities of typical 2D superconductors. These results revolutionized the field of "twistronics". Other groups have shown correlated insulating states at ¾ filling[519,520], ferromagnetism[520–522], and twist angle control of 0.02° over a span of 10 µm[523]. Electric field-tunable correlated states have now been observed in different graphene-related heterostructures, including twisted bilayered-bilayered graphene[524–526], monolayered-bilayered graphene[527], and trilayered graphene-h-BN[528,529], as well as in twisted "beyond graphene" materials such as TMDs that will be detailed below.

*1. Twisted TMD Homobilayers*

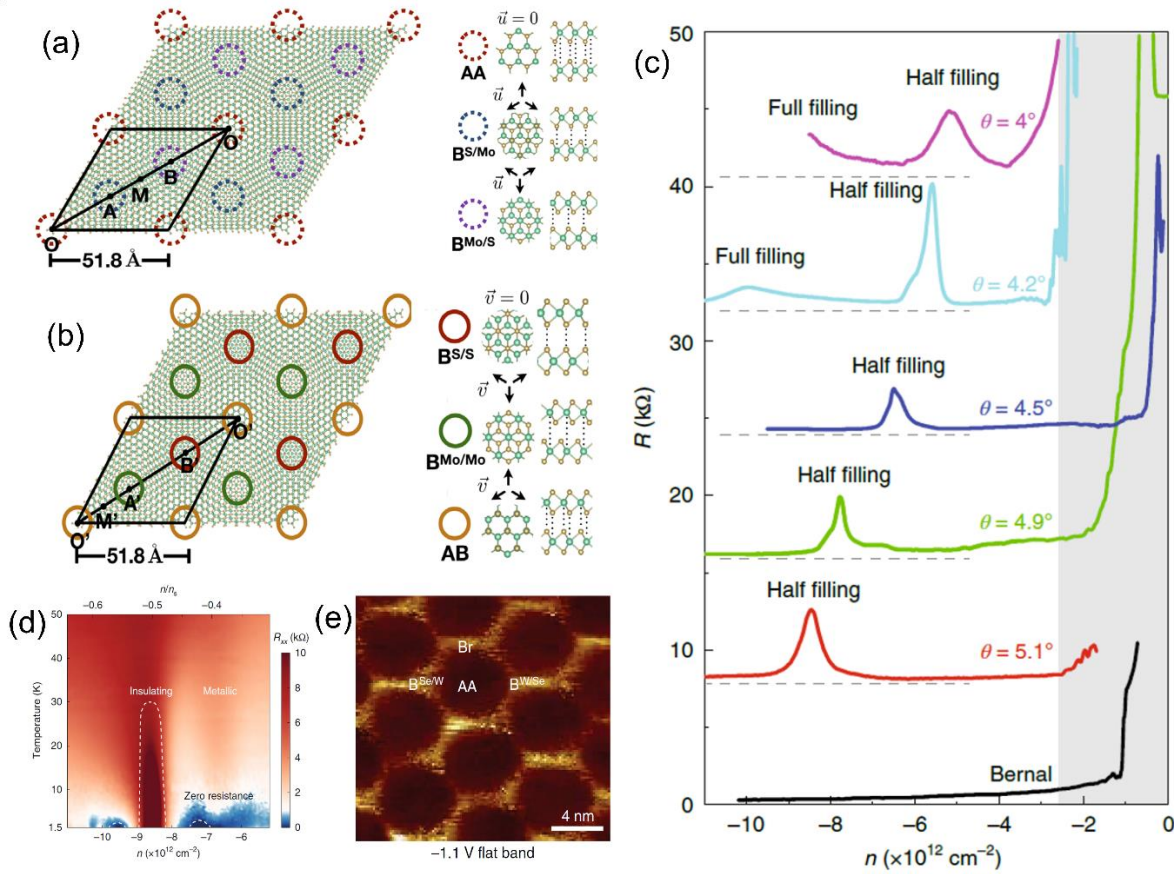

**FIG. 28.** Comparing moiré superlattices for twisted bilayered $MoS_2$ at (a) 3.5° and (b) 56.5° twist angle, respectively. Stacking sequences at high symmetry points are shown in the circles. Reprinted figure with permission from [530]Copyright 2018 by the American Physical Society (c) Resistance versus carrier density for five different twisted bilayered $WSe_2$, as well as Bernal stacking (equivalent to 60° twist). Curves are offset for clarity and measured at different displacement fields to capture the peak in resistance. Dashed lines indicate zero resistance for each curve. (d) Resistance as a function of carrier density and temperature for $tWSe_2$ ($\theta = 5.1°$, $V_{tg}$ = -12.25 V). When the device is doped away from the correlated insulating state, two zero-resistance domes appear, suggesting possible superconducting behaviour. Reprinted figure with permission from [534] (e) Local density of states mapping at the flat-band energy for $tWSe_2$ ($\theta \approx 3°$), showing the flat band is localized on the hexagonal network separating AA sites. Reprinted figure with permission from [533].

With all the fascinating physics revealed in tBLG, researchers have begun to look towards twisting other 2D materials. The "magic angle" nature of tBLG, where correlated states were observed at angles only slightly larger than 1°, poses a significant experimental challenge of growing large area heterostructures without introducing significant strain or layer reconstruction. Thus, another 2D twisted system to study correlated physics by backgating but without the constraint of specific magic angles is highly desired. Theoretical reports have shown promising results for twisted "beyond graphene" materials, predicting ultraflat bands with narrow bandwidths in twisted bilayers of $MoS_2$, $WS_2$, $MoSe_2$, $WSe_2$, $MoTe_2$, and h-BN[530–532]. Unlike tBLG, twisted bilayers of TMDs (tTMDs) form two distinct moiré patterns at small twist angles near 0° and 60°, where inversion symmetry is broken (present) for 0° (60°), as shown for $MoS_2$ in Figure 28(a) and 28(b), providing an additional degree of control[530,533]. For homobilayered tTMDs, exotic phenomena was predicted to occur by applying a vertical electrical field, such as quantum anomalous Hall insulators, Mott insulating phases, spin-liquid states, or topological insulating phases[530,531]. For twisted bilayers of h-BN[532], multiple flat bands were predicted to appear that were well isolated from other bands, in contrast with the single pair of flat bands in MAtBLG

Low-energy flat bands and strongly correlated electrons have now been observed in twisted bilayered $WSe_2$ (tWSe$_2$) using both transport[534] and STM/STS[533] measurements. Lei Wang *et al.*[534] studied the resistance as a function of carrier density for five tWSe$_2$ devices with twist angles ranging from 4° to 5.1°, all of which displayed a resistive peak at one hole per moiré unit cell (half filling of the moiré sub-band) at slightly different densities for the five different twist angles [Figure 28(c)]. This resistive peak was not seen for a naturally occurring Bernal stacked bilayer, and is not expected in a single-particle band structure, indicating the presence of strong correlations within the low-energy band. By varying both the top gate and back gate voltages, it was found that the insulating state was only present for a finite range of displacement fields. Inspired by the results on MAtBLG, they also investigated whether the strong correlations could result in superconductivity in tTMDs. At densities near the correlated insulator state for the 5.1° twist angle, there were regions where the resistance reached below the instrumentation noise level (~10 Ω), suggesting a possible superconducting transition below 3 K [Figure 28(d)]. Unfortunately, the large contact resistance due to Schottky barriers in tTMDs hampers direct verification of the superconducting state. In STM/STS measurements, Z. Zhang et al.[533] also measured flat bands in tWSe$_2$ and showed they are highly localized within the moiré superlattice. Specifically, the spatial distribution is different for 3° vs. 57.5° twist angles, where the filled flat band for 3° tWSe$_2$ is localized on the hexagonal network separating the AA sites, as shown in Figure 28(e), whereas for 57.5° the first flat band is localized on the AB sites.

These early observations on homobilayered tTMDs have already confirmed theoretical predictions that the gate-tunable correlated states in tTMDs would be present for a continuum of twist angles, as opposed to the "magic angle" observed in tBLG. Although this research field is just in its infancy, we expect significant advancements on tTMDs in the next few years, following in the footsteps of tBLG. One benefit to using tTMD homobilayers, as compared to heterobilayers that will be discussed next, is that the "tear and stack" technique for fabricating twisted heterostructures allows for more precise control over of the relative twist angle.

## 2. TMD Heterobilayers

When bilayer heterostructures are made from two different TMDs, no twist angle is necessary to create a moiré superlattice due to the lattice mismatch between the layers. For example, the 4% lattice mismatch between WSe$_2$ and WS$_2$ results in a moiré pattern with an 8 nm period when the two layers are stacked at a near-zero twist angle. One of the exciting discoveries of these heterobilayer systems is they can simulate the Hubbard model[535,536] much better than tBLG. In the Hubbard model[537], which is a simple model to describe interacting particles in a crystal lattice, there are only two terms in the Hamiltonian: a kinetic hopping term ($t$) for particles tunnelling between sites and a potential term ($U$) for on-site Coulomb repulsion, where the physics of the system is governed by the ratio of $U$ and $t$. The Hubbard model accurately captures the essential physics of metal-insulator phase transitions, high temperature superconductors, and other quantum many-body ground states[538,539]. Despite its simplicity, the Hubbard model is difficult to solve for dimensions higher than one. Thus, for the 2D case researchers are searching for solid-state systems that can simulate the Hubbard model to help elucidate the physics of many strongly interacting particles.

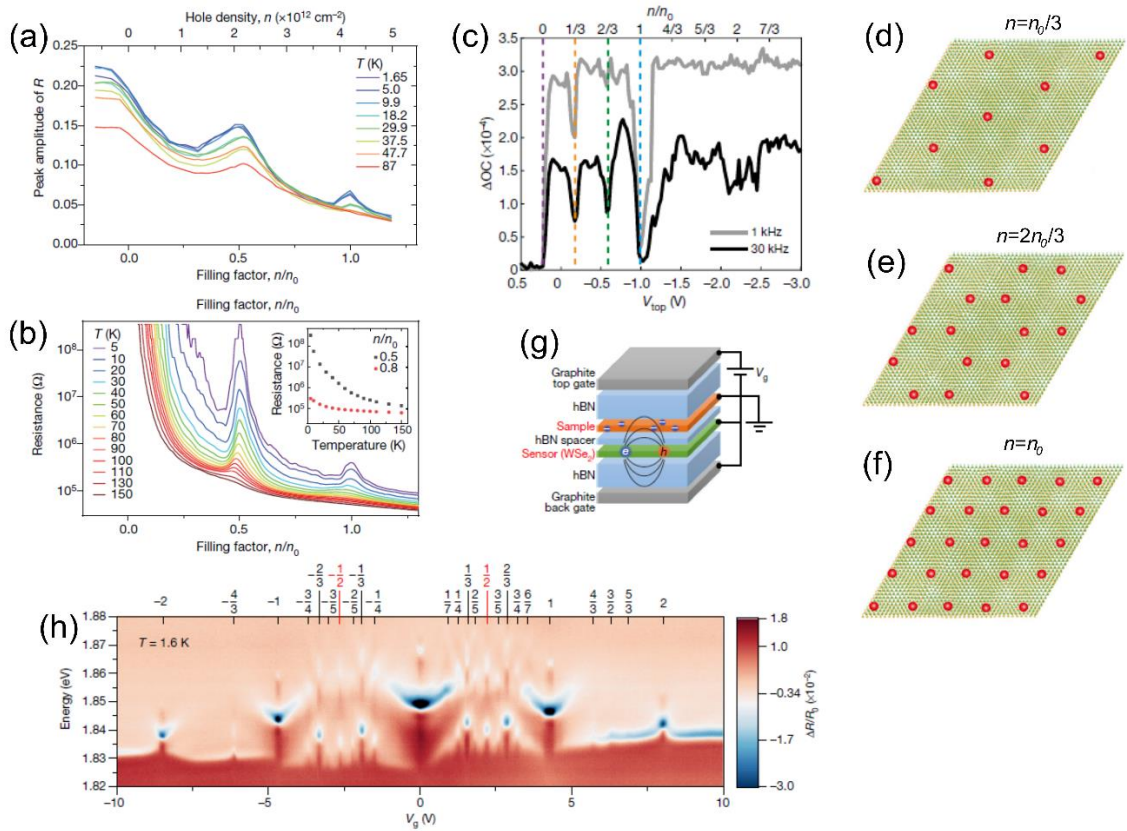

**FIG. 29.** Correlated insulating states in WSe$_2$/WS$_2$ angle-aligned heterostructures. (a) Reflectance at the lowest-energy exciton resonance versus filling factor (bottom) and hole density (top) for various temperatures. (b) Resistance (two-terminal) versus filling factor for the same heterostructure as in (a). Both show two insulating peaks. Reprinted figure with permission from [536] (c) Optically detected resistance and capacitance signal with increased hole doping, displaying gap-like features at n = n$_0$/3, n = 2n$_0$/3, and n = n$_0$. Schematic of generalized Wigner crystal (d and e) and Mott insulating state (f) in the WSe$_2$/WS$_2$ moiré superlattice, where red dots illustrate holes. Reprinted figure with permission from [540] (g) Cartoon illustration of device structure where monolayered WSe$_2$ is used as a sensor to probe insulating states in In 2018, it was theoretically shown that flat, isolated moiré bands in TMD heterobilayers could be described by a generalized triangular lattice Hubbard model when the chemical potential was within the topmost valence bands, where both $U$ and $t$ could be tuned by varying the twist angle[535]. For the specific case of WS$_2$/WSe$_2$ heterostructures, the highest energy moiré valence band in the WSe$_2$ provided a realization of the triangular lattice Hubbard model if the twist angle was smaller than 3.5 degrees. When the carrier density was tuned to half-filling, they predicted spin-liquid states were likely to occur. Almost two years later, experiments have now caught up with the theoretical predictions, where the phase diagram of the 2D triangular lattice Hubbard model has been obtained using a combination of optical spectroscopy of moiré excitons and transport measurements[536]. As expected in the strongly-interacting regime of the Hubbard model, they observed a Mott insulating state in both optical and transport measurements [Figure 29(a) and 29(b)] with AFM Curie-Weiss behaviour at half filling (one hole per moiré site) of the first hole moiré band. The ability to gate-tune the carrier density allowed for the phase space of the Hubbard model to be explored over a wider range of carrier densities and temperatures than what is normally possible in cold atoms in optical lattices.

Additional correlated states have been identified in angle-aligned (near zero twist angle) WSe$_2$/WS$_2$ heterostructures. We define ±n$_0$ to be one electron (+n$_0$) or one hole (-n$_0$) per moiré unit cell, which is half filling of the moiré miniband (full filling is two electrons or two holes per moiré unit cell since the TMD heterostructure has a degeneracy of two from spin-valley locking). Taking advantage of strong light-matter interactions in TMDs, E. Regan *et al.*[540] utilized an optically detected resistance and capacitance (ODRC) technique to identify not only the Mott insulating state at $n$ = -n$_0$ ($n$ is carrier density), but also surprising insulating states at -1/3n$_0$ and -2/3n$_0$ [Figure 29(c), 29(d), 29(e) and 29(f)], which they assigned to generalized Wigner crystallization on the underlying lattice. The presence of these generalized Wigner crystal states revealed the need for an extended Hubbard model that included long-range interactions. Yang Xu *et al.*[541] identified nearly two dozen correlated insulating states in WSe$_2$/WS$_2$ heterostructures at fractional fillings by designing an optical sensing technique that used a monolayer of WSe$_2$ separated from the heterostructure by a thin h-BN spacer [Figures 29(g) and 29(h)]. The excitons in the monolayered WSe$_2$ were very sensitive to their dielectric environment, where their resonance energy and oscillator strength could be used to measure correlated states in the WSe$_2$/WS$_2$ in close proximity. They proposed the different insulating states were due to charge-ordering using classical Monte Carlo simulations, including generalized Wigner crystal states and charge density waves. A scanning microwave impedance microscopy technique was utilized by X. Huang *et al.*[542] to identify many correlated insulating states at fractional fillings, some even persisting up to 120 K. With Monte Carlo simulations, they also proposed the charge ordering of the states and revealed surprisingly strong, long-range interactions exceeding nearest-neighbor interactions.

Other works have also reported the observation of numerous correlated insulating states at fractional fillings using scanning microwave impedance microscopy[542] and optical anisotropy/electronic compressibility measurements[543]. Table 1 summarizes the correlated insulating states observed in recent work. The optical techniques recently used/developed to study correlated insulating states as a function of electric field-tuned carrier density in these TMD heterostructures circumvent the large contact resistance (due to the large bandgap) that hinders direct electrical transport measurements, especially for low temperatures and low carrier densities.

**TABLE I.** Correlated insulating states observed in twisted-TMD structures

| Heterostructure | $n$ of correlated insulating state | Assignment of correlated insulating state |
|---|---|---|
| tWSe$_2$ ($\theta$ = 4°, 4.2°, 4.5°, 4.9°, 5.1°) | $n$ = -1n$_0$ for all [534] | Mott insulating phase |
| | $n$ = -2n$_0$ for 4.2° and 4.5° [540] | Full filling of the moiré subband |
| WSe$_2$/WS$_2$ ($\theta$ ≈0°, ≈60°) | $n$ = +1n$_0$ [541–543], -1n$_0$ [536,540–542] | Mott insulating state [536,540,541,543] with AFM Curie-Weiss behaviour [536] |
| | $n$ = -0.6n$_0$ [536] | Possible phase transition from AFM to weak FM state [536] |
| | $n$ = 1/6n$_0$ [542] | Stripe phase [542] |
| | $n$ = +1/4n$_0$ [541–543], -1/4n$_0$ [541,542] | Generalized Wigner crystal states [541], triangular phase [542], stripe crystal state [543] |
| | $n$ = +1/3n$_0$ [541–543], -1/3n$_0$ [540–542] | Generalized Wigner crystal states [541], triangular phase [542], isotropic electron crystals [543] |
| | $n$ = +2/5n$_0$ [541,543], -2/5n$_0$ [541] | Close to CDW state [541] stripe crystal state [543] |
| | $n$ = +1/2n$_0$ [541–543] | Close to generalized Wigner crystal states [541], stripe phase [542], stripe crystal [543] |
| | $n$ = -1/2n$_0$ [541,542] | Stripe phase [542] |
| | $n$ = +5/9n$_0$ [542] | |
| | $n$ = +3/5n$_0$ [541], -3/5n$_0$ [541] | Close to CDW state [541] stripe crystal state [543] |
| | $n$ = 2/3n$_0$ [541–543], -2/3n$_0$ [540–542] | Generalized Wigner crystal state [540,541], isotropic electron crystals [543] |
| | $n$ = +3/4n$_0$ [541,542] | Close to generalized Wigner crystal states [541] |
| | $n$ = -3/4n$_0$ [541,542] | Close to CDW state [541] |
| | $n$ = -7/9n$_0$ [542] | |
| | $n$ = -5/6n$_0$ [542] | |
| | $n$ = +6/7n$_0$ [542] | |
| | $n$ = -8/9n$_0$ [542] | |
| | $n$ = +4/3n$_0$ [541], -4/3n$_0$ [541] | |
| | $n$ = +3/2n$_0$ [541,542], -3/2n$_0$ [542] | |
| | $n$ = +5/3n$_0$ [541] | |
| | $n$ = +2n$_0$ [541,542], -2n$_0$ [536,541] | Complete filling of conduction miniband [541,542] |

### *3. Future Directions in moiré Magnets and Magnon Bands*

As was discussed above in Subsection IV B, 2D magnetism has been a rapidly growing research area and provides an additional platform to study moiré effects. It has been shown that magnons, which are quantized magnetic excitations, can be used in magnetic insulators to transfer spin angular momenta without the movement of charge, with potential applications in future magnonic devices. It is natural to wonder if twisting two magnetic layers on top of each other would result in tunable magnetic phenomena and magnons, and what that would mean for transport phenomena, magnon-magnon or magnon-phonon interactions, or if nontrivial magnon bands can be generated. In addition, if flat magnon bands do exist, the physics governing them would not be a straightforward comparison to flat bands in electronic systems, considering that magnons are bosons and statistically distinct from electrons. Within the last year, researchers have started to delve into these questions with theoretical predictions of how magnetic phenomena is tuned by the relative twist angles between two magnetic layers, showing the potential for exotic phenomena such as flat and topological magnon bands[544–546].

K. Hejazi *et al.*[544] have provided a theoretical framework to study moire structures of 2D magnets that allowed for an understanding of magnetic structures and excitations without the need for quasiperiodicity. As an example, they performed detailed calculations for the case of a Néel AFM on a honeycomb lattice, such as $MnPS_3$ or $MnPSe_3$, showing that twisting of two of these layers could lead to noncollinear spin textures depending on the twist angle and parameters such as the spin stiffness, uniaxial anisotropy, and the interlayer exchange interaction. In the noncollinear phase and at small twist angles, they also predicted some of the magnons will have flat dispersion curves. For twisted bilayered $CrI_3$, they input experimentally determined parameters to predict that three separate spin ordering phases could be found in this heterostructure depending on the twist angle.

Magnons and their dispersions for twisted magnetic bilayers have been theoretically investigated for both collinear FM[545,546] and AFM ordering[545], predicting flat magnon bands and non-zero Chern numbers for a continuum of twist angles. In particular, the presence of the Dzyaloshinskii-Moriya (DM) interaction in 2D magnets with broken inversion symmetry[547] plays a key role in determining the magnon properties in the twisted bilayer magnet. For two FM layers and with negligible DMI, the situation was analogous to tBLG, with discrete magic angles where the moiré magnon bands are flat. The introduction of even a weak DMI resulted in numerous topological flat magnon bands over a continuum of angles, as opposed to discrete "magic" angles, where they estimated the topological nature of the bands should be present for angles less than 10°. The magnon bands valley Berry curvature, valley Chern numbers, and the valley thermal magnon Hall and Nernst conductivities could be tuned *via* the twist angle. Y.-H. Li and R. Cheng[545] extended this to include four different cases of interlayer and intralayer spin ordering, finding that the flat magnon bands and angle-tunable topological Chern numbers are not limited to FM bilayer magnets. If experimental works can verify the topological nature of the magnon bands through the thermal magnon Hall response, we expect moiré magnets to open an intriguing playground where twist-angle could serve as a knob to tune topology. Since magnons and magnon transport are tunable *via* the application of an electric field, it will be interesting to see how modifying moiré magnets will lead to new opportunities.

## V. CONCLUSIONS

This review provides a survey and future outlook of the multitude of compelling electronic and optical properties that can be electrically tuned in 2D materials beyond graphene. The atomically thin nature of individual layers that can be stacked and twisted into synthetic heterostructures provide unique opportunities for the creation of a variety of devices for applications that span memory and logic to energy harvesting and sensing. The review also provides figures-of-merit and benchmarking targets to highlight exciting areas of opportunity that will require further research and development efforts in both the device structure and materials design. With the opening of the design space to new synthetic 2D materials beyond what has been covered in this review, it will be possible to realize uncharted physical properties of materials that will surely lead the re-imagining of functionalities in existing electrically tunable technologies. There is a large movement in the community to implement these materials in electronics, particularly with the success in the development of wafer scale growth schemes and the fabrication of in-plane, out-of-plane and mixed dimensional heterostructures. In addition, there has been a significant advancement in large scale wafer transfer and integration schemes with conventional bulk materials. What will bring these 2D materials into the forefront of next generation technology is the ability to not only control the defect density and conductivity type of each individual layer during synthesis, but also to synthesize any polymorph phase and stacking order with desired twist angles into complex heterostructures, particularly in strongly correlated systems. This will lead to new exotic phenomena that can be electrically reconfigured and tuned on demand for applications that have yet to be identified. These 2D materials can also withstand high levels of mechanical strain and deformation which will serve useful in electrically tunable flexible devices that integrate interconnects, active and passive components all based on 2D materials. This review highlights several representative and noteworthy research directions in the use of electrical means to tune the aforementioned physical and structural properties, with an emphasis on discussing major applications of these materials in devices in the past few years. Moreover, the coupling of electrical with external optical and magnetic means will provide additional degrees of device tunability and modulation to further enhance functionalities in these atomically thin materials. This will enable access to stronger signatures of quantum phenomena and new regimes of electronic and optical property control which have only begun to be implemented and explored.

**ACKNOWLEDGEMENTS**

This project has received funding from the European Union's Horizon 2020 research and innovation program under grant agreement GrapheneCore3 785219 number. OK TV, EC acknowledge support from the Department for Business, Energy and Industrial Strategy through NMS funding (2D Materials Cross-team project). DJ acknowledges primary support from the Air Force Office of Scientific Research (AFOSR) contracts FA9550-21-1-0035 and FA2386-20-1-4074 and U.S. Army Research Office (ARO) contract W911NF-19-1-0109. D.J. also acknowledges partial support from National Science Foundation (NSF)-funded University of Pennsylvania Materials Research Science and Engineering Center (MRSEC)(DMR-1720530) and NSF DMR-1905853. S.S. acknowledges support from Fulbright-Nehru Postdoctoral Fellowship supported by the US-India Educational Foundation. ZA acknowledges support of this work by the Laboratory Directed Research and Development Program of Lawrence Berkeley National Laboratory under U S Department of Energy Contract No. DE-AC02- 05CH11231.

Tom Vincent: https://orcid.org/0000-0001-5974-9137
Eli G. Castanon: https://orcid.org/0000-0003-4316-4796
Amber McCreary: https://orcid.org/0000-0002-9988-5800
Olga Kazakova: https://orcid.org/0000-0002-8473-2414
Deep Jariwala:  https://orcid.org/0000-0002-3570-8768
Zakaria Y Al Balushi: https://orcid.org/0000-0003- 0589-1618

## DATA AVAILABILITY

Data sharing is not applicable to this article as no new data were created or analyzed in this study.